\documentclass[useAMS,usenatbib]{mn2e}
\usepackage{epsfig}
\usepackage{natbib}
\usepackage{soul}
\usepackage{color}
\title[Determining the Age of Young Embedded Clusters]{Determining the Age of Young Embedded Clusters}
\author[]{J.J. Stead$^{1}$\thanks{E-mail:
phy2j2s@leeds.ac.uk; m.g.hoare@leeds.ac.uk}, M.G. Hoare$^{1}$\footnotemark[1]\\
$^{1}$School of Physics and Astronomy, University of Leeds, Leeds, LS2 9JT\\
}

\def\Bes{\rm Besan$\c{c}$\rm on}
\def\arcmin{^{\prime}}
\begin{document}

\date{}

\pagerange{\pageref{firstpage}--\pageref{lastpage}} \pubyear{2002}

\maketitle

\label{firstpage}

\begin{abstract}
A new Monte Carlo method has been developed in order to derive ages of young embedded clusters within massive star forming regions where there is strong differential reddening. After foreground and infrared excess source candidates are removed, each cluster candidate star is individually dereddened. Simulated clusters are constructed using isochrones, an IMF, realistic photometric errors, simulated background field populations and extinction distributions. These synthetic clusters are then dereddened in the same way as the real data, obtained from a deep near infrared survey, and used to derive the ages of 3 embedded clusters. Results were found to be consistent with those determined using spectrophotometric methods. This new method provides way to determine the ages of embedded clusters when only photometric data are available and there is strong differential reddening.
\end{abstract}


\section{Introduction}
\label{sec:intro}

Stellar clusters are born embedded within giant molecular clouds (GMCs) and so they are often obscured by many magnitudes of visual extinction. The study of the youngest objects using optical data is often impossible, and therefore infrared data are required to penetrate the natal cloud. Clusters contain groups of stars, covering a large range of stellar masses, confined to a small volume of space. As all cluster members have been formed more or less simultaneously from the same precursor cloud \citep{lada03}, their study plays an important role in understanding many aspects of stellar evolution. \\
$\indent$Theoretical isochrones, developed using stellar models and the current understanding of pre-main sequence (PMS) evolution, can be used to estimate cluster ages in several different ways. Such an approach is possible as pre-main sequence, main sequence, giant and supergiant stars have very different intrinsic magnitudes and colours, therefore they occupy very different positions on colour-magnitude diagrams (CMDs). The age of a cluster can therefore be estimated by fitting theoretical isochrones to real data in colour-magnitude space.\\
$\indent$The life cycle of a typical cluster may include some or all of the following steps. First a cluster of stars form after the gravitational collapse of a dusty molecular cloud. Gas and dust will fragment throughout the cloud and will eventually form individual stars and systems of stars. In the earliest stage, most stars may still be accreting material and will therefore be obscured by many magnitudes of visual extinction. The cluster at this stage is referred to as an embedded cluster. As the most massive young stars join the main sequence, the subsequent intense UV radiation will begin to ionise the surrounding gas to form an HII region. Through a combination of winds, massive outflows and expanding HII regions, the surrounding natal cloud is slowly dispersed. Once dispersed, all cluster members will have stopped accreting material, may be visible optically and the final mass of all stars in the cluster is set. Such a cluster is referred to as an open cluster.\\
$\indent$Many authors \citep{bica04,ortolani05,bonatto09} use the theoretical intrinsic magnitudes and colours from computed isochrones to make comparisons with real photometry on apparent colour-magnitude diagrams (CMDs). As open clusters have existed long enough to have dispersed their natal clouds, it can be assumed that all cluster members will suffer from a similar amount of extinction as there is very little, differential reddening between cluster members. \citet{naylor06} developed an isochrone fitting technique that can be applied in to clusters where a constant amount of reddening can be assumed. More recently work by \citet{dario10} developed a CMD fitting technique that does allow for some differential reddening between cluster members. They applied their method to pre-main sequence stars in the Magellanic Clouds using Hubble Space Telescope data. However in deeply embedded clusters, due to surrounding gas/dust, the extinction towards cluster members can vary over tens of visual magnitudes. This is particularly true towards massive star forming regions where the extinction inside a cluster can be highly differential. \\
$\indent$The extreme differential reddening, between embedded cluster members, means that it is often impossible to determine the age of an embedded cluster using the traditional techniques of comparing apparent colours and magnitudes with isochrones. However, as dense young clusters probably have nearly all their original stellar population, and low mass members are brighter than at any other time during their main sequence evolution so are easier to count \citep{lada03}, young clusters offer unique opportunities to study the initial mass function (IMF) and its variation in space and time. Despite the difficulty, for these reasons the study of young embedded clusters are considered very worthwhile. \\
$\indent$Ideally we would spectral type each star in a field of view as the spectral typing of stars can remove the problem of differential reddening. This is because if the spectral type is known, a star can be placed directly onto a Hertzsprung-Russell diagram in order to derive the age. In practice however, due to the observational time required, it is not possible to obtain spectra for the hundreds of cluster members in each of the thousands of clusters contained within the Milky Way. It is however possible to obtain photometry of all of these hundreds of thousands of cluster members, using deep survey data such as the United Kingdom Infrared Telescope Deep Sky Survey (UKIDSS) Galactic Plane Survey \citep[GPS,][]{lucas08}, and then apply techniques developed using spectroscopy in a systematic fashion to determine parameters such as cluster distance and age.\\
$\indent$In this paper an original approach is used to determine the ages of three embedded clusters, via Monte Carlo simulations, whereby real photometry is dereddened in colour-colour space and then compared in dereddened colour-magnitude space with the photometry of synthetic clusters, generated from theoretical isochrones. In section \ref{sec:data} we describe the sources of the real data used in this paper. In section \ref{sec:isochrones_magscolours} we discuss the isochrones used in this paper and the choice of intrinsic colours. In section \ref{sec:artif_clus} we generate an artificial cluster in order to highlight certain cluster properties and to test the dereddening method used in this paper. In section \ref{sec:deredProcess} the dereddening process is described. In section \ref{sec:AgeDet} the age determination process is described and applied to the artificial cluster generated in section \ref{sec:artif_clus}. The age determination process is extended to three real embedded clusters in section \ref{sec:real_cluster_ages} and we conclude our results in section \ref{sec:conclusion}. 

\section{Data}
\label{sec:data}

\subsection{The UKIDSS GPS}

The UKIDSS GPS covers the region of the Galactic Plane, accessible by the United Kingdom Infrared Telescope (UKIRT); $15^{o} < l < 107^{o}$ and 141$^{o} < l < 230^{o}$, $|$b$|$ $< 5^{o}$ and $-2^{o} <$ l $< 15^{o}$, $|$b$|$ $< 2^{o}$, in the J(1.248$\mu$m), H(1.631$\mu$m) and K(2.201$\mu$m) filters. Classically JH\&K is taken to be JH\&K on the Johnson-Glass system. However, unless otherwise explicitly stated, JH\&K in this paper refers to the UKIDSS photometric system \cite[for the individual filter profiles and transformation equations to other systems, please see][]{hewett06}. \\
$\indent$The median 5$\sigma$ depths (Vega system) are J=19.77, H=19.00 and K=18.05 in the second data release of the GPS \citep{warren07}. However, the survey depth is spatially variable and typically decreases longitudinally towards the Galactic centre and latitudinally
 towards the Galactic Plane, due to crowding caused by the higher density of stars found towards these regions of the Galaxy. The modal depths in uncrowded fields, determined by \citet{lucas08}, are J=19.4 to 19.65, H=18.5 to 18.75 and K=17.75 to 18.0. Finally, the GPS data are obtained from the WFCAM Science Archive \citep{hambly08}. 

\subsubsection{Point Spread Function Fitting Photometry}

The regions of sky containing young embedded clusters are some of the most crowded regions within the GPS. Furthermore, such regions are likely to contain highly spatially variable nebulosity. For these regions we have performed our own Point Spread Function (PSF) photometry on the UKIDSS images using the IRAF DAOPHOT program. DAOPHOT specialises in obtaining photometric measurements of stars in crowded fields. Zeropoint and aperture correction are not required as the DAOPHOT photometry has been bootstrapped to the UKIDSS photometry of stars in less crowded regions of the same field. \\
$\indent$To test the reliability of the DAOPHOT procedure, for a given region, the residual flux, i.e. the remaining flux left over after a given star has been subtracted from the UKIDSS image using DAOPHOT, was measured as a fraction of the original flux as a function of measured magnitude. For the majority of the stars the fractional difference did not contribute significantly to the measured photometric error, suggesting that the DAOPHOT program performed adequately. \\
$\indent$Performing the PSF fitting was deemed necessary as, in the most crowded UKIDSS fields, many overlapping stars are not present in the UKIDSS catalogue after data cuts have been applied to select the most reliable data \citep[][contains a list and short description of each of the aforementioned data cuts]{lucas08}. 

\subsubsection{The RMS Survey}

In this paper we determine the age of three embedded clusters. Each cluster has been selected because it contains an object in the Red MSX Source (RMS) survey \citep{urquhart07}. The RMS survey, with its primary basis from the Midcourse Space Experiment (MSX), is an ongoing multi-wavelength observational programme designed to return a well-selected sample of massive young stellar objects (MYSOs), and ultra compact (UC) HII regions. Initially, a large sample of MYSO/UCHII candidates were colour-selected from the MSX point source catalogue \citep{egan03} by \citet{lumsden02}.\\
$\indent$Each source in the RMS survey has a computed kinematic distance from $^{13}$CO observations \citep{urquhart08}, and these have been used to determine the luminosity of each MYSO candidate \cite{mottram11a}. Kinematic distances are derived by applying a Galactic rotational model to measured kinematic velocities. Excluding localised velocity perturbations, the kinematic velocity is the projection of the orbital velocity of a molecular cloud, about the Galactic centre, along the line of sight. Therefore for clouds within the Solar circle, there is not a unique solution to the derived distance and such clouds will possess a near/far kinematic ambiguity. One of the clusters in our sample is well documented in the literature and has a, independently measured, spectrophotometric distance. The remaining two clusters also have independently measured distances, derived from extinction measurements, presented in \citet{stead10}. It is these non-kinematic cluster distances that have been adopted for this paper. \\
$\indent$As each cluster has been selected from the RMS survey, each will contain at least one massive star ($>$8 M$_{\odot}$).

\subsection{Choice of Intrinsic Colours}
\label{choiceofcolours}

$\indent$For the purpose of the work in this paper a new set of UKIDSS intrinsic colours has been determined from previous infrared observations. Several hundred stars of every spectral sub-type in the table of the MK spectral types from \citet{Allen00}, over the range O8V to M0V, were extracted from the All-Sky Compiled Catalogue \citep[ASCC;][]{kharchenko09}. The spectral classes were obtained from several catalogues, most notably the Tycho-2 Spectral Type catalogue \citep[compiled by][]{wright03}. The spectral sub-types of \citet{Allen00} were used so the sample of stars selected also have reliable optical intrinsic colours. Knowledge of the intrinsic colours makes it possible to determine the amount of reddening each star suffers, specified as E(B-V). \\
$\indent$For each star a value of A$_J$ is determined from the derived value of E(B-V) using the ratio A$_J$/A$_V$=0.2833, calculated using the \citet{cardelli89} extinction curve with R$_V$=3.1 (obtained from the Trilegal website http://stev.oapd.inaf.it/cgi-bin/trilegal). Next, a spectral type specific reddening track is used to compute corresponding values of A$_H$ and A$_K{_S}$. These extinction values are used to deredden the 2MASS apparent magnitudes of each star in order to determine their intrinsic magnitudes. Stars are grouped into their respective spectral sub-types and plotted on an H-K vs J-H CCD. The mean H-K and J-H position of each spectral sub-type is then computed. The data are continuously clipped at the 2$\sigma$ level until no further reductions are made. The final number of stars in each sample is presented in Table \ref{tabz:mainsequence} alongside these new intrinsic colours. These new 2MASS colours are then converted to UKIDSS colours using appropriate colour transformation equations. \\
$\indent$Stellar spectra can be convolved through specific photometric filters in order to derive a set of colour transformation equations. For the colour equations presented in this paper, the stellar models of \citet[][hereafter CK04]{castelli04} have been convolved through the UKIDSS and 2MASS filter profiles. Synthetic photometry is therefore produced for the JHK$_{UKIDSS}$ and JHK$_{2MASS}$ filters. Colour equations can then be derived by creating CCDs in the format (2MASS$_{MAG_1}$ - 2MASS$_{MAG_2}$) vs (UKIDSS$_{MAG_{1or2}}$ - 2MASS$_{MAG_{1or2}}$). In order to compute the colour transformation equations, lines of best fit are then made to intrinsic data on each CCD. On an H-K vs J-H CCD two distinct trends in the data form, splitting at H-K=0. It is for this reason that two sets of colour equations have been derived. Only stars as late as M0V have been considered. The resultant equations are presented below: \\
\\
For blue stars with H-K $<$ 0\\
J$_{UK}$ = J$_{2M}$ -- 0.042$\pm$0.010 * (J$_{2M}$--H$_{2M}$) -- 0.003$\pm$0.005 \\
H$_{UK}$ = H$_{2M}$ + 0.255$\pm$0.102 * (H$_{2M}$--K$_{2M}$) -- 0.001$\pm$0.004 \\
K$_{UK}$ = K$_{2M}$ + 0.113$\pm$0.109 * (H$_{2M}$--K$_{2M}$) -- 0.002$\pm$0.004 \\
\\
For red stars with H-K $>$ 0\\
J$_{UK}$ = J$_{2M}$ -- 0.025$\pm$0.065 * (J$_{2M}$--H$_{2M}$) + 0.000$\pm$0.007 \\
H$_{UK}$ = H$_{2M}$ + 0.042$\pm$0.091 * (H$_{2M}$--K$_{2M}$) + 0.000$\pm$0.007 \\
K$_{UK}$ = K$_{2M}$ -- 0.040$\pm$0.091 * (H$_{2M}$--K$_{2M}$) + 0.000$\pm$0.007 \\
\\

\subsection{The Stellar Population Synthesis Model of the Galaxy}
\label{sec:modelcdd-Aj}

Work in this paper makes use of the Stellar Population Synthesis Model of the Galaxy, developed in $\Bes$ (hereafter referred to as $\Bes$ data). \citet{robin03} contains a complete description of the $\Bes$ model inputs, however we summarise it here for completeness. Four populations of stars are modelled, the thin disc, thick disc, bulge and spheroid, and each is described by a star formation rate history, an initial mass function, an age or age range, metallicity characteristics, kinematics, a set of evolutionary tracks, and includes a white dwarf population. The extinction, modeled in terms of visual magnitudes per kiloparsec (mag kpc$^{-1}$), is caused by a diffuse thin disc. It is also possible to insert discrete clouds with a specified A$_V$ and distance. Using the extinction-distance relationships provided by \citet{marshall06}, the distribution of giant stars is modelled, with respect to both distance and extinction, along specific lines of sight. 

\subsection{Isochrones}
\label{sec:isochrones_magscolours}

\citet[][hereafter S00]{Siess00} use stellar models to determine how effective temperature and luminosity vary with age and mass. They model PMS stars in the mass range 0.1 to 7.0 M$_{\odot}$. These values are converted to near infrared intrinsic magnitudes and colours, presented in the Johnson-Cousins-Glass (JGC) system, using a conversion table presented in \citet{kenyon95}. The largest star modelled in the S00 isochrones is 7.0 M$_{\odot}$ star. Such a star has a spectral type B3 upon arrival on the zero-age main sequence  (ZAMS). S00 define the ZAMS to be the point in the life of a star when a minimum of 99\% of the stellar luminosity is provided by nuclear fusion. Only modelling stars with masses below 7.0 M$_{\odot}$ is not sufficient for the purpose of this paper considering that each cluster contains an RMS candidate. \\
$\indent$The isochrones of \citet[][hereafter G02]{girardi02} contain much more massive stars, but unlike the isochrones of S00, the G02 isochrones do not model pre-main sequence stars, and furthermore the youngest G02 isochrone available has an age of 4.0 Myr. Work in this paper uses the varying luminosities of pre-main sequence stars to determine the age of young clusters. However, the 4.0 Myr isochrone agrees well with the ZAMS track of S00, so assuming the larger mass stars have arrived on the ZAMS, we use the G02 isochrones to provide a way to analyse stars with masses larger than 7.0 M$_{\odot}$.  \\
$\indent$To combine the G02 4.0 Myr isochrone with a S00 isochrone requires estimating which stars in the S00 isochrone, of a particular mass, have and have not joined the ZAMS. This is not an issue for S00 isochrones with ages $<$0.5 Myr, as the most massive, 7.0 M$_{\odot}$, star has colours and magnitudes consistent with a 7.0 M$_{\odot}$ ZAMS star from
 an age of 0.5 Myr onwards. Note the  7.0 M$_{\odot}$ star does not reach the ZAMS until $\sim$1.15 Myr, however the colours and magnitudes differ by $<$0.01 mag between this age and 0.5 Myr. For all of the combined isochrones with ages as young as 0.2 Myr, the path between the S00 7.0 M$_{\odot}$ star and the G02 9.0 M$_{\odot}$ star is interpolated. However, the youngest combined isochrone used in this paper has an age of 0.1 Myr. For such an isochrone, it is the path between the S00 7.0 M$_{\odot}$ star and the G02 20.0 M$_{\odot}$ star that has been interpolated.\\
$\indent$The colours of the S00 1 Myr isochrone and the S00 ZAMS track are found to be very similar when plotted on a colour-colour diagram. This is despite the fact that the stellar luminosities of young stars vary dramatically with age. The reason for this is because \citet{Siess00} use a single temperature-colour relationship from \citet{kenyon95} and do not account for the different surface gravities. In reality the M stars are likely to have colours more similar to those of giant stars than those of the main sequence dwarfs. What effect choosing a different temperature-colour relationship will have on the results is also discussed in more detail later (section \ref{sec:AgeDetSyn}). \\
$\indent$As previously mentioned, the intrinsic colours of the S00 isochrones are presented the JGC system. The colours must therefore be converted to the UKIDSS photometric system for a reliable comparison with the observed data. This conversion process was done using the same method as in section \ref{choiceofcolours}, whereby the 2MASS colours were converted to the UKIDSS system, with one exception. It is not possible to produce JHK colour equations to convert the isochrones into the new UKIDSS colours using the new data alone. This is because a series of new colours have been generated, as opposed to independent intrinsic magnitudes which are required to produce the colour equations. Later in the paper J-H vs K CMDs will be used to derive the ages of clusters. For this reason the K$_{JGC}$ magnitude of the young stars can be converted to a K$_{UKIDSS}$ magnitude using the same process as in section \ref{choiceofcolours}. This K$_{UKIDSS}$ intrinsic magnitude can then be used to determine the individual J and H magnitudes from the new intrinsic colours. The K filter profile presented in \citet{bessell88} is used to make this first transformation between the JGC and UKIDSS systems. Next, stars of the same spectral type are matched between the new UKIDSS intrinsic colours and the S00 isochrones. In a similar manner as before, a second set of equations have been derived and are presented below:  \\
\\
For blue stars with H-K $<$ 0\\
K$_{UK}$ = K$_{JGC}$ + 0.001$\pm$0.103 * (H$_{JGC}$--K$_{JGC}$) -- 0.029$\pm$0.005 \\
J$_{UK_{(New)}}$ = J$_{JGC}$ -- 0.256$\pm$0.133 * (J$_{JGC}$--H$_{JGC}$) -- 0.035$\pm$0.010\\
H$_{UK_{(New)}}$ = H$_{JGC}$ + 0.088$\pm$0.111 * (H$_{JGC}$--K$_{JGC}$) -- 0.006$\pm$0.010\\
\\
For red stars with H-K $>$ 0\\
K$_{UK}$ = K$_{JGC}$ -- 0.045$\pm$0.098 * (H$_{JGC}$--K$_{JGC}$) -- 0.029$\pm$0.005\\
J$_{UK_{(New)}}$ = J$_{JGC}$ -- 0.197$\pm$0.019 * (J$_{JGC}$--H$_{JGC}$) -- 0.015$\pm$0.007\\
H$_{UK_{(New)}}$ = H$_{JGC}$ + 0.028$\pm$0.019 * (H$_{JGC}$--K$_{JGC}$) -- 0.009$\pm$0.006\\
\\

\section{The Dereddening Process}
\label{sec:deredProcess}

\begin{figure}
\begin{center}
\begin{tabular}{cc}
\resizebox{65mm}{!}{\includegraphics[angle=0]{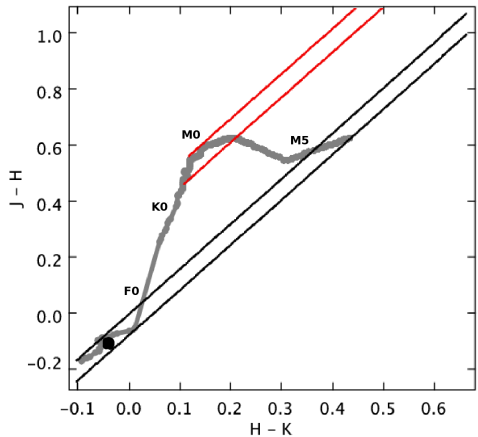}}
\end{tabular}
\caption[]{\small A schematic diagram illustrating the dereddening method. Several different coloured A0V reddening tracks have been used to illustrate how dereddened stars intersect the intrinsic colour locus (see text). In this particular case a 2 Myr isochrone has been used to illustrate the intrinsic colours.}
\label{fig:deredschem}
\end{center}
\end{figure}

Individual cluster members are dereddened, along spectral type specific reddening tracks \citep{stead09}, to their point of intersection with the intrinsic colour locus. This section details this process. The slope of the intrinsic colour locus relative to the reddening tracks, that are used to deredden the individual cluster members, gives rise to ambiguities and uncertainties in specific regions. \\
$\indent$As spectral type specific reddening tracks are used to deredden individual stars, such a process therefore requires an estimate of the spectral type of each star. To do this, first every star in a sample is dereddened along an A0V reddening track to the point of intersection with the intrinsic locus which provides an estimation of spectral type. The length of the track traversed is related to the amount of extinction suffered by the young star. The difference between using a B0V reddening track in preference to an M0V reddening track, at a mean extinction of A$_J$=4.3 mag, corresponds to a difference of E(H-K)$\sim$0.03 mag, which is a small but systematic error. \\
$\indent$Fig. \ref{fig:deredschem} contains a schematic diagram illustrating the dereddening process. For early-type stars, as there are two points of intersection with the intrinsic locus, each star is instead dereddened to a single point on the locus, created by averaging the intrinsic colours over the range of stars for which there are two possible points of intersection. Any star with a spectral type earlier than A5V possesses this ambiguity. \\
$\indent$Due to the shape of the late type sequence there is also a second ambiguity when dereddening the lowest mass stars, occurring because stars with spectral types later than M1V begin to curve back on the intrinsic locus. Such stars are dereddened to the earlier part of the intrinsic locus, thereby overestimating the amount of extinction suffered which transforms into unrealistically small  J-H colours (see section \ref{sec:AgeDet}, Fig. \ref{fig:sudoCMD}). However, due to their low masses, such stars are very faint and are therefore unlikely to be detected in UKIDSS data. \\
$\indent$The slope of the locus covering stars with spectral types between K2 and M0, contained within the two red A0V reddening tracks in Fig. \ref{fig:deredschem}, is very similar to the slope of the A0V reddening tracks. For this reason, the point of intersection made with the intrinsic locus is less certain. Finally, stars whose photometric errors move their position on the CCD above the upper reddening track are dereddened to a line perpendicular to the reddening track, placed at the upper tip of the intrinsic locus (dashed line). \\
$\indent$The dereddening of individual cluster members makes it possible to determine the intrinsic colours and magnitudes of each star, if the distance is known. Therefore through dereddening, it is possible to compare the measured intrinsic values with those of the theoretical isochrones in colour-colour space, which in turn allows the age of the embedded cluster to be determined.

\section{Artificial Clusters}
\label{sec:artif_clus}

In this section we create an artificial cluster to test the robustness and determine the limitations of the methods, presented in following sections, that are used to deredden the individual cluster members and determine cluster ages, in a controlled manner.
 
\subsection{Initial Cluster Parameters}
\label{sec:intial_cluster_param}
 
To create a realistic artificial cluster, several cluster parameters must be modeled accurately. These include a realistic estimation of the cluster size and description of the stellar distribution, or initial mass function, and a realistic estimation of the differential reddening. The former allows for accurate modeling of the intrinsic colours and magnitudes, the latter allows for accurate modeling of the apparent colours and magnitudes. \\
$\indent$The IMF of a cluster describes how the mass contained within a cluster was distributed at the time of formation. The IMF can be described using the following equation that allows for the calculation of the number of stars, N, in a given mass range:
\begin{equation}
\int dN = \int_{m_1}^{m_2}c.m^{-\alpha} dm,
\label{eq:YSOno1}
\end{equation}
where c is a constant and $\alpha$ describes the IMF in terms of a power law. The literature contains work on the IMF by numerous authors that often describes $\alpha$ as a function of mass, i.e. the power law changes depending on the mass range studied. \citet{salpeter55} derived a power law for stars with masses greater than 0.4 M$_{\odot}$ of $\alpha$=2.35 and this value has changed very little over the past few decades. The power law used in this thesis is that of \citet{kroupa01} that assumes $\alpha$=2.3 for masses greater than 0.5 M$_{\odot}$ and $\alpha$=1.3 for masses below. The power law of \citet{kroupa01} changes again for stars with masses lower than 0.08 M$_{\odot}$, however such stars would not be observed in a typical embedded cluster, using UKIDSS data, and so they are of no consequence to work in this paper. \\
$\indent$As stars form from the collapse of a clumpy molecular cloud, the concentration of material, and therefore the line of sight extinction, will be variable over small spatial scales. Furthermore, depending on whether or not stars are at the near or far side
 of the cluster, the amount of extinction suffered by neighbouring cluster members will be very different. To replicate the effect of differential extinction, a realistic skewed Gaussian cluster extinction distribution with A$_J$=3.1$_{-0.2}^{+0.6}$ mag has been applied to the artificial clusters created in this section. However the extinction distribution applied has been clipped at A$_J$=2.8 to replicate the sudden jump in extinction along the line of sight to the cluster (see section \ref{sec:G042.1268-0.6224_cluster}). Finally, the cluster has been assumed to be at a realistic distance of 4.3 kpc (section \ref{sec:G042.1268-0.6224_cluster}).

\subsection{Production of an Artificial Cluster}
\label{sec:production_artif_clus}

\begin{figure*}
\begin{center}
\begin{tabular}{cc}
\resizebox{80mm}{!}{\includegraphics[angle=0]{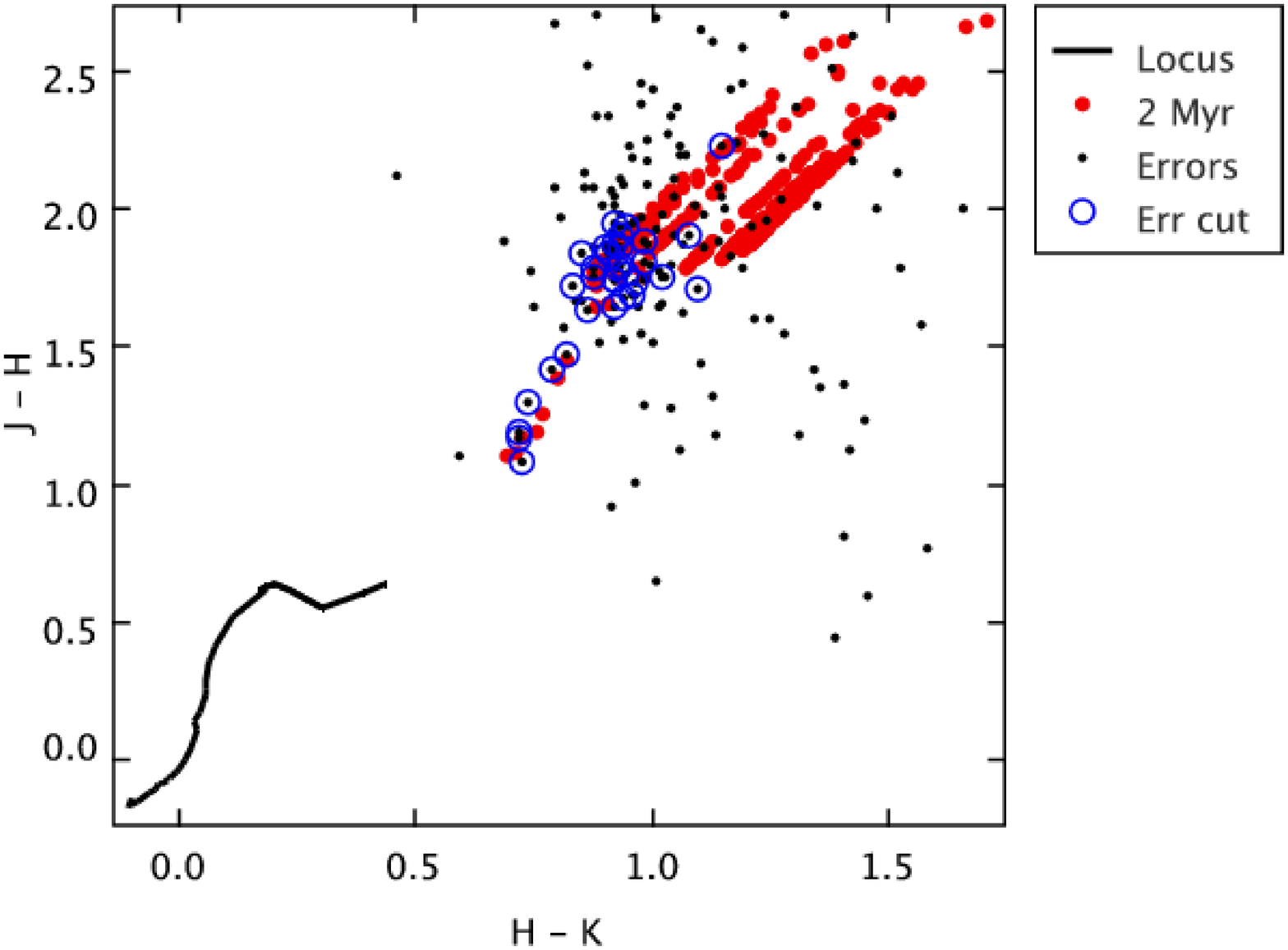}}&
\resizebox{80mm}{!}{\includegraphics[angle=0]{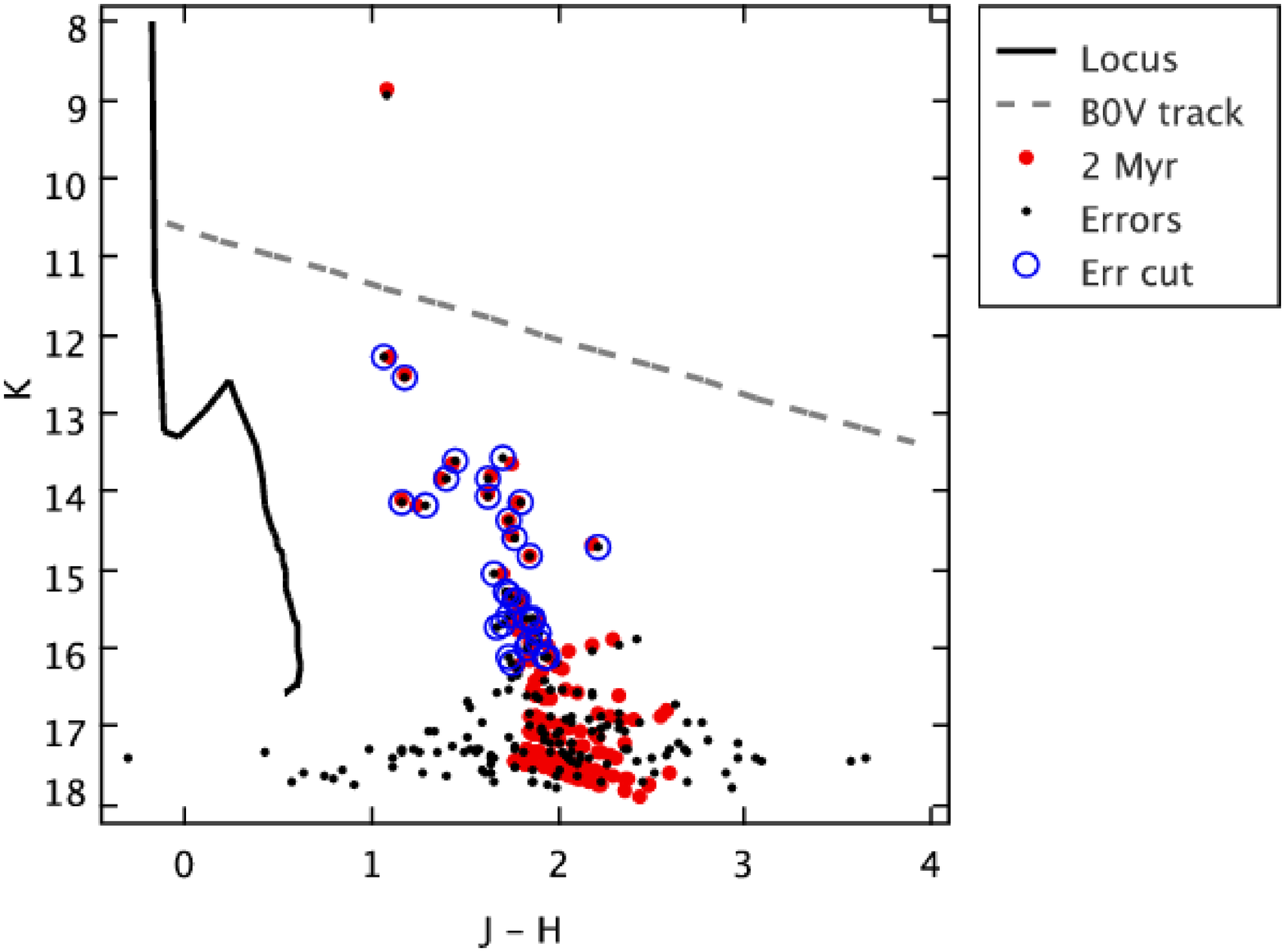}}
\end{tabular}
\caption[A 1 Myr artificial cluster with included photometric errors]{\small a (Top): A CCD of a 2 Myr artificial cluster illustrating the four different production stages described in the text. First intrinsic data (black line) are reddened using a realistic extinction distribution (red circles). Realistic photometric errors are applied to the apparent magnitudes (black dots) and because some very faint stars have apparent magnitudes so large that they would not be observed by UKIDSS, the assigned error causes a large scatter in the data. Finally these new magnitudes are clipped to be consistent with typical observed apparent magnitudes (blue circles) and they occupy realistic places on the CCD. b (Bottom): A J-H vs K CMD illustrating the same four production stages described in (a), however the intrinsic locus has had a distance of 4.3 kpc assumed. Also plotted as a dashed grey line is a B0V reddening track.}
\label{fig:sudoCCD}
\end{center}
\end{figure*}

In order to replicate the differential reddening of a real cluster, all artificial stars presented in this section have been reddened along spectral type specific reddening tracks using a reddening law with a slope of $\beta$=2.14 \citep{stead09}. Realistic observational photometric errors, determined from UKIDSS data, are also included in the derived apparent magnitudes. This is done by using a Gaussian distribution and an exponential function that describes the relationship between apparent magnitude and photometric error, e.g. Jerr = A + exp(B x Jmag - C), where A, B and C are coefficients specific to each photometric system. These new apparent magnitudes are then clipped by applying an initial photometric error cut of 0.1 mag in each band. As will be shown later, during the age determination process there is a tradeoff between the number of sources used to determine the fit and the reliability of the data used. Therefore the photometric cut shall be left as a free parameter when determining the ages of real clusters. For the worked example of this artificial cluster, the initial 0.1 mag cut has been chosen because, on an H-K vs J-H CCD, there is a $\sim$0.25 mag difference along the H-K axis between a B0V and an M0V reddening track placed at corresponding ends of the intrinsic data. Therefore the applied 0.1 mag photometric error cut allows a generic star, that is positioned in the middle of the B0V and M0V reddening tracks, to remain between both reddening tracks when the 0.1 mag photometric errors are considered. \\
$\indent$Fig. \ref{fig:sudoCCD} (a) and (b) contain a CCD and CMD of a 2 Myr artificial cluster illustrating the four different production stages described above. In the CCD the black line represents the intrinsic magnitudes, red circles are the reddened apparent magnitudes without errors, black points are the apparent magnitudes with the included magnitude specific photometric errors, and the blue circles are the clipped apparent magnitudes with included errors. Initially, from the assumption that there would be at least one star with a mass larger than 15.0 M$_{\odot}$, equivalent to a very early B-type star \citep{Allen00}, the artificial cluster contained 178 stars spanning a mass range 0.3 M$_{\odot}$ to 34.8 M$_{\odot}$, the mass of the largest star created. After the photometric error and saturation limit cuts are applied, the cluster contains 33 stars spanning a mass range 1.0 M$_{\odot}$ to what was originally the second largest star with a mass 11.0 M$_{\odot}$.

\section{The Age Determination Process}
\label{sec:AgeDet}

Due to the nature of the dereddening method it is not tractable to analytically determine photometric error bars on a dereddened CMD. If an artificial cluster can be created that can be considered a true representation of a real cluster, i.e. the correct number of observed stars, cluster IMF, the distribution of individual stellar extinction and photometric errors are modeled correctly, then it is possible to model the CMD photometric errors through the use of Monte Carlo simulations. Therefore the age of a cluster can be constrained by comparing model and real data in dereddened colour-magnitude space.

\subsection{Dereddened Colour-Magnitude Diagrams}

\begin{figure*}
\begin{center}
\begin{tabular}{ccc}
\resizebox{55mm}{!}{\includegraphics[angle=90]{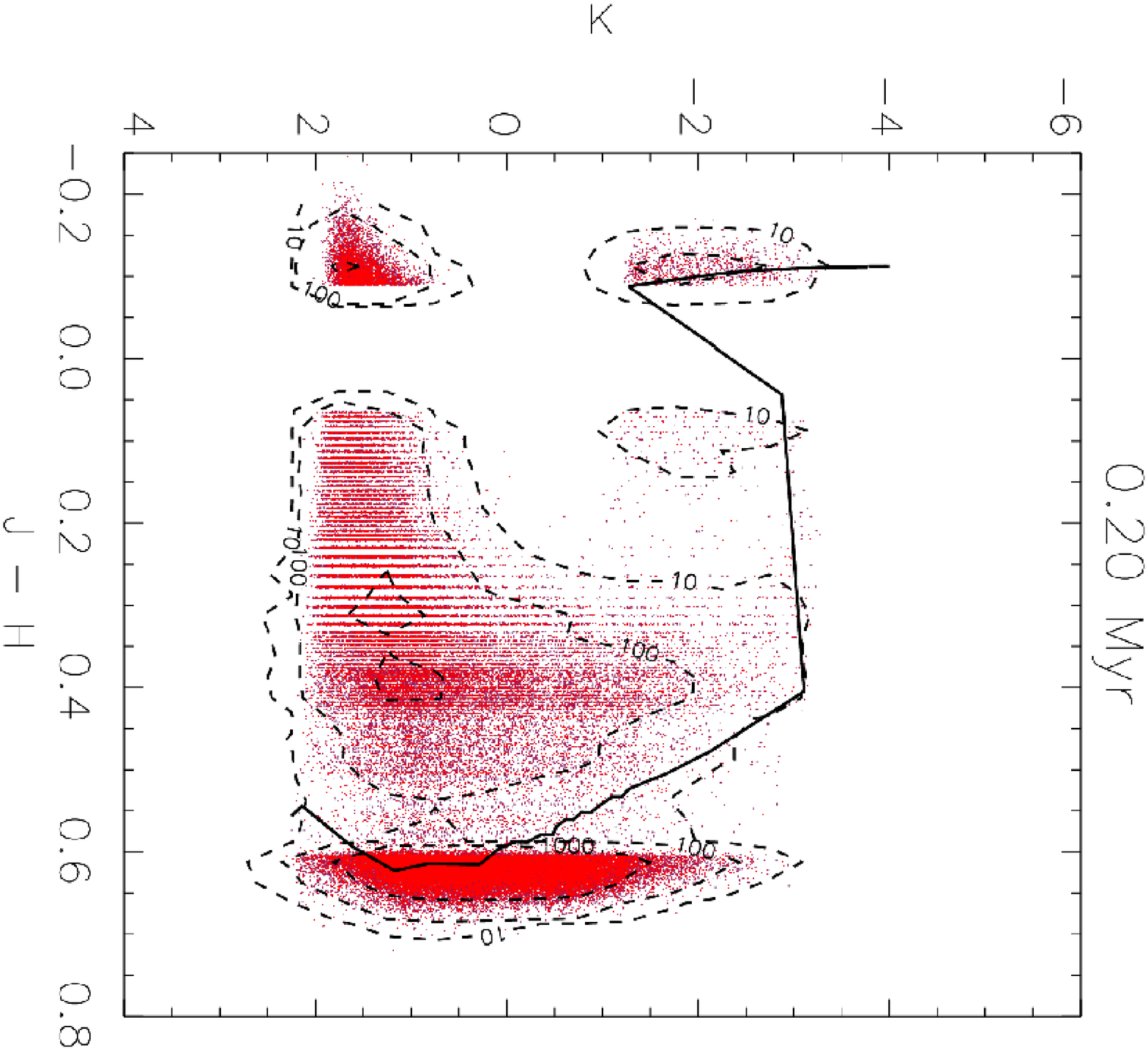}}&
\resizebox{55mm}{!}{\includegraphics[angle=90]{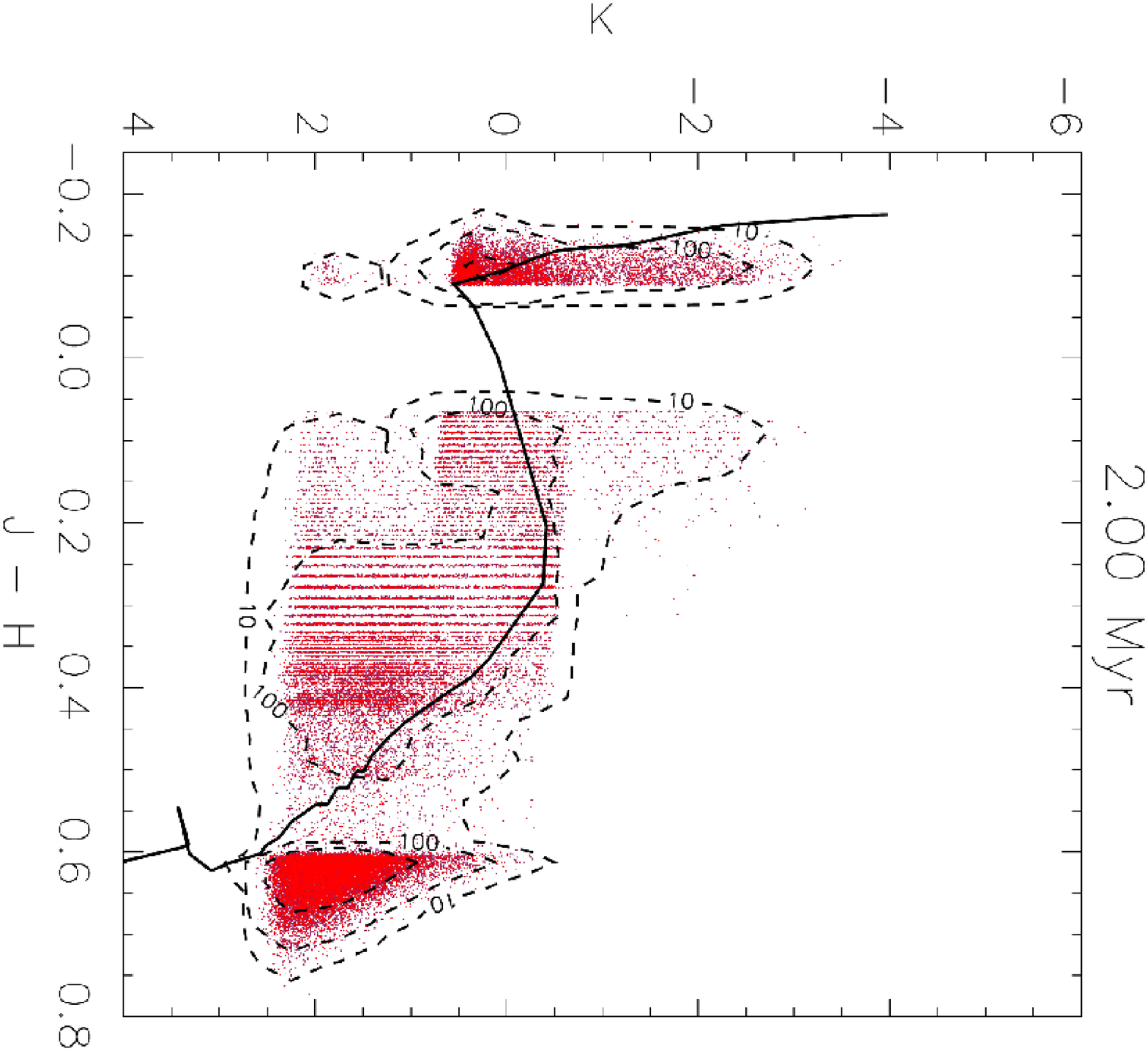}}&
\resizebox{55mm}{!}{\includegraphics[angle=90]{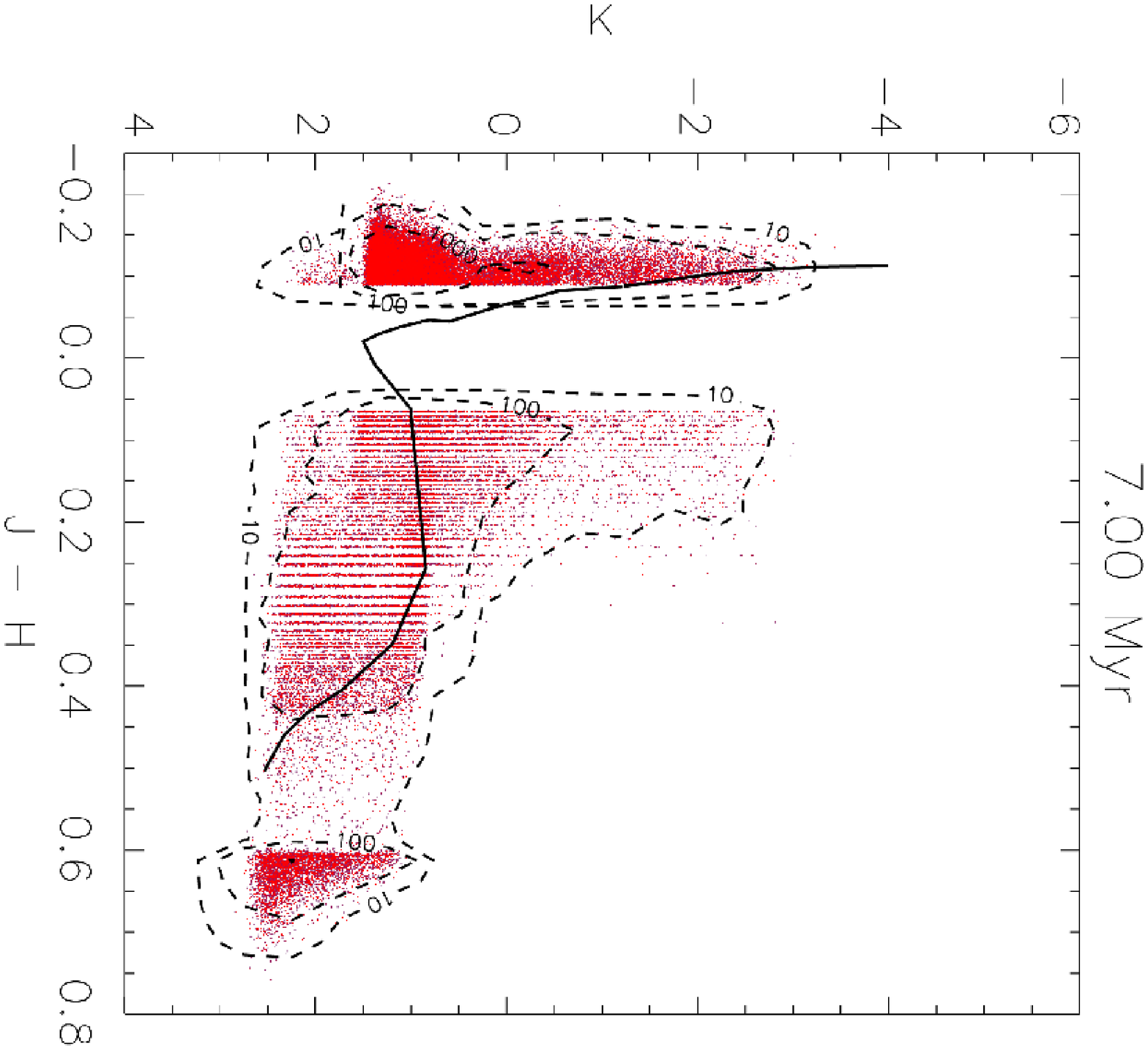}}
\end{tabular}
\caption[]{\small a (Left): A J-H vs K (intrinsic) dereddened CMD of the synthetic stars generated from a 0.2 Myr isochrone. The 0.2 Myr isochrone (in the UKIDSS system) has been overlaid as a solid line. Contours show where red points are above 10, 100, and 1000 per 0.025 mag x 0.5 mag regions. b (Centre): Same as (a) except with a 2.0 Myr isochrone. c (Right): Same as (a) except with a 7.0 Myr isochrone.}
\label{fig:sudoCMD}
\end{center}
\end{figure*}
The process by which the artificial cluster was created has been extended, thereby creating a very large artificial cluster containing many tens of thousands of stars. These numerous artificial cluster members can be processed, in the same manner as in section \ref{sec:artif_clus}, and then dereddened to create a large sample of dereddened stars. From this large sample of data many small subsets, equal in size to the number of real observed stars in a cluster, are randomly drawn to compare with real data in dereddened colour-magnitude space, to model the errors that arise in the dereddening method. \\
$\indent$Figs. \ref{fig:sudoCMD} (a), (b) and (c) are CMDs containing large samples of dereddened stars created using the 0.2, 2.0 and 7.0 Myr isochrones respectively with contour lines illustrating where the dereddened stars are most concentrated. The simulated stars do not deredden to their original location on the isochrone. This is due to a combination of random photometric errors and systematic errors in the dereddening process. There are several features worth noting. In the top-left of each CMD is a vertical strip of high-mass early type stars that stretch up from the main sequence `turn-on' point. Such a grouping of stars is therefore a good age indicator. The bottom-left grouping of stars contains late M-type stars that have been wrongly dereddened to the early type sequence. Note that in the 7.0 Myr CMD no such stars are present. This is because at this older age the M-type stars are too faint to be `detected' using the specified parameters. The broad gap in each of the CMDs, centred at J-H$\sim$0.0, occurs because early type stars have been dereddened to a single point on the intrinsic locus. \\
$\indent$The far right group of stars on each CMD occur because stars whose photometric errors move their position on a CCD above the upper reddening track, placed at the upper tip of the intrinsic locus, are dereddened to a line perpendicular to the reddening track. Therefore the scatter of such stars along the dereddened J-H axis has been underestimated. \\
$\indent$In the middle of each CMD there is a low density region of stars which we split into two further groups, 0.06 $<$ J-H $<$ 0.4 and 0.4 $<$ J-H $<$ 0.6. The latter, lower density, region occurs because the slope of the intrinsic locus, over this same range in colour-colour space, is very similar to the slope of the reddening track, thereby reducing the possible range of positions where the track can intersect the locus. The former region is made up largely by stars whose errors have allowed them to scatter from either the early type or late type sequences. Very few stars are actually found in this position on a real CMD as stars quickly cross from one side to the other during their evolution. For example, on the 2 Myr isochrone this range of J-H values only covers a mass range 2.7 M$_{\odot}$ to 3.3 M$_{\odot}$. Finally, there is a vertical striping effect in all CMDs. The vertical striping occurs due to the discrete nature of the intrinsic locus used to deredden the young stars.

\subsection{Age Determination of Synthetic Clusters}
\label{sec:AgeDetSyn}

\begin{figure}
\begin{center}
\begin{tabular}{c}
\resizebox{70mm}{!}{\includegraphics[angle=0]{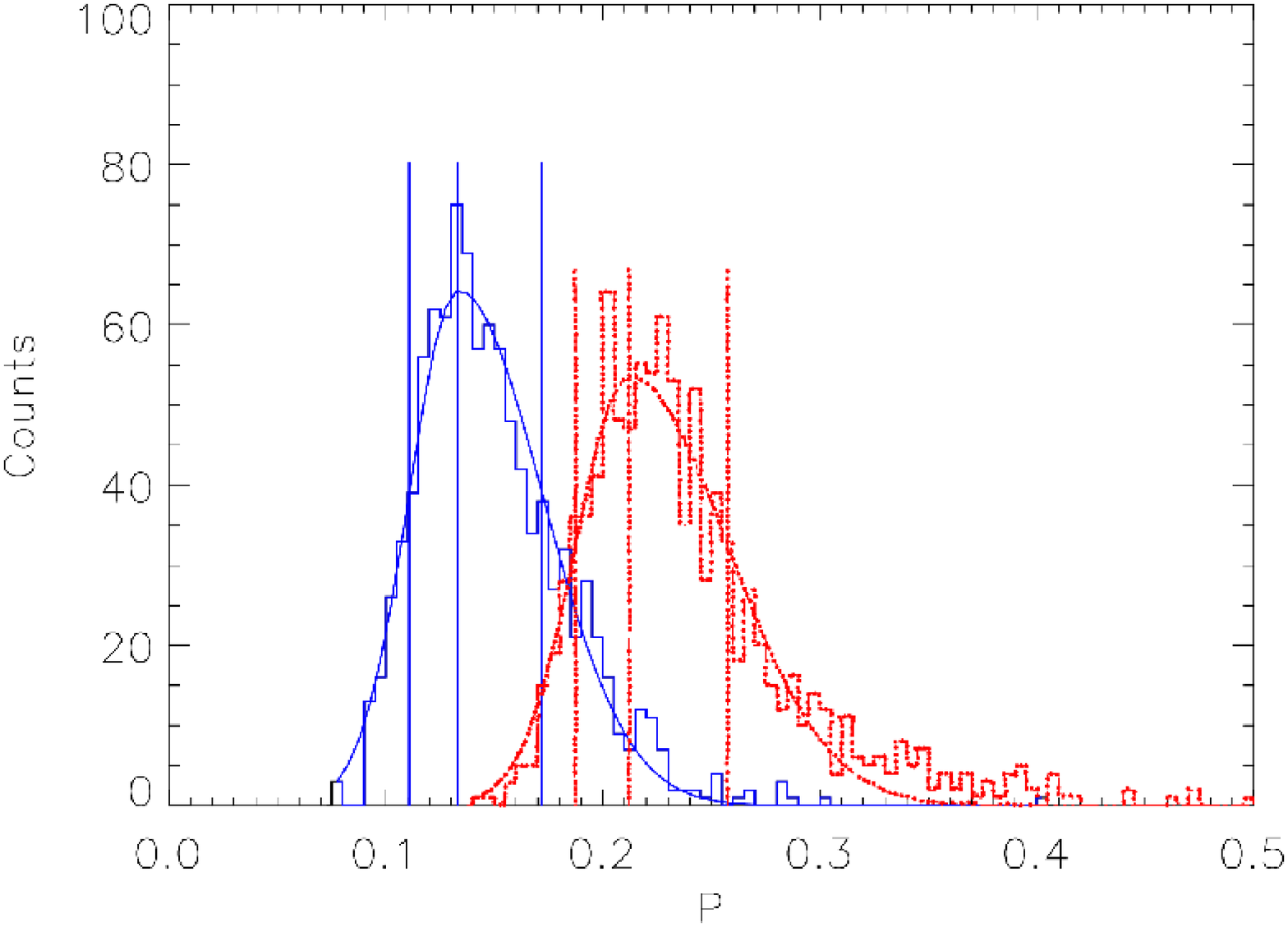}}\\
\resizebox{70mm}{!}{\includegraphics[angle=0]{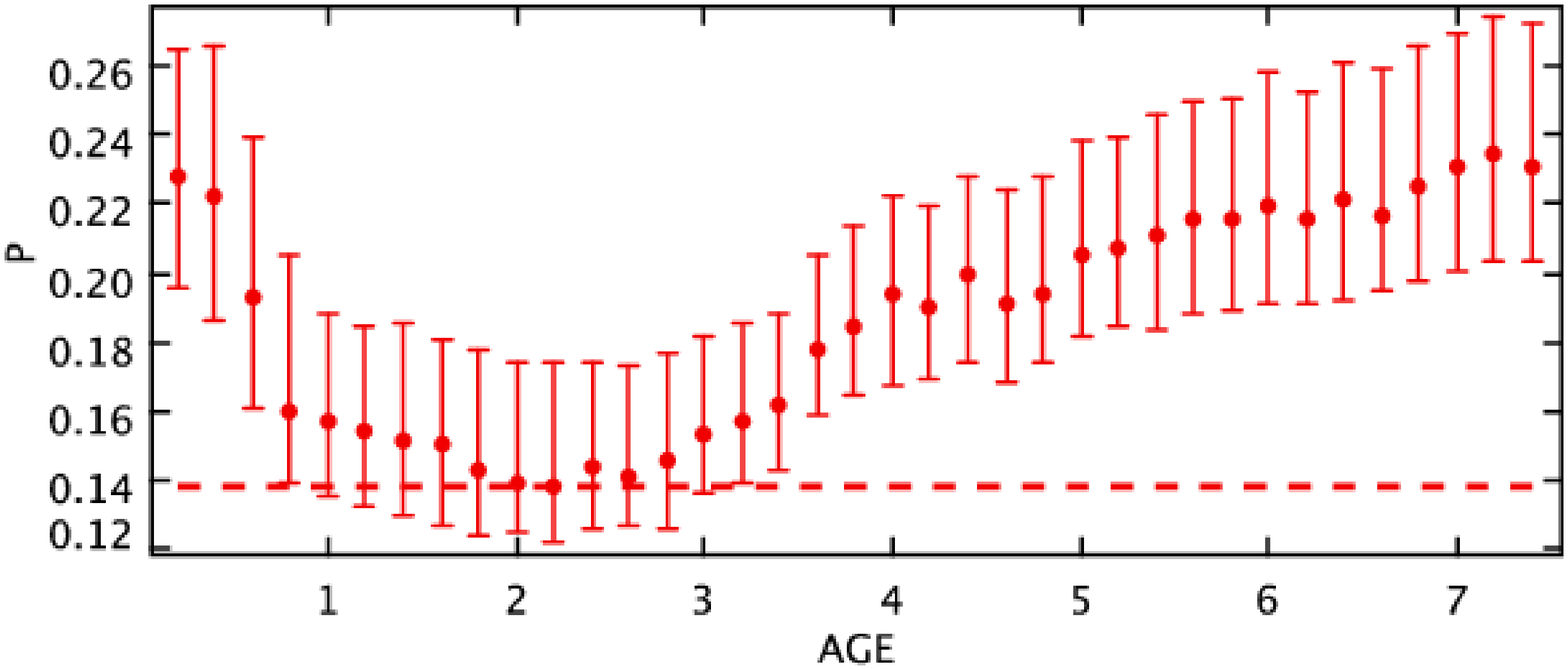}}
\end{tabular}
\caption[]{\small a (Top): The best fit 2.2 Myr isochrone is shown as a solid histogram and skewed Gaussian, the solid lines represent the median and 1$\sigma$ deviations. The dashed histogram and skewed Gaussian are generated via the 1 Myr isochrone. b (Bottom): The values of P where the 1$\sigma$ deviation error bars pass below the dashed line (smallest median) defines the error in the determined age. The error bars on each point are derived from the 1$\sigma$ deviations of the histograms and skewed Gaussians as shown in (a).}
\label{fig:AGE_2Myr_Pdev}
\end{center}
\end{figure}

We have tested
 our method by recovering the age of our realistic artificial cluster with 33 stars, as described in section \ref{sec:artif_clus}. In the same process as described above, we have produced many more large artificial clusters, with the assumption that each contains at least 1,000 stars with masses larger than 15.0 M$_{\odot}$, covering the age range 0.2 to 7.6 Myr, in 0.2 Myr intervals. The following equation can be used to compare a real cluster in dereddened colour-magnitude space with several different sets of artificial data, derived from isochrones of varying age: 
\begin{equation}
P =\frac{1}{n}\sum^{n}_{i=1}(S_{(J-H)_{i}}-R_{(J-H)_{i}})^{2}+(S_{(K)_{i}}-R_{(K)_{i}})^{2})^{\frac{1}{2}},
\label{eq:tauYSO}
\end{equation}
where P is defined as the mean separation, in colour-magnitude space, between each real star (R) and its nearest synthetic star neighbour (S). The above equation is asymmetric, with more emphasis being placed on the magnitude. However tests with several different types of scaling used had a negligible effect on the derived ages and errors. \\
$\indent$Monte Carlo simulations are performed to find the age of the isochrone used to create the artificial data that minimises P, and therefore yield the cluster age. The process is repeated 1,000 times to determine the distribution of P and determine the associated 1$\sigma$ error in the derived age. As an example, the age of the 2 Myr artificial cluster, created in section \ref{sec:artif_clus}, has been calculated by minimising equation \ref{eq:tauYSO} using data from each of the large artificial clusters. Monte Carlo simulations have been performed by comparing one thousand randomly selected different sets of data each equal in size to the target cluster (33 stars). Fig. \ref{fig:AGE_2Myr_Pdev} (a) illustrates this process. This fitting method is identical to the method presented in \citet{stead09}. Fig. \ref{fig:AGE_2Myr_Pdev} (b) shows that the best fit was made using data from the 2.2 Myr large artificial cluster and the dashed line shows the value of P that produces the best fit. This line can be used to determine the 1$\sigma$ error in the age by linearly interpolating the lower error bars to their point of intersection with the dashed line. This implies a cluster age = 2.2$_{-1.3}^{+0.9}$ Myr, which is within agreement with the real cluster age.\\
$\indent$As previously mentioned, PMS stars may have a different temperature-colour relationship than main sequence stars of a similar type. Different to the data provided in the S00 isochrones, the higher luminosity PMS stars actually lie along the giant locus, not the ZAMS line. To take this effect into consideration, we generated new artificial clusters whereby the higher luminosity PMS stars follow a giant locus. The ages derived for such clusters were consistent with the results previously determined in this paper.

\subsection{Field Star Contaminants}
\label{sec:field_contamination}

\begin{figure*}
\begin{center}
\begin{tabular}{cc}
\resizebox{80mm}{!}{\includegraphics[angle=0]{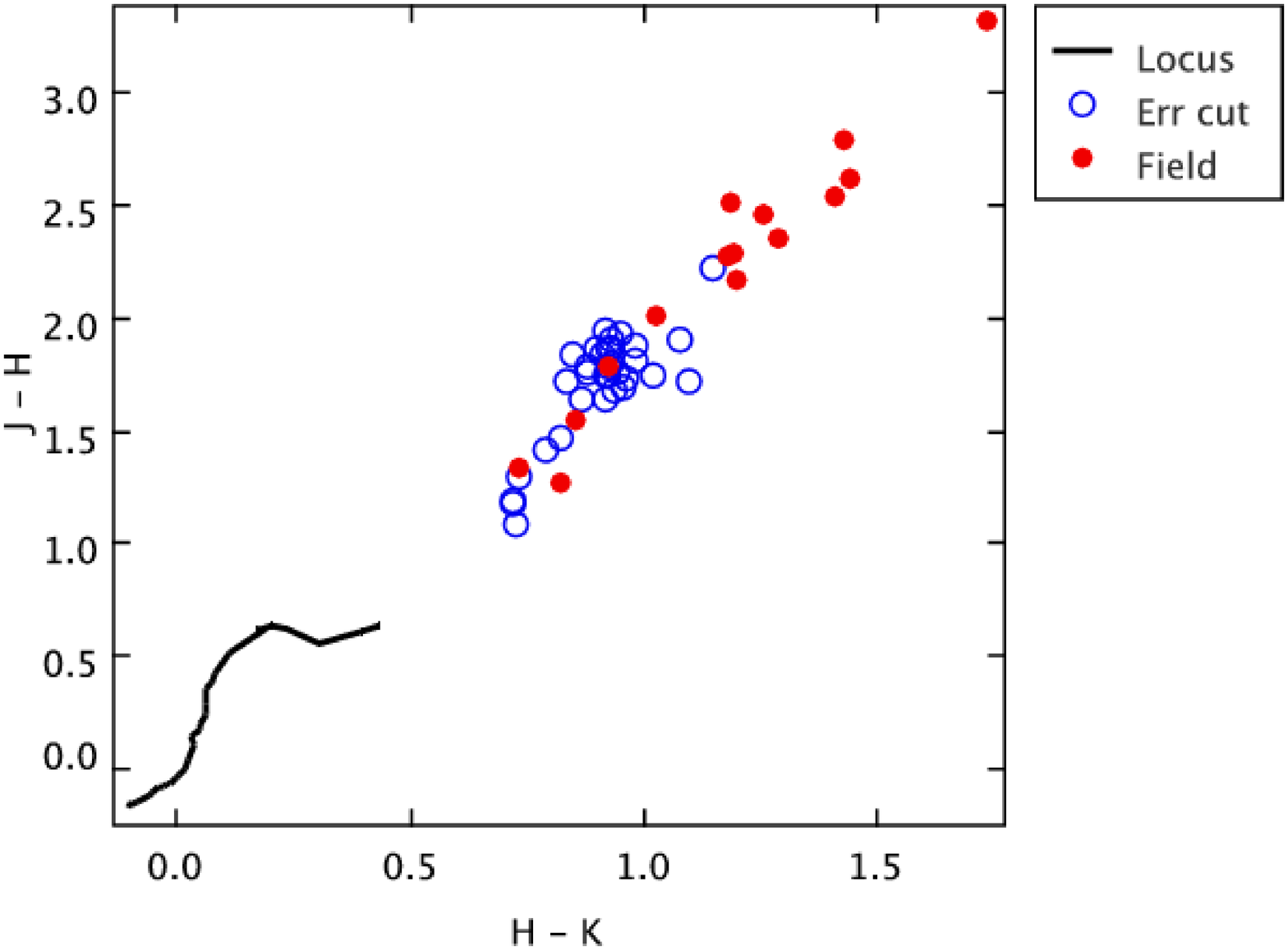}} &
\resizebox{80mm}{!}{\includegraphics[angle=90]{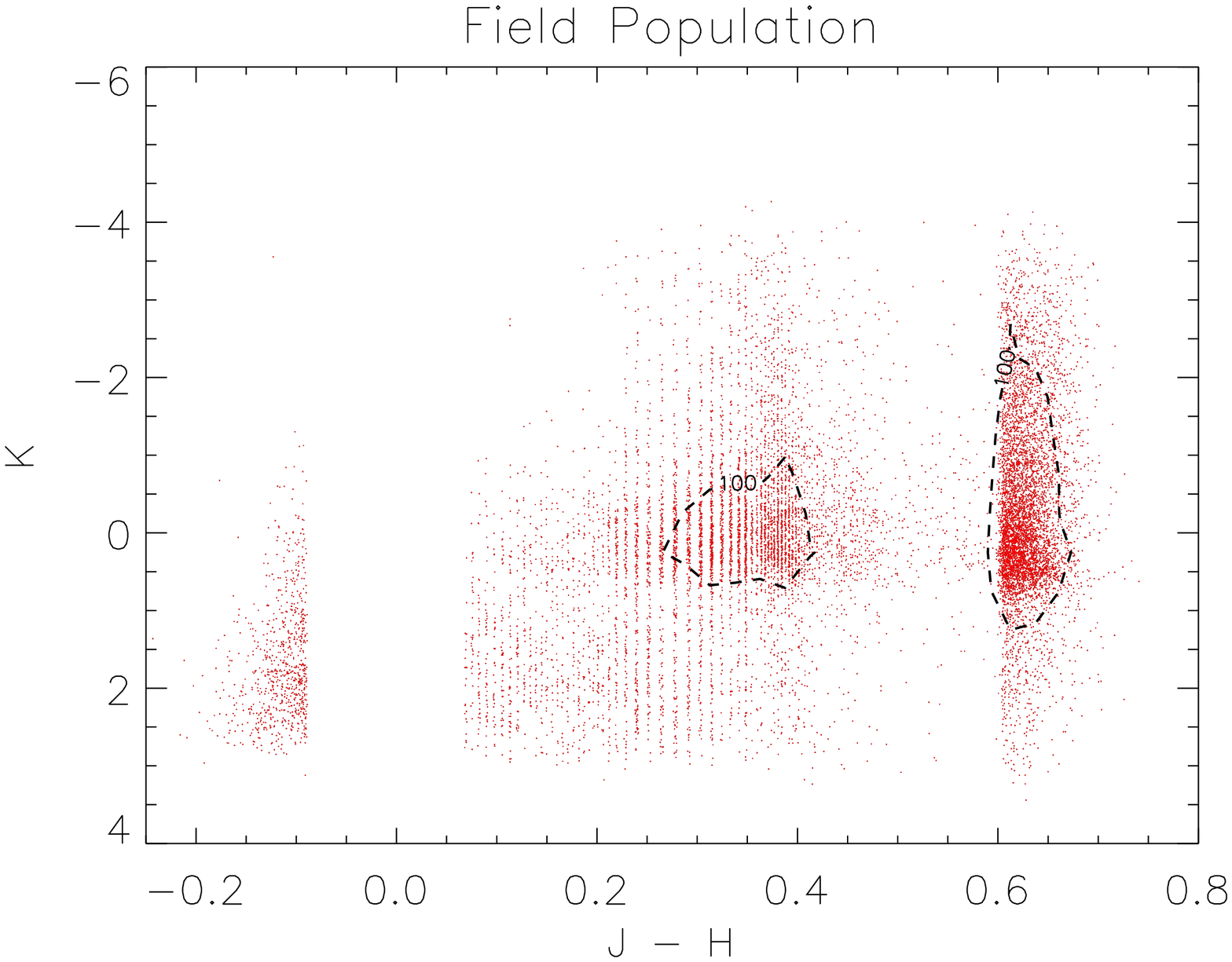}}
\end{tabular}
\caption[]{\small a (Left): A CCD of the 2 Myr artificial cluster from Fig. \ref{fig:sudoCCD}. The 33 artificial cluster members (blue circles) have been overlaid against a random field population of 15 background stars (red circles). b (Right): A J-H vs K (intrinsic) dereddened CMD created using $\Bes$ data that have been manipulated to replicate a typical field population. Contours show where data points are above 100, and 1000 per 0.025 mag x 0.5 mag regions.}
\label{fig:sudoCCD_field}
\end{center}
\end{figure*}

\begin{figure}
\begin{center}
\begin{tabular}{c}
\resizebox{70mm}{!}{\includegraphics[angle=0]{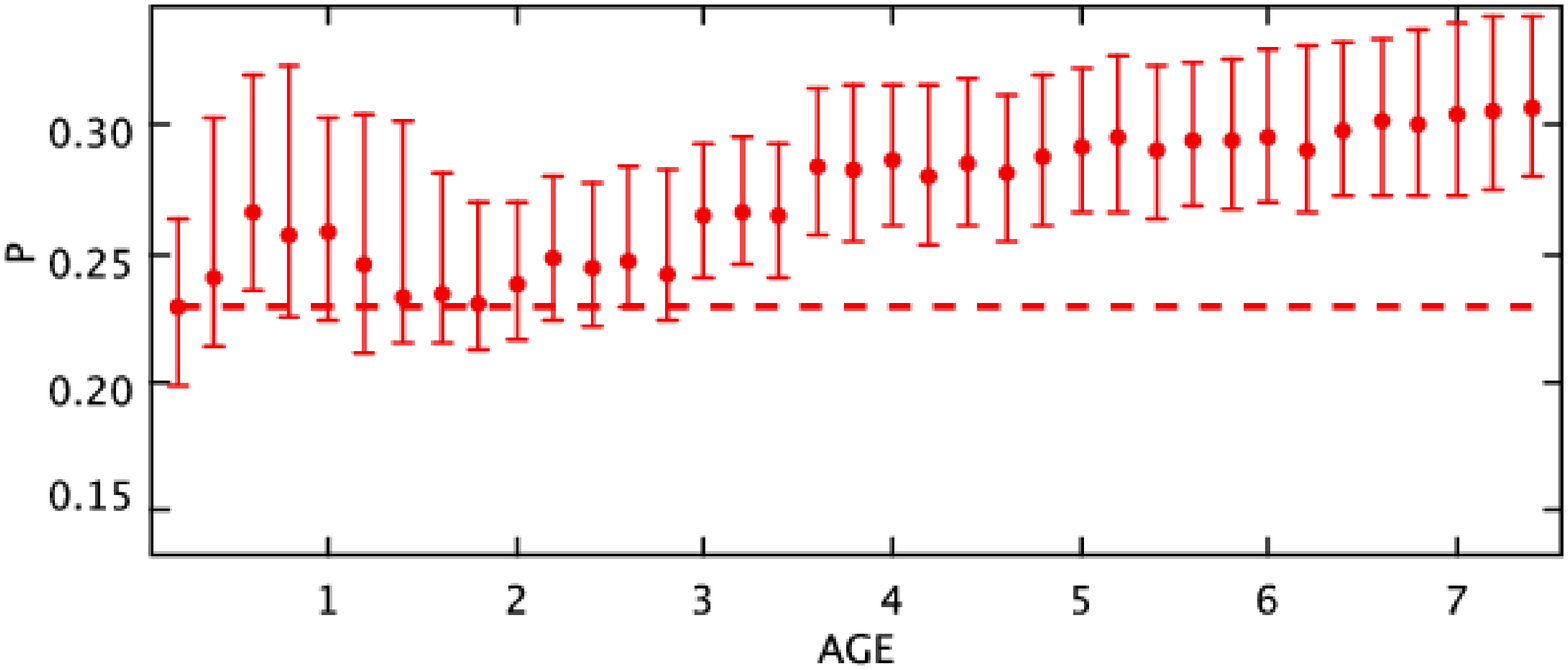}}\\
\resizebox{70mm}{!}{\includegraphics[angle=0]{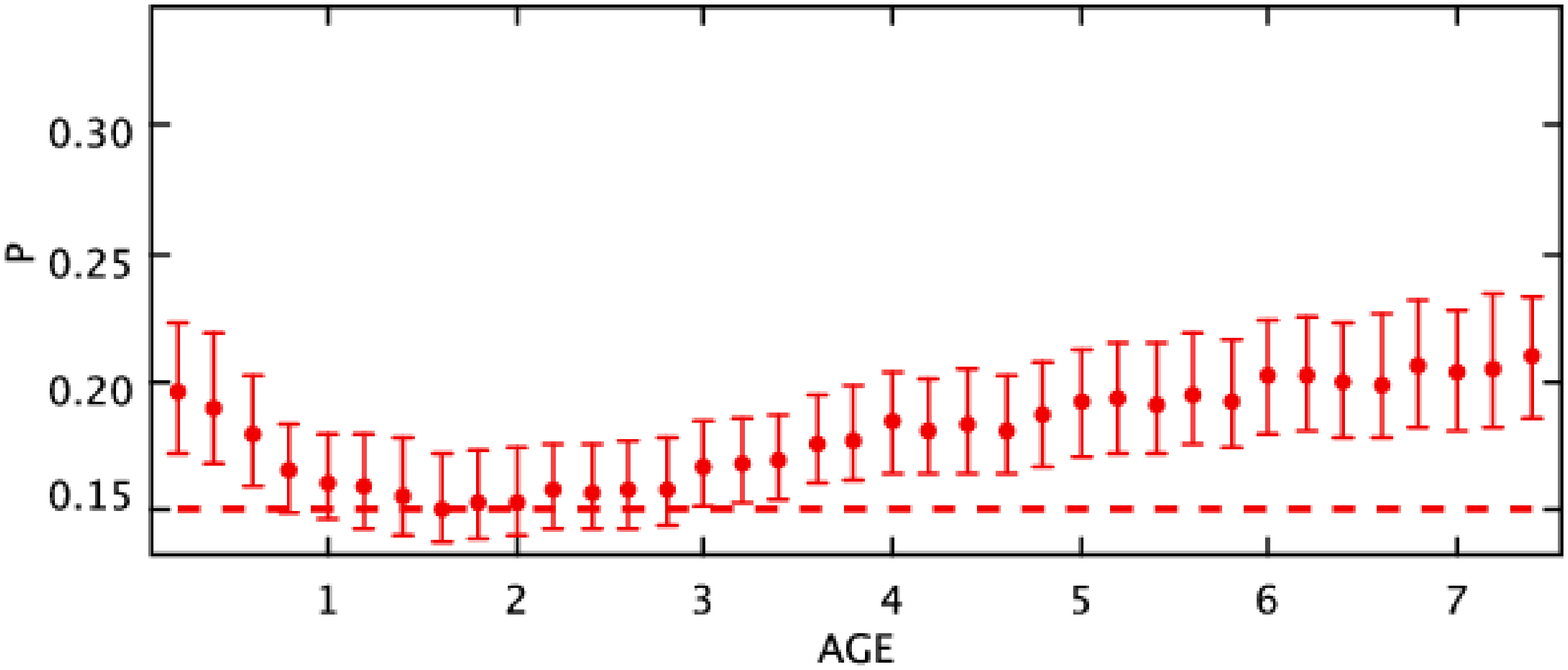}}
\end{tabular}
\caption[]{\small a(Top): The age of the 2 Myr cluster has been derived again, however 10 field stars have also been included in the sample, totaling 48 cluster candidates. b (Bottom): The same as (a) except the 15 field stars have been modelled in the simulations. Both figures have been plotted along the same y-axis as in Fig. \ref{fig:AGE_2Myr_Pdev} to aid comparison between the best-fit values of P.}
\label{fig:AGE_2Myr_Pdev_Field}
\end{center}
\end{figure}

When extracting cluster members from a small area of the sky, the sample will almost always be contaminated with field stars. This is particularly true for clusters that are situated very close to the Galactic Plane. Photometric selection (as will be shown in the following section) can be applied to remove the majority of foreground stars from a sample. However, without obtaining radial velocities or performing a proper motion analysis, etc, any sample of cluster candidates will still contain background stars. If the fraction of background stars within a cluster field can be estimated, through comparison with offset fields, simulations can include typical field stars when determining the age of a cluster to take their effect into consideration. Fig. \ref{fig:sudoCCD_field} (a) shows a CCD containing the same 33 artificial cluster members, created from the 2 Myr isochrone, presented in Fig. \ref{fig:sudoCCD}. A random field population has also been included using $\Bes$ data that have been converted to UKIDSS colours and magnitudes, with included photometric errors, and given a reliable extinction-distance relationship using \citet{marshall06} data. The field population contains 15 stars representing a contamination of $\sim$30\%. This value is slightly larger than the field star contamination in the real cluster fields, presented in the following sections, and should therefore represent a `worst-case' scenario.  \\
$\indent$To include background stars in the Monte Carlo simulations, first a very large amount of $\Bes$ data are produced as above. This large sample is then dereddened in the same manner as the real cluster members to create a large sample of dereddened field stars. From this large sample, small subsets are randomly drawn and included in the Monte Carlo simulations, alongside data from the large artificial clusters. Similar to Fig. \ref{fig:sudoCMD}, Fig. \ref{fig:sudoCCD_field} (b) contains a dereddened CMD created using $\Bes$ data. There are $\sim$13,000 dereddened field stars, after the 0.10 mag photometric error cut has been applied, and contour lines illustrate where the dereddened stars are most concentrated. As the majority of the field population are late type giants, there is a large collection of dereddened stars on the CMD at J-H $>$ 0.6 but very few near the expected cluster turn-on point. The intrinsic magnitudes of the field stars in the dereddened CMD should not be taken literally, instead they represent what the dereddened population would look like on a dereddened CMD, after assuming the stars to be cluster candidates, and therefore to be at the cluster distance. \\
$\indent$The 33 artificial cluster members and 15 field stars from Fig. \ref{fig:sudoCCD_field} have been combined together to produce 48 cluster candidates. These 48 cluster candidates have been dereddened and used to determine the age of the cluster. This has been done in two different ways. First, in the same manner as above, all 48 cluster candidates have been treated as genuine cluster members and are compared with data drawn from the cluster models only. Next, the field star contamination has instead been considered in the Monte Carlo simulations. This is done by comparing the 48 cluster candidates with a mix of cluster model data ($\sim$70\%) and field star model data ($\sim$30\%). Fig. \ref{fig:AGE_2Myr_Pdev_Field} (a) and (b) show the minimisation of P for the 48 cluster candidates using both of the aforementioned methods. In the former case, where only cluster model data were used, there is a localised dip at the correct cluster age in the measured values of P. However the best fit is made using the 0.2 Myr cluster model giving rise to the derived age = 0.2$_{-0.2}^{+2.7}$ Myr. In the latter case, where the field stars were considered in the simulations, the derived age = 1.6$_{-0.9}^{+1.4}$ Myr, which is once again within agreement to the real age. By including the field star population in the simulations, the true cluster age has once again been derived within errors, this time giving no bias towards younger ages. The minimised value of P is also smaller than when no field stars are assumed, suggesting a better fit. However, the presence of these field stars has reduced the ability to constrain the result, compared to the test when only the 33 cluster members were considered (2.2$_{-0.9}^{+1.3}$ Myr), as shown in several ways. The error in the derived age is larger, the error bars of the older cluster models are within much better agreement to the minimised value of P, and finally the minimised value of P is also larger. Therefore field stars will effect the age determination process, however their presence can be accounted for and modelled in the simulations. 

\subsection{Binary Star Contaminants}
\label{sec:binary_contamination}

Another strong cluster contaminant is that of binarity. At a distance of 4.3 kpc, assuming an angular separation 0.3$\arcsec$, it would only be possible to resolve a binary system if the two stars orbited each other at a distances larger than $\sim$1,000 AU. As \citet{mathieu94} suggests that at least 50\% of PMS stars are found in multiple systems, it can therefore be considered that within the cluster sample there will be unresolved binaries. \\
$\indent$The contamination of a dereddened CMD, caused by unresolved binary systems, is dependent on the mass ratio of the binary system being dereddened. In the most extreme case, two equal mass stars
 have the same colours as a single star, and therefore intersect the same point on the intrinsic locus when dereddened, however the measured magnitude is $\sim$0.75 mag brighter.  \\
$\indent$To test this assumption, the age of the 2 Myr artificial cluster has been derived again, with several of the stars considered to be equal mass unresolved binaries. i.e. their measured magnitudes were reduced by $\sim$0.75 mag. When 16 of the 33 stars (i.e. 50\%) were considered to be equal mass unresolved binaries the derived age = 2.4$_{-0.9}^{+2.4}$ Myr. This value remains consistent with the true value, however the errors have been slightly broadened. As binary ratio and the mass distribution is unknown, and as they have been shown to have a negligible effect on the derived results, the issue of binarity has been ignored for the remainder of this paper. \\

\subsection{Infrared Excess Source Contaminants}
\label{sec:IRXS_contamination}

Young stars often possess large amounts of circumstellar dust. This dust absorbs the optical and UV radiation emitted by the star, causing it to re-radiate the absorbed energy as thermal IR radiation. Such sources are known as infrared excess (IRXS) sources. This excess radiation can be detected by UKIDSS predominantly in the K band, therefore an IRXS source shifts to the right of the bulk of the data on an H-K vs J-H CCD. However, it is plausible that certain IRXS sources will not have sufficiently large K band excess to be identified on an H-K vs J-H CCD. Such sources still possess atypical stellar colours and it is necessary to remove them from any final samples. As the spectrum of the heated circumstellar material peaks at longer wavelengths than covered by UKIDSS, CCDs constructed using GLIMPSE data \citep{benjamin03}, where the effect of an IRXS is more pronounced, can therefore be used to help identify and remove IRXS sources. 

\section{Determining the Age of Real Embedded Clusters}
\label{sec:real_cluster_ages}
\begin{figure*}
\begin{center}
\begin{tabular}{cc}
\resizebox{65mm}{!}{\includegraphics[angle=0]{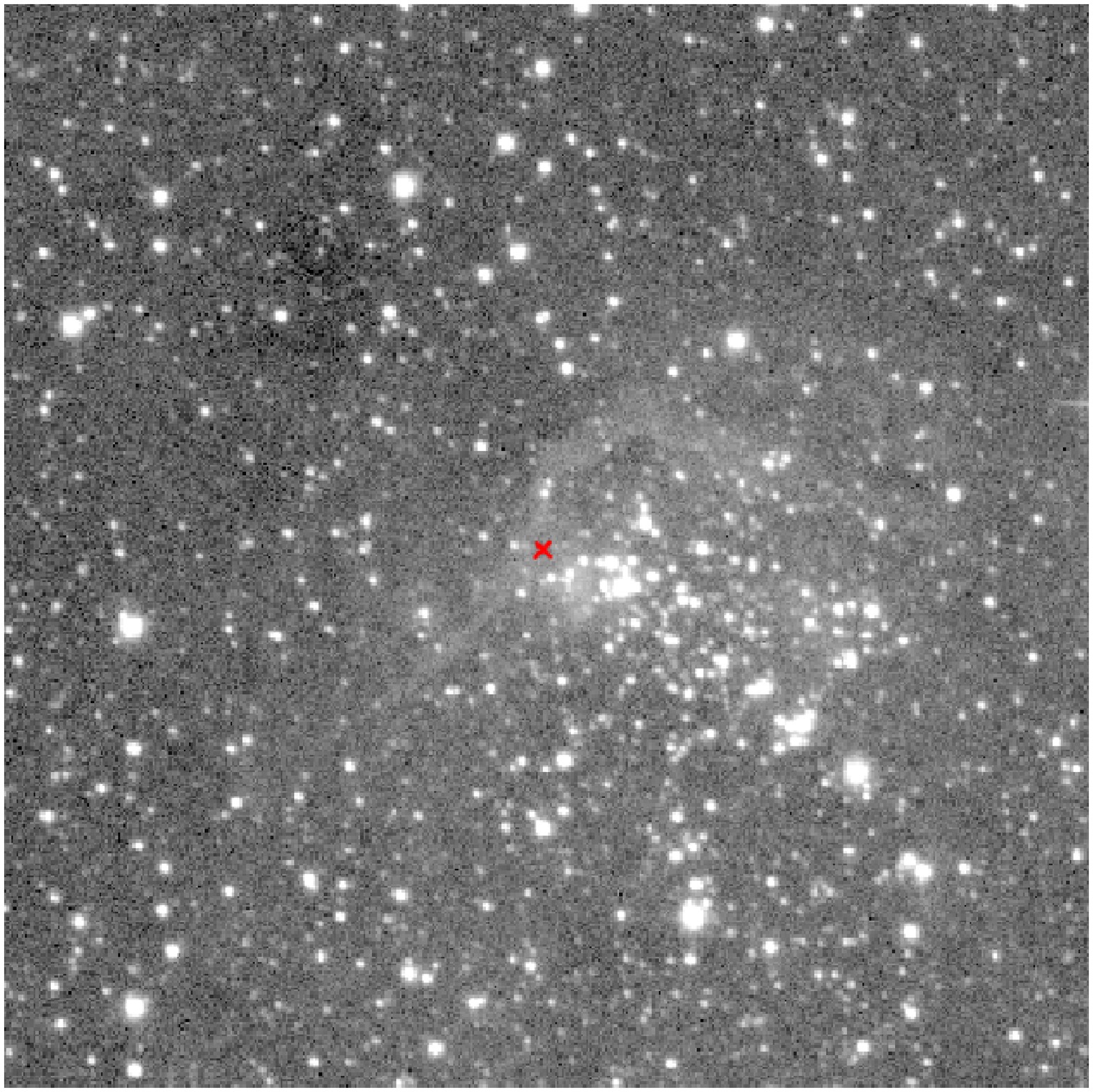}} &
\resizebox{75mm}{!}{\includegraphics[angle=90]{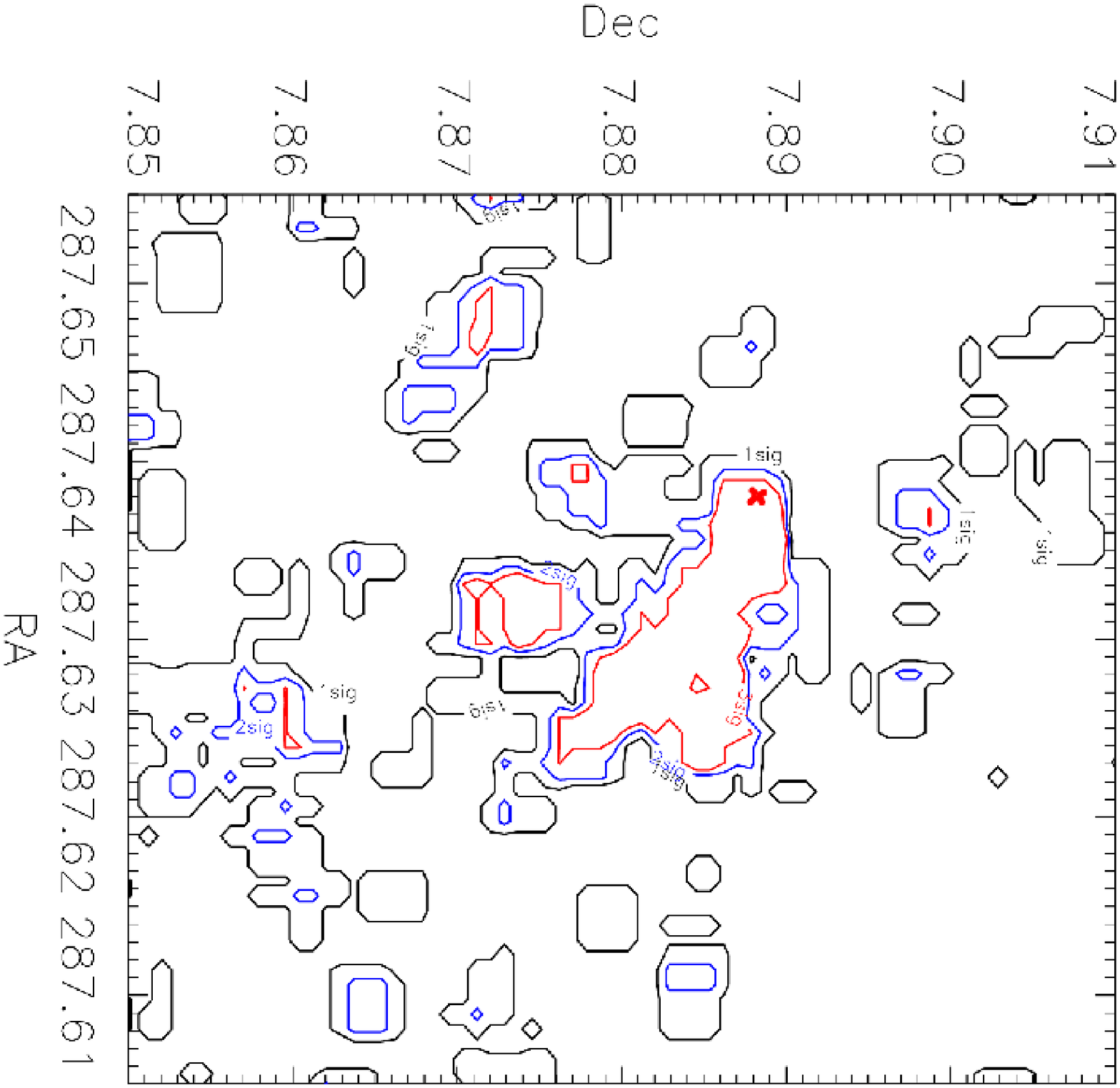}}
\end{tabular}
\caption[]{\small a (Left): A 3$\arcmin$x3$\arcmin$ UKIDSS K band image of the area of sky containing the RMS source G042.1268-0.6224 (indicated with a cross). North is up and east is left. b (Right): A 3$\arcmin$x3$\arcmin$ stellar surface density map centred close to the RMS source (indicated with a cross) G042.1268-0.6224. Black, blue and red contour lines represent regions where star counts exceed the 1, 2 and 3$\sigma$ levels. The closed 3$\sigma$ contour towards the centre of the map has been used to extract cluster candidates.} 
\label{fig:G42.1268_SpatialMap} 
\end{center}
\end{figure*}

The aforementioned methods are now applied to 3 real clusters. As a full worked example, young embedded cluster members associated with the RMS HII region G042.1268-0.6224 have been identified, dereddened and used to determine the age of the cluster. Fig. \ref{fig:G42.1268_SpatialMap} (a) shows the region of sky containing the cluster. G042.1268-0.6224 provides an easy test of the method on a real cluster because it is not too deeply embedded. As will be shown, it is even possible to see a sequence of stars in a CCD (Fig. \ref{fig:offsetVScluster}), allowing an age determination of the cluster using more traditional methods. This cluster was originally identified as [BDS2003] 131 \citep{bica03b}.

\subsection{G042.1268-0.6224}
\label{sec:G042.1268-0.6224_cluster}

The RMS source G042.1268-0.6224 is flagged as a diffuse HII region in the RMS survey. \citet{urquhart08} have derived a a kinematic distance, via $^{13}$CO observations, of 5.3/7.6$\pm$1.1 kpc. This ambiguity was resolved by \citet{stead10} who derived a distance from extinction measurements, that is independent to kinematic methods, D = 4.3$\pm$0.5 kpc. This result is consistent with that of \citet{roman09} who used both HI self-absorption techniques and 21 cm continuum sources to resolve the ambiguity, and is used in this section to determine the age of the cluster.\\
$\indent$Clusters can usually be identified as an overabundance of stars on a relatively smooth field population. The spatial extent of a cluster can therefore be determined from the production of stellar surface density maps. Photometric constraints are used to remove as many field stars as possible without excluding the numerous and faint smaller mass cluster members. The actual photometric constraints used depend on many factors including cluster richness, Galactic location and line of sight extinction. \\
$\indent$Fig. \ref{fig:G42.1268_SpatialMap} (b) contains a 3$\arcmin$x3$\arcmin$ spatial map centred close to the RMS source G042.1268-0.6224. The star count was determined for small regions 15$\arcsec$x15$\arcsec$ in size, over a total area of 10$\arcmin$x10$\arcmin$. Contour lines show regions where the star count is above various $\sigma$ levels, where $\sigma$ is the standard deviation of the mean, and the smooth 3$\sigma$ contour line has been chosen as the cluster boundary. All following work on G042.1268-0.6224 considers only stars within the region traced by the 3$\sigma$ contour, covering an area 0.35 arcmin$^2$ in size. From this region, 44 cluster candidates have been extracted using a 0.05 mag error cut (see section \ref{sec:age_det_G42}).  \\
$\indent$To obtain a realistic estimation of the field population when using data as deep as UKIDSS, an offset field must be chosen with a similar line of sight extinction as the cluster. The line of sight extinction can be measured by producing molecular cloud and far line of sight extinction maps \citep[MCX and FLSX maps respectively, in][]{stead10}. Using both MCX and FLSX maps, three offset fields, equal in size to the 3$\sigma$ contour, have been used to extract field stars, each with a similar line of sight extinction as the cluster. Furthermore, GLIMPSE 8.0$\mu$m images were used to ensure that the selected regions of sky were away from any nebulosity that could be associated with the cluster. All of the offset fields were within $\sim$6$\arcmin$ from the cluster centre.\\
\begin{figure}
\begin{center}
\begin{tabular}{c}
\resizebox{80mm}{!}{\includegraphics[angle=0]{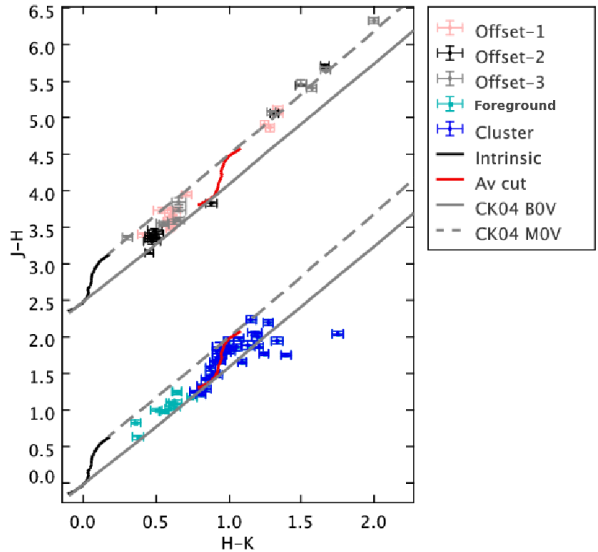}}
\end{tabular}
\caption[]{\small An H-K vs J-H CCD of the 44 cluster candidates. Also plotted, offset along the J-H axis by 2.5, are the stars extracted from each of the three offset fields (see text). The intrinsic colour curve has been reddened to A$_V$$\sim$9 \citep[this value has been obtained from the line of sight extinction measurements made in][]{stead10} and is used to split the cluster candidates, and offset fields, into two different populations (see text).}
\label{fig:offsetVScluster} 
\end{center}
\end{figure}
$\indent$Fig. \ref{fig:offsetVScluster} contains an H-K vs J-H CCD of the 44 cluster candidates plotted alongside the stars extracted from the three offset fields. Both the cluster candidates and the offset field stars show a gap in the data at H-K$\sim$0.75. It is plausible that the two populations represent the beginning of the molecular cloud that contains G042.1268-0.6224, and therefore the sudden increase in extinction splits the population into two. Using the line of sight extinction measurements made in \citet{stead10} to the cloud containing G042.1268-0.6224, an intrinsic colour curve has been reddened to A$_V$$\sim$9, representing the beginning of the molecular cloud containing G042.1268-0.6224. As the reddened curve is positioned around the gap in data at H-K$\sim$0.75, stars to the left of the curve are likely to be foreground stars, and stars to the right, a mixture of genuine cluster members and background stars. After exclusion of the foreground stars, 32 cluster candidates remain. The background stars are indistinguishable amongst the cluster members, however, as previously suggested, the background population can be modelled in the simulations, after estimating the source count from the three offset fields. As there are a total of 13 field stars in the offset fields, it can be assumed that the cluster candidate sample contains 4.3$\pm$1.2 background stars.\\
\begin{figure*}
\begin{center}
\begin{tabular}{cc}
\resizebox{80mm}{!}{\includegraphics[angle=0]{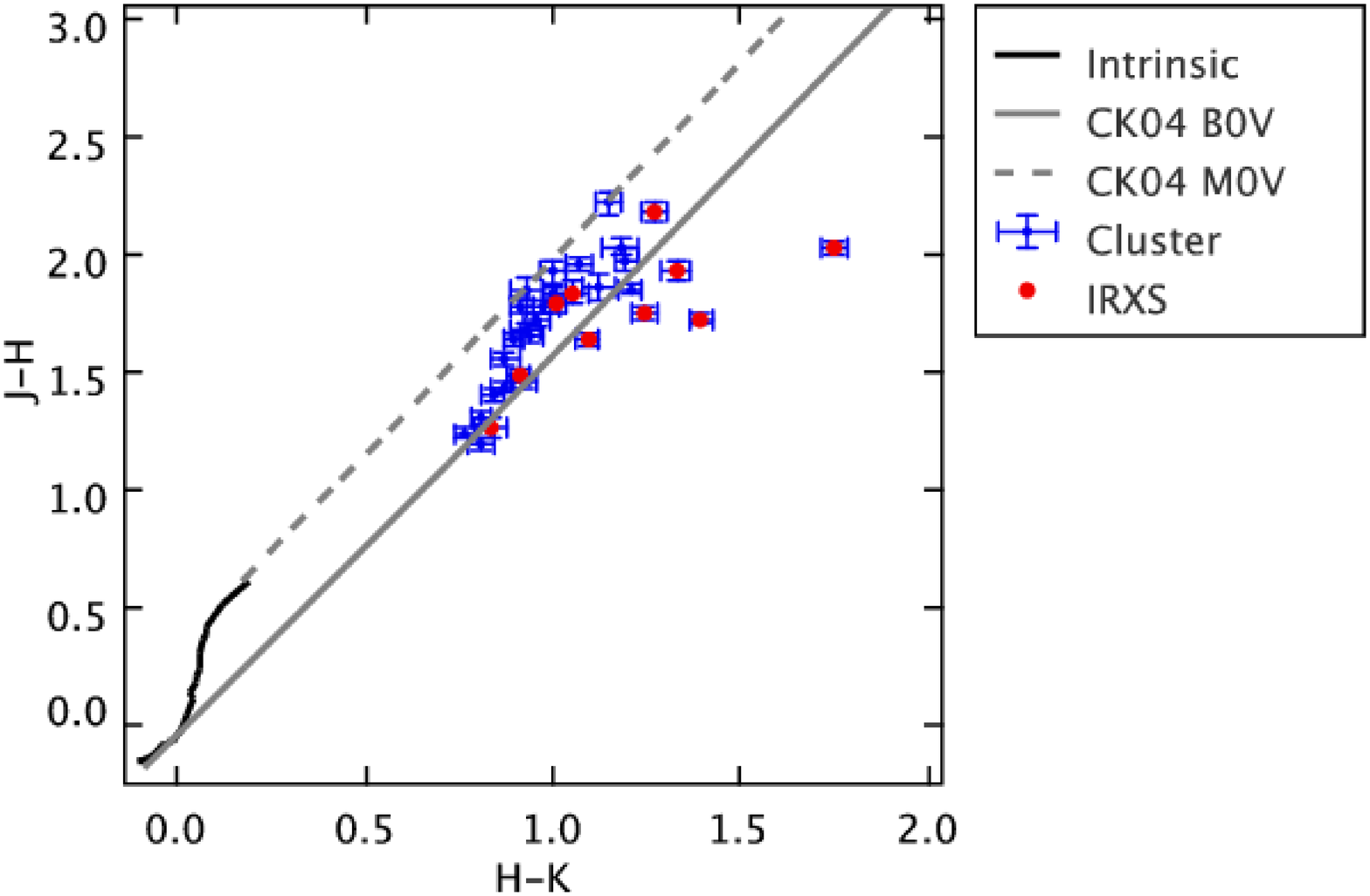}}&
\resizebox{80mm}{!}{\includegraphics[angle=0]{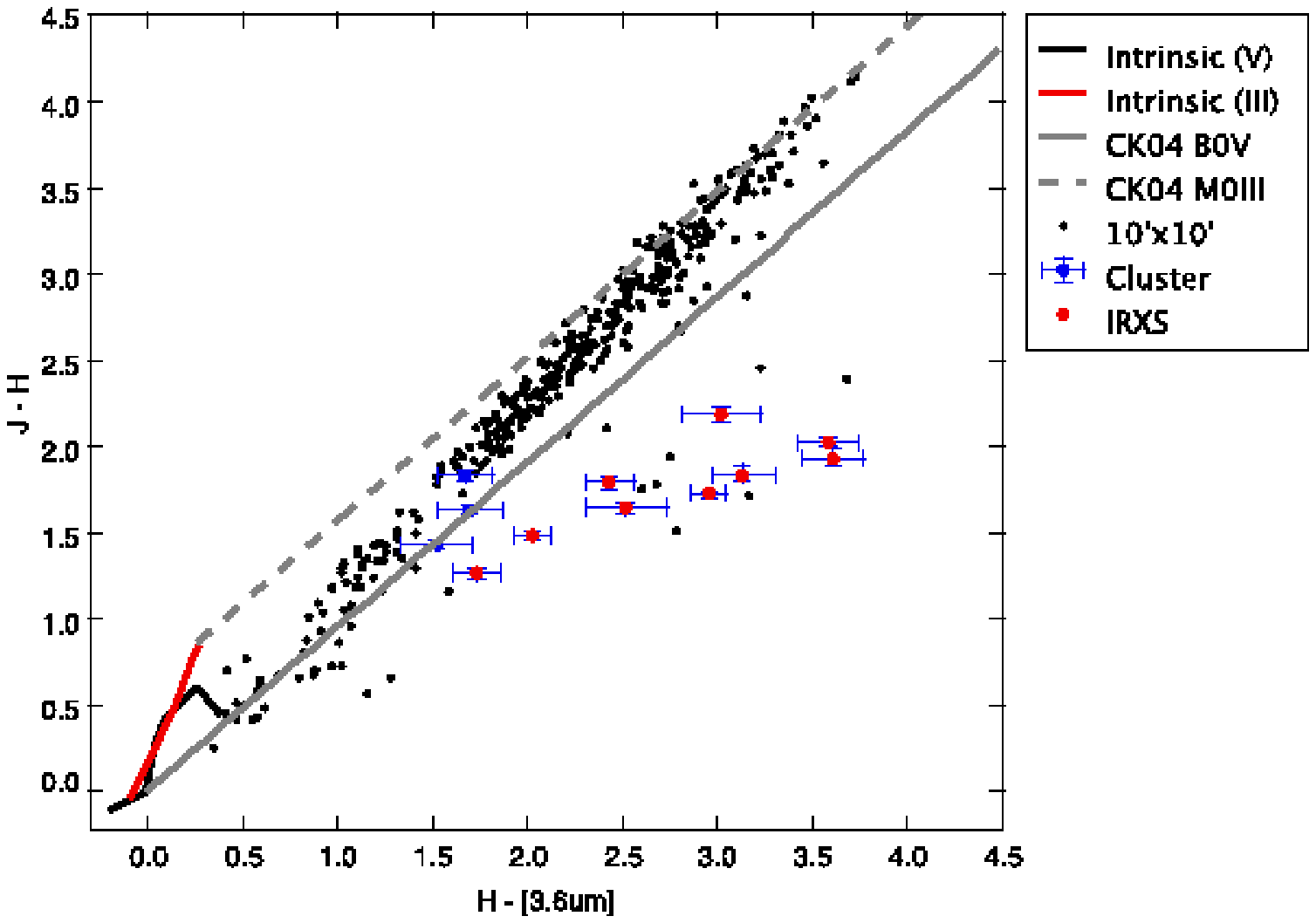}}
\end{tabular}
\caption[]{\small a (Left): An H-K vs J-H CCD of the remaining 32 cluster candidates surrounding G042.1268-0.6224 (blue error bars). The 10 sources that were identified as IRXS candidates using GLIMPSE data in (b) are shown as red circles. Note: Some of these sources are not positioned below the B0V reddening track and would not have been identified as IRXS candidates using UKIDSS data only. b (Right): An H-[3.6$\mu$m] vs J-H CCD of the most reliable UKIDSS (err$\le$0.05 mag) and GLIMPSE (err$\le$0.08 mag) data (grey points), extracted from a 10$\arcmin$x10$\arcmin$ region centred on the RMS source, highlighting the general shape of the field population on a UKIDSS-GLIMPSE CCD. A total of 13 GLIMPSE sources were extracted from within the 3$\sigma$ contour (blue error bars), of these 13 sources, 10 (red circles) stand below the B0V reddening track and have therefore been identified as IRXS candidates (one GLIMPSE IRXS source was identified using the 4.5$\mu$m filter on an H-[4.5$\mu$m] vs J-H CCD). Also plotted are the CK04 III and V intrinsic colours and the M0III reddening track. Due to the changing population of the background stars, the dominant intrinsic colours shift causing the loci of grey points to move away from the B0V track towards the M0III track.}

\label{fig:IRXS_CCD}
\end{center}
\end{figure*}

$\indent$Fig.\ref{fig:IRXS_CCD} (a) contains an H-K vs J-H CCD of the remaining 32 cluster candidates. Overlaid are 10 IRXS candidates, identified on an H-[3.6$\mu$m] vs J-H CCD using GLIMPSE data, in Fig. \ref{fig:IRXS_CCD} (b), where [3.6$\mu$m] represents the GLIMPSE photometric band centred at 3.6$\mu$m. A pairing radius of 0.7$\arcsec$ has been used to match UKIDSS and GLIMPSE data. This corresponds to the 3$\sigma$ deviation of all measured UKIDSS-GLIMPSE offsets, and should therefore limit the number of false matches. Of the 10 IRXS candidates identified, 5 are not positioned below the B0V reddening track on the UKIDSS CCD in Fig. \ref{fig:IRXS_CCD} (a). Therefore had the GLIMPSE data been unavailable, these 5 IRXS candidates would not have been identified among the cluster members. These 10 sources have been removed from the final sample, leaving 22 cluster candidates. None of the stars in the offset fields were identified as IRXS candidates, therefore the estimated number of field stars remains unchanged at 4.3$\pm$1.2, or $\sim$20\% contamination.\\

\subsubsection{Age Determination of G042.1268-0.6224}
\label{sec:age_det_G42}

\begin{figure}
\begin{center}
\begin{tabular}{c}
\resizebox{70mm}{!}{\includegraphics[angle=90]{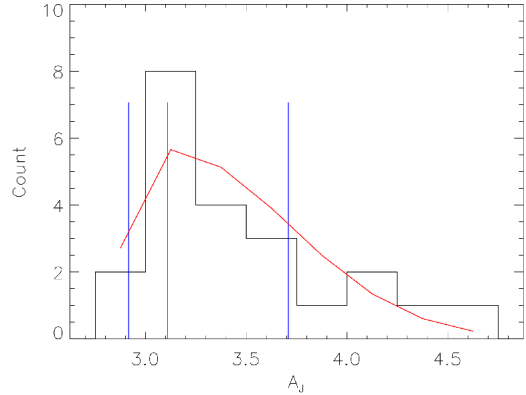}}
\end{tabular}
\caption[]{\small An A$_J$ histogram of 22 stars extracted from the cluster associated with the RMS source G042.1268-0.6224. The fitted skewed Gaussian and median are plotted as red lines and the two 1$\sigma$ deviations as blue lines: A$_J$=3.1$_{-0.2}^{+0.6}$ mag.}
\label{fig:Cluster_Extinction}
\end{center}
\end{figure}
\begin{figure*}
\begin{center}
\begin{tabular}{ccc}
\resizebox{55mm}{!}{\includegraphics[angle=0]{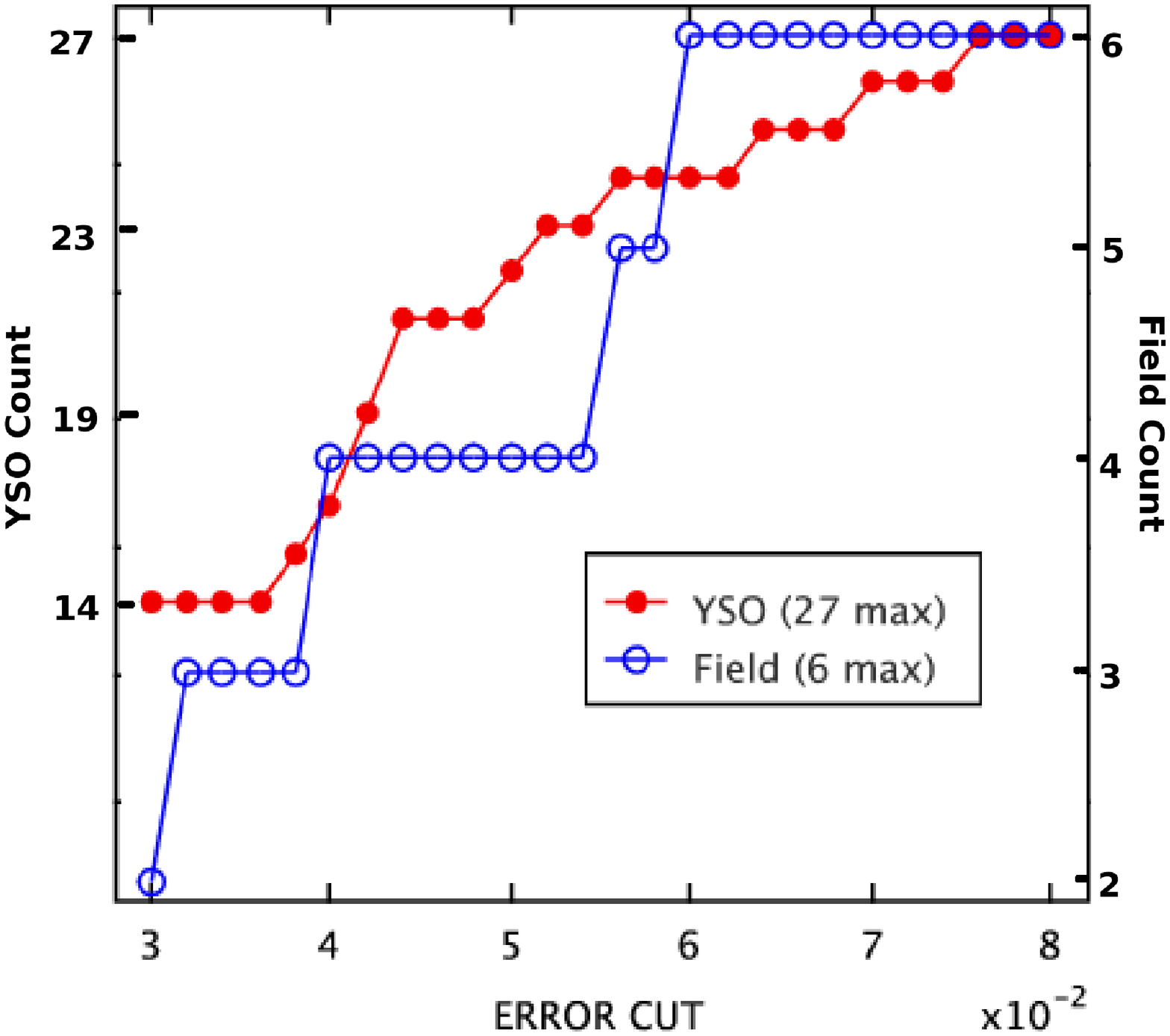}}&
\resizebox{55mm}{!}{\includegraphics[angle=0]{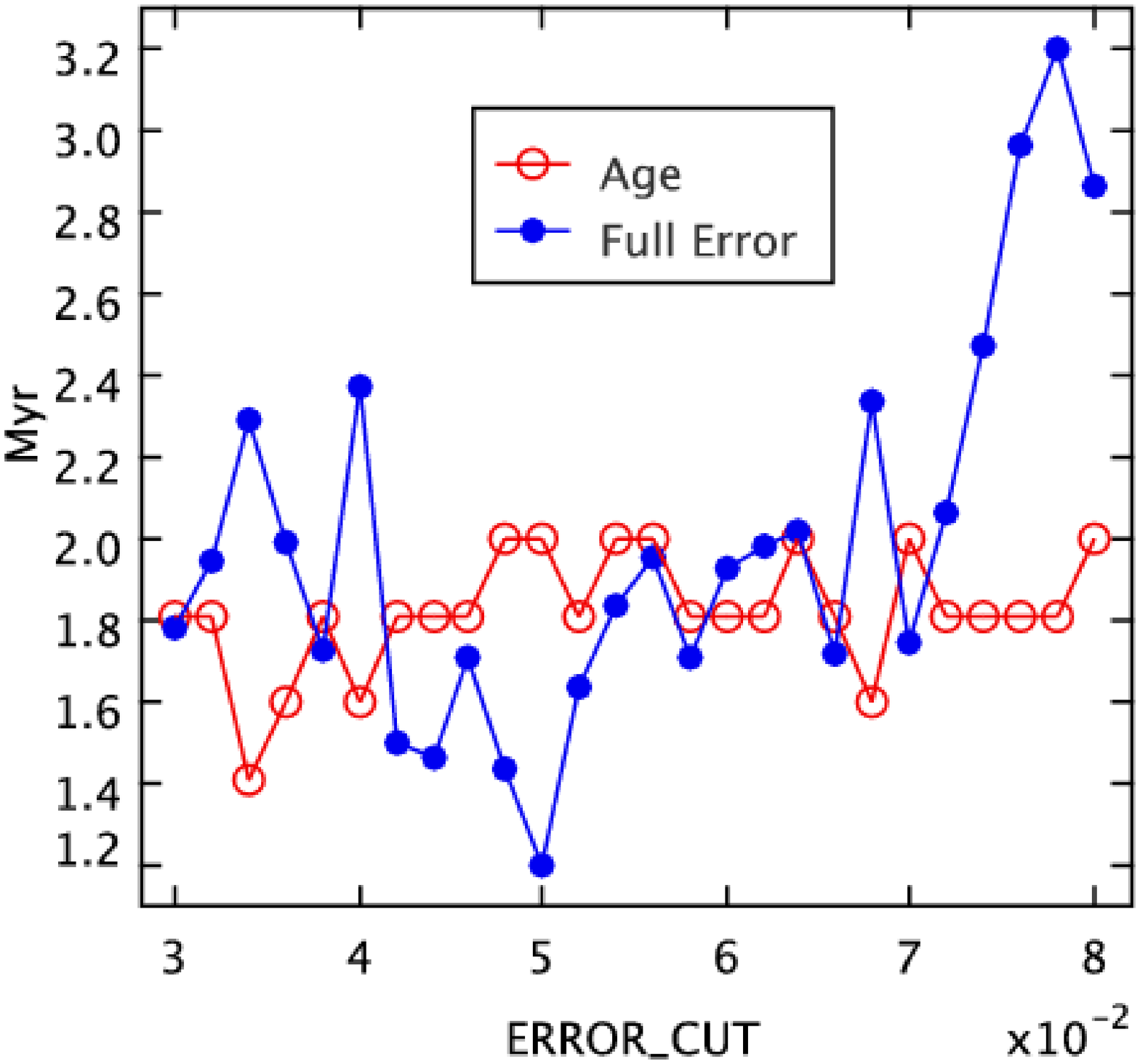}}&
\resizebox{55mm}{!}{\includegraphics[angle=0]{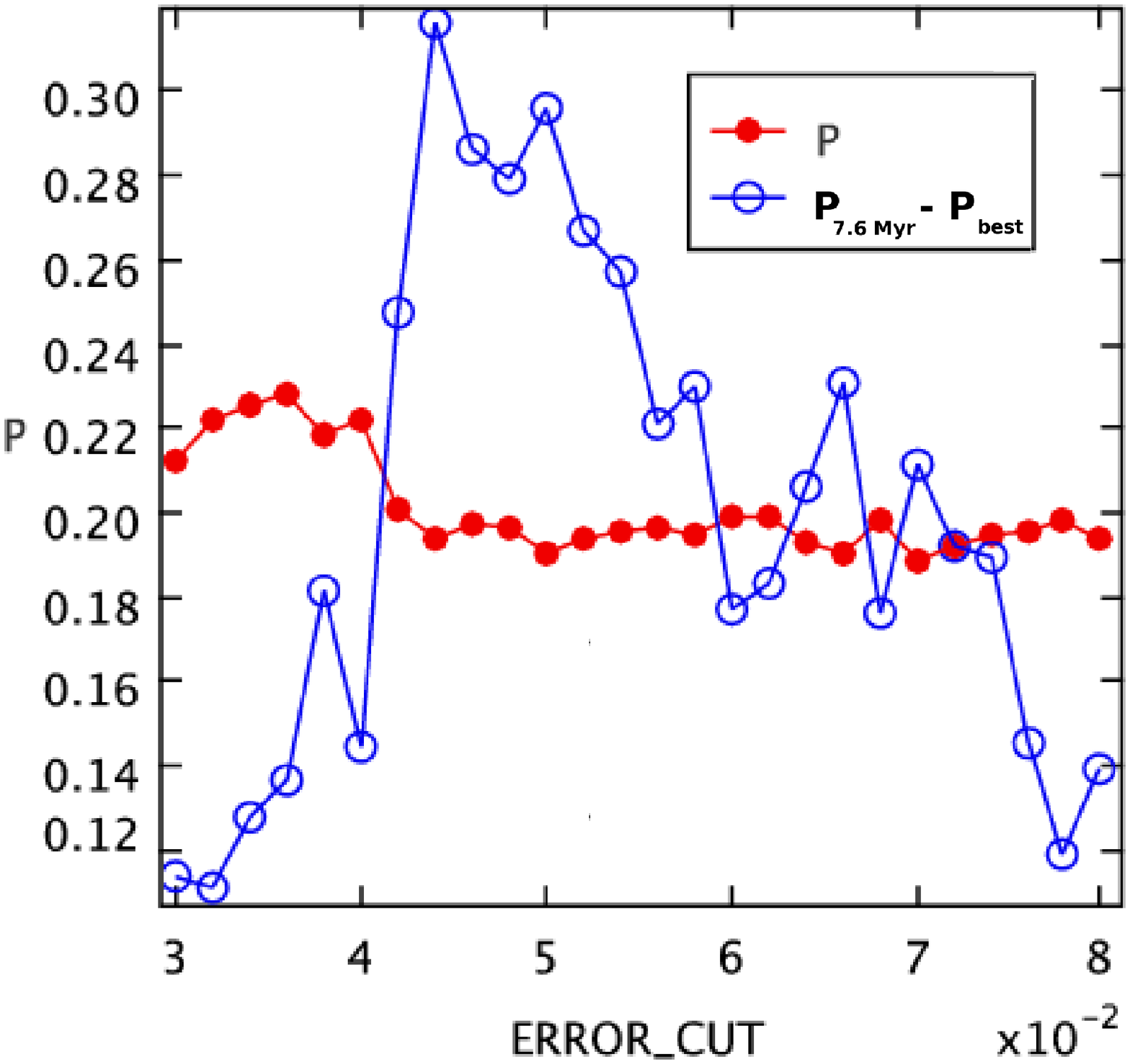}}
\end{tabular}
\caption[]{\small a (Left): The number of cluster candidates (closed circles) and field stars (open circles) changes as a function of the photometric error cut changes. b (Centre): Shows how the change in the photometric error cut effects the derived age and the size of the derived error in the age, i.e. the difference between the min and max isochrones. Using an error cut of 0.05 mag yields the smallest error in the age. c (Right): Shows how the change in the error cut effects the best fit value of P and the difference between the value of P for the best fit isochrone and the oldest, 7.6 Myr, available isochrone (see text).}
\label{fig:G42_errorparams}
\end{center}
\end{figure*}
\begin{figure*}
\begin{center}
\begin{tabular}{ccc}
\resizebox{55mm}{!}{\includegraphics[angle=90]{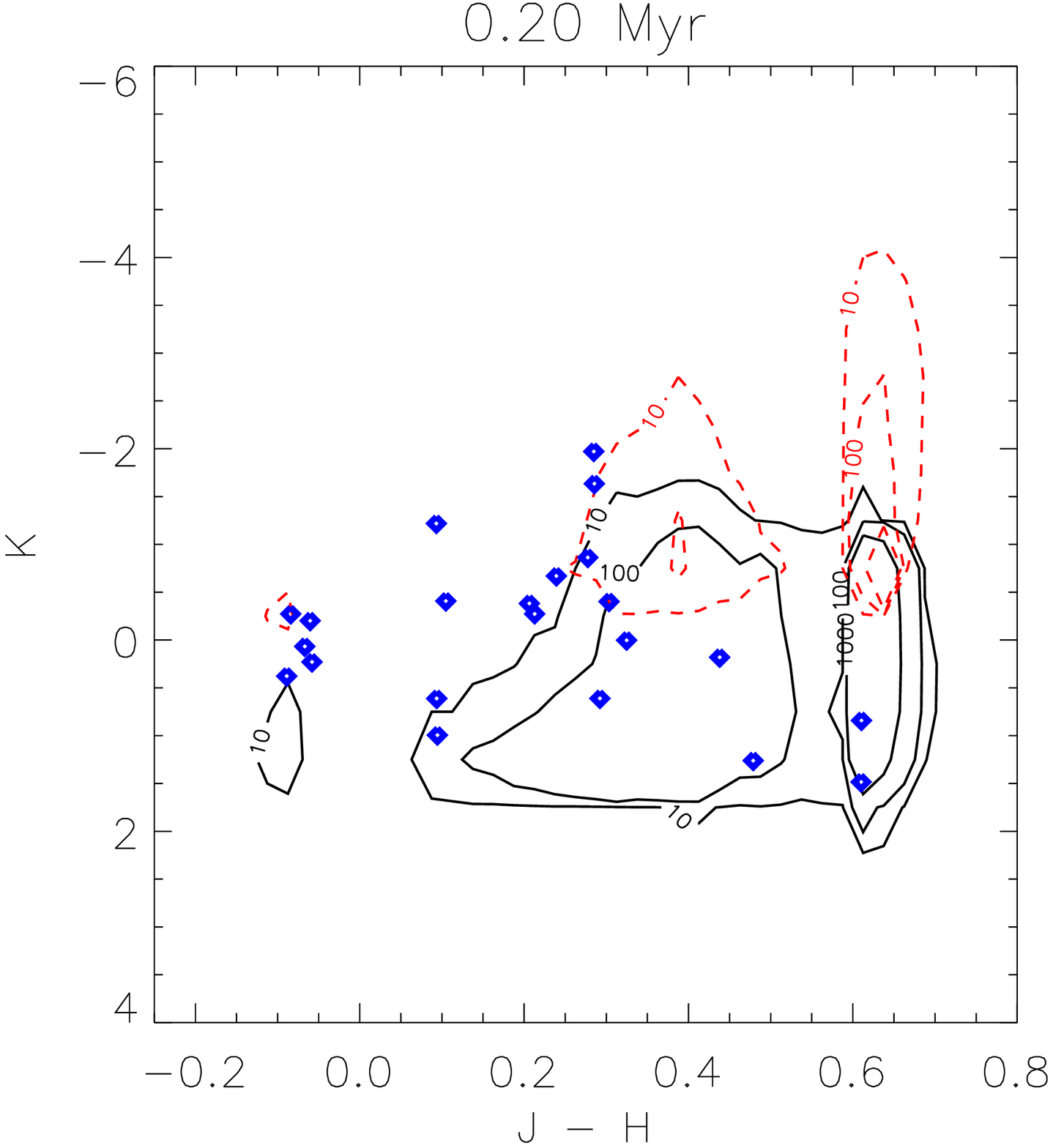}}&
\resizebox{55mm}{!}{\includegraphics[angle=90]{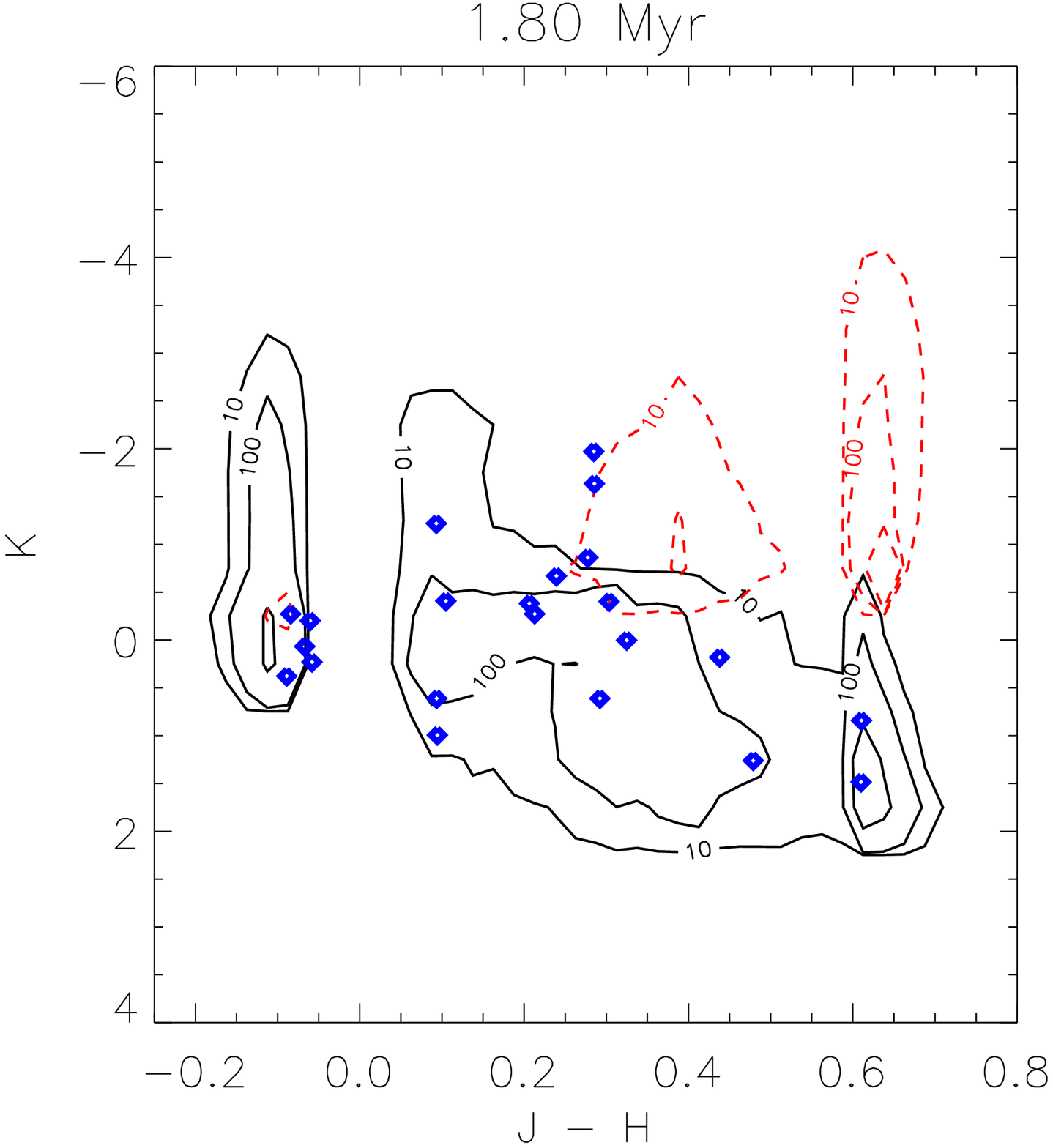}}&
\resizebox{55mm}{!}{\includegraphics[angle=90]{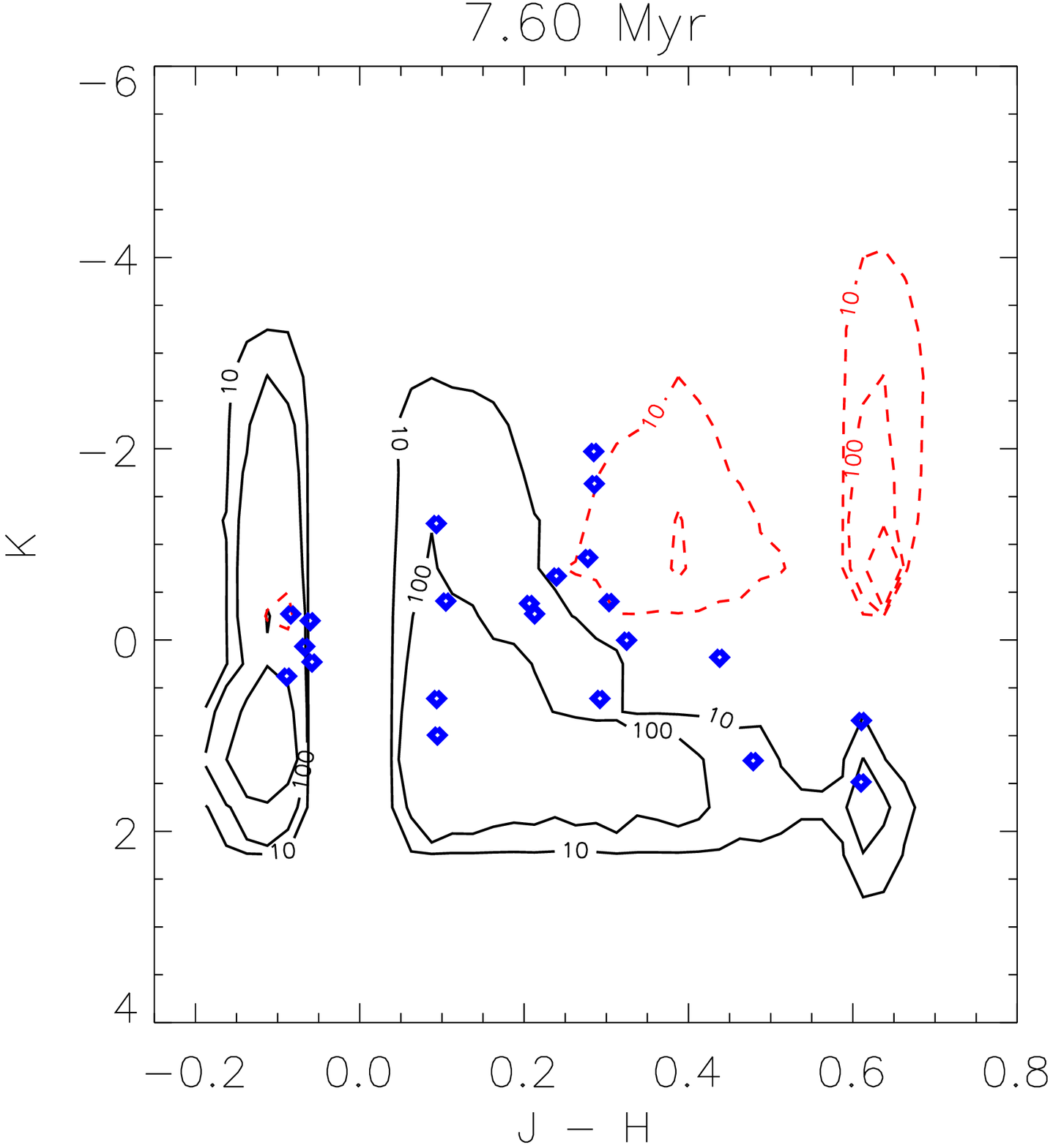}}
\end{tabular}
\caption[]{\small a (Left): A J-H vs K (intrinsic) dereddened CMD showing the position of the 22 cluster candidates extracted from the cluster associated with the RMS source G042.1268-0.6224 (blue diamonds), no binary stars have been included in the plot. Overlaid are both the solid contours of the 0.2 Myr cluster model and the dashed contours of the field population (provided by the $\Bes$ data, see Fig. \ref{fig:sudoCCD_field} (b)), showing where the artificial data are above 10, 100, and 1000 per 0.025 mag x 0.5 mag regions, after the 0.05 mag photometric error cut has been applied. b (Centre): Same as (a) except with the best fit 1.8 Myr cluster model. c (Right): Same as (a) except with the 7.6 Myr cluster model.}
\label{fig:0.2_2.0_7.6_G42_CMD}
\end{center}
\end{figure*}

The final 22 cluster candidates, extracted from the 3$\sigma$ contour surrounding the RMS source G042.1268-0.6224, have been dereddened and used to create the A$_J$ histogram presented in Fig. \ref{fig:Cluster_Extinction}. A skewed Gaussian has been fit to the A$_J$ histogram to estimate the extinction distribution of the cluster to be A$_J$=3.1$_{-0.2}^{+0.6}$ or A$_V$$\sim$10.9$_{-0.7}^{+2.1}$ mag. However, as mentioned previously in section \ref{sec:intial_cluster_param}, to replicate the sudden jump in extinction along the line of sight to the cluster the extinction distribution applied has been clipped at A$_J$=2.8, the smallest A$_J$ histogram bin. \\
\begin{figure}
\begin{center}
\begin{tabular}{c}
\resizebox{70mm}{!}{\includegraphics[angle=0]{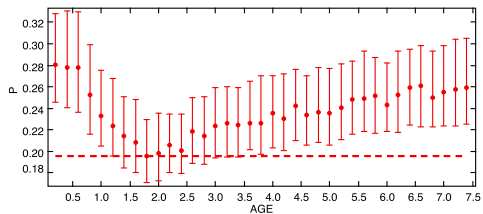}}
\end{tabular}
\caption[]{\small The minimisation of P for the 22 stars extracted from the cluster associated with the RMS source G042.1268-0.6224. P was mimimised at an age of 1.8 Myr. The presence of the 4 (4.3$\pm$1.2) field stars has also been modeled in the simulations.}
\label{fig:AGE_G42}
\end{center}
\end{figure}
$\indent$As shown in the previous section, the sample of cluster candidates also contains 4.3$\pm$1.2 background stars. For this reason, four field stars have also been included in the Monte Carlo simulations. Fig. \ref{fig:AGE_G42} shows that the best fit was made using the 1.8 Myr cluster, hence the derived age = 1.8$_{-0.6}^{+1.6}$ Myr, assuming the distance to be 4.3 kpc. Using this fitting method there is a trade-off between the number of sources in the real sample used to perform the fit and the photometric errors on the sources themselves. This is because it is possible to remove the more unreliable data, thus reducing the scatter in the data, but this will also decrease the sample size, thus increasing the uncertainty in the fit. A photometric error cut of 0.05 mag, applied to both the real and synthetic data, was used to perform the above analysis. This value was chosen after performing three different tests while changing the size of the error cut, measuring the size of the derived error in the age, the size of the value of P for the best fit isochrone, and the difference between the values of P for the best fit isochrone and the oldest isochrone available. The effect of the later test is apparent in Fig. \ref{fig:AGE_G42}, the oldest isochrone available, 7.6 Myr, does not fit well with the data, hence the poor fit. Therefore as the error cut is optimised, the difference between the value of P for the 7.6 Myr isochrone and the best fit isochrone should increase. Figs. \ref{fig:G42_errorparams} (a), (b) and (c) illustrate the effects of changing the photometric error cut. As a function of photometric error cut, the smallest error derived in the age (Fig. \ref{fig:G42_errorparams} (b)) and the sharpest minimum, i.e. the largest difference between the value of P for the best fit isochrone and the oldest, 7.6 Myr, available isochrone (Fig. \ref{fig:G42_errorparams} (c)), occur at an error cut of $\sim$0.05 mag. It is for this reason that an error cut of 0.05 mag that has been used for G042.1268-0.6224.\\
$\indent$Fig. \ref{fig:0.2_2.0_7.6_G42_CMD} (a), (b) and (c) contain dereddened CMDs showing the positions of the 22 cluster candidates. Overlaid are the contours of the 0.2, 1.8 and 7.6 Myr cluster models respectively, after the 0.05 mag photometric error cut has been applied. It is clear that the data fit the 1.8 Myr contours more closely than the 0.20 and 7.6 Myr contours. The 1.8 Myr contours do not cover all of the data which may indicate an age spread in the cluster. A few of the outliers could also be background giants as they align with the field star contours.\\
$\indent$To determine the error in the age caused by the error in the assumed distance to the cluster, D = 4.3$\pm$0.5 kpc, the cluster models were reproduced assuming the full 1$\sigma$ error in the distance. This has the effect of including/excluding some of the faintest and brightest stars into the cluster models, as at smaller distances the former are more likely to be observed and the latter more likely to be saturated, and vice versa. Assuming a distance of 3.8 and 4.8 kpc, applying a 0.05 mag error cut and including four (4.3$\pm$0.7) field stars in the simulations, the derived ages were 2.0$_{-0.6}^{+1.7}$ Myr and 1.4$_{-0.8}^{+1.4}$ Myr respectively. Both of these results are consistent with the age at the assumed distance.\\
\begin{figure}
\begin{center}
\begin{tabular}{c}
\resizebox{75mm}{!}{\includegraphics[angle=0]{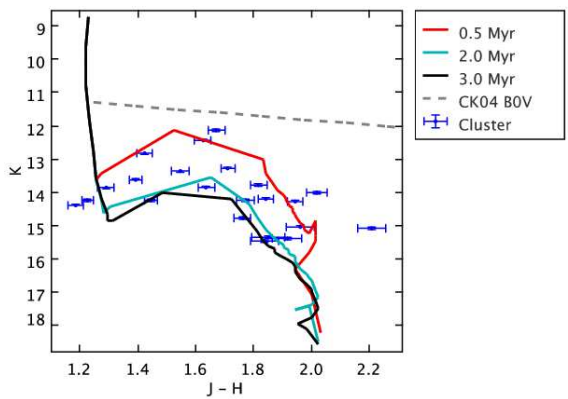}}
\end{tabular}
\caption[]{\small A J-H vs K (apparent colours and magnitude) CMD of the 22 cluster candidates (blue error bars) from the cluster containing G042.1268-0.6224. All foreground and IRXS stars have been removed from the sample, leaving only genuine cluster members and an estimated 4.3$\pm$0.7 field stars. The 0.5, 2.0 and 3.0 Myr isochrones are plotted as red, light blue and black lines respectively. Each has been placed at the assumed cluster distance and reddened to the median measured extinction (A$_J$=3.1 mag). Due to the limited differential reddening, the cluster candidates align well with the isochrones. A CK04 B0V reddening track has also been plotted for reference to the direction of reddening.}
\label{fig:G42_CMDnondered}
\end{center}
\end{figure}
$\indent$The derived cluster age of 1.8$_{-0.6}^{+1.6}$ Myr suggests that star formation in the cluster has finished. This is consistent with the identification of G042.1268-0.6224 in the RMS survey as a diffuse HII region. The cluster associated with G042.1268-0.6224 is not very embedded as the range in the A$_J$ histogram covers only 2 mag, corresponding to a differential reddening of A$_V$$\sim$7 mag between cluster members. Again this is consistent with the older age for the cluster if it is assumed that an older cluster has had more time to disperse its natal cloud. The limited differential reddening is very apparent in the CCD (Fig. \ref{fig:IRXS_CCD} (a)) that has had foreground stars removed, as the remaining data trace out the shape of a reddened intrinsic colour curve. These stars have been plotted in apparent colour-magnitude space
 in Fig. \ref{fig:G42_CMDnondered}. The 0.5, 2.0 and 3 Myr isochrones have been placed at the assumed cluster distance and reddened to the median extinction, A$_J$=3.1, using extinction values from an A0V reddening track. From inspection of this apparent CMD, it is clear that the age of the cluster is somewhere between 0.5 and 3.0 Myr. This is therefore within agreement with the previously derived result but much less certain.

\subsection{A Further Two Embedded Clusters}

Two more embedded clusters are examined in this paper, centred on the RMS sources G048.9897-0.2992 and G010.1615-0.3623. A brief description of each source is given below and figures for each cluster, containing UKIDSS K band images, stellar surface density maps, CCDs illustrating the removal of both field stars and IRXS sources, age determination plots and dereddened CMDs showing both cluster candidates and contours illustrating the best-fit cluster models, are presented in Appendix \ref{sec:otherfigures} (available online only).

\subsubsection{G048.9897-0.2992}

\begin{figure}
\begin{center}
\begin{tabular}{c}
\resizebox{65mm}{!}{\includegraphics[angle=0]{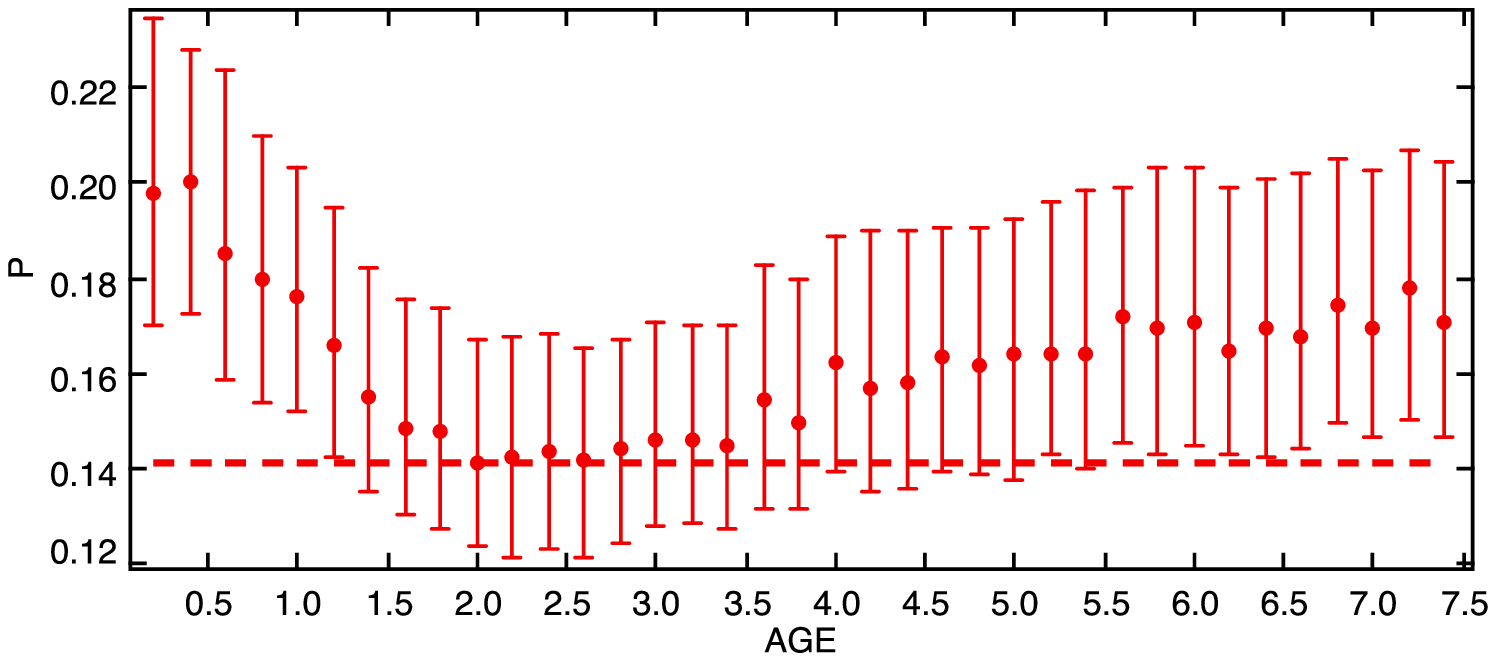}}\\
\resizebox{65mm}{!}{\includegraphics[angle=90]{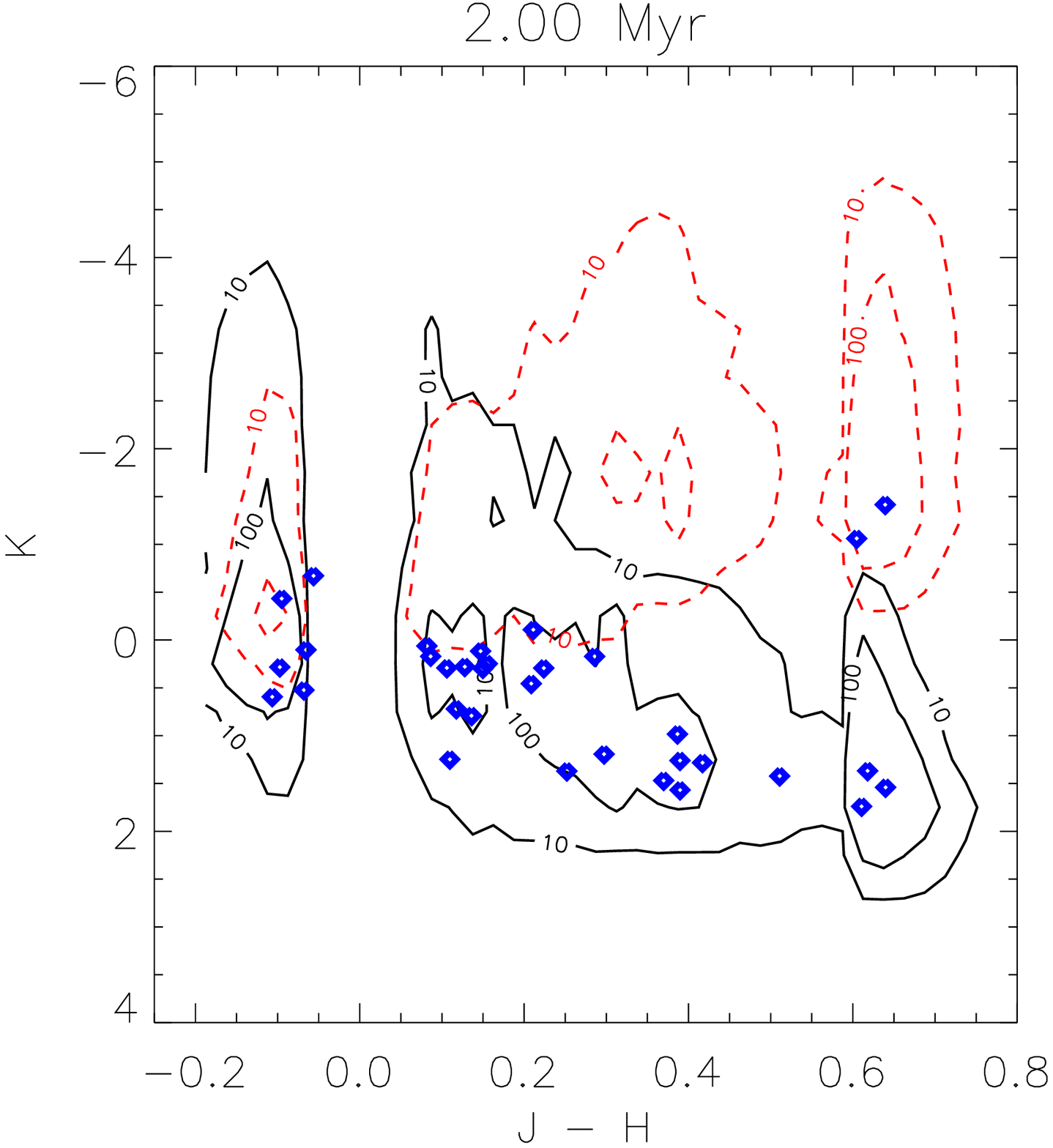}}
\end{tabular}
\caption[]{\small a (Top): Shows the minimisation of P (normalised to one) for the 33 cluster candidates extracted from the region containing the RMS source G048.9897-0.2992. P was mimimised at an age of 2.0 Myr. The presence of the 7 (7.0$\pm$1.9) field stars has also been modeled in the simulations. b (Bottom): A J-H vs K (intrinsic) dereddened CMD showing the position of the 33 cluster candidates, no binary stars have been included in the plot. Overlaid are both the solid contours of the 2.0 Myr cluster model and the dashed contours of the field population (provided by the $\Bes$ data), showing where the artificial data are above 10, 100, and 1000 per 0.025 mag x 0.5 mag regions, after the 0.064 mag photometric error cut has been applied. Only two stars are not contained within the 2.0 Myr contours, at J-H$\sim$0.6, K$\sim$-1, as they are contained within the field star contours it is highly plausible that they are indeed background giants, however this same position is also occupied by very young stars.} 
\label{fig:AGE_G48} 
\end{center}
\end{figure}
\begin{figure}
\begin{center}
\begin{tabular}{c}
\resizebox{75mm}{!}{\includegraphics[angle=0]{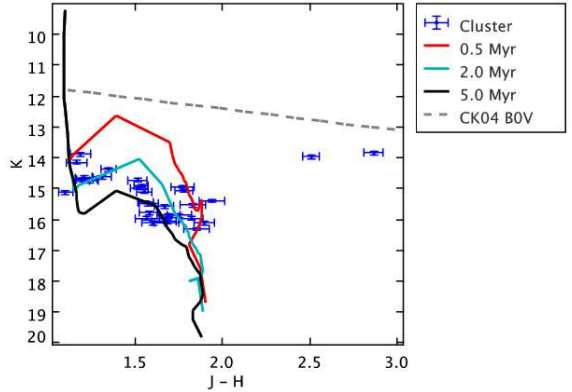}}
\end{tabular}
\caption[]{\small A J-H vs K (apparent colours and magnitude) CMD of the 33 cluster candidates (blue error bars) from the cluster containing G048.9897-0.2992. All foreground and IRXS stars have been removed from the sample, leaving only genuine cluster members and an estimated 7.0$\pm$1.9 field stars. The 0.5, 2.0 and 5.0 Myr isochrones are plotted as red, light blue and black lines respectively. Each has been placed at the assumed cluster distance and reddened to the median measured extinction (A$_J$=2.8 mag). Due to the limited differential reddening, the cluster candidates align well with the isochrones. A CK04 B0V reddening track has also been plotted for reference to the direction of reddening. Two cluster candidates have very red colours and are most likely giant stars, incidentally they are also discussed in Fig. \ref{fig:AGE_G48} (b) where it has also been suggested that they are indeed background giants. }
\label{fig:G48_CMDnondered}
\end{center}
\end{figure}

The RMS source G048.9897-0.2992 has both an MYSO and a UCHII region. It has a RMS kinematic distance of 5.2$\pm$0.2 kpc, is located on the tangent point so therefore there is no ambiguity, and has a derived distance from extinction measurements, that is independent to kinematic methods, D = 5.6$\pm$0.5 kpc \citep{stead10}. The latter distance, with associated errors, is used in this section to determine the age of the cluster. This cluster was originally identified as [BDS2003] 139 \citep{bica03b} and is also studied by \citet{kumar04}. The latter author study several clusters contained within the W51 GMC and derive ages for some of them. Although they do not give an estimate of an age for G49.0-0.3 (the cluster containing G048.9897-0.2992), they state that it and a neighbouring cluster, G48.9-0.3, are considered the youngest of all clusters inside W51 as they have a high fraction of young stellar objects. As they quote G48.9-0.3 as having an age of 1.4 Myr and the remaining clusters typically 2-3 Myrs, it can probably be considered that G048.9897-0.2992 has an age younger than 2 -3Myrs and has a similar age to G48.9-0.3. \\
$\indent$A stellar surface density map was used to select a $\sim$0.56 arcmin$^2$ region of the sky from which to extract a total of 101 stars, using a photometric error cut of 0.064 mag. After removal of 57 foreground stars and 11 IRXS candidates, 3 of which would have remained unidentified without GLIMPSE data, 33 cluster candidates remain. These 33 cluster candidates have been dereddened to estimate the extinction distribution towards G048.9897-0.2992 to be A$_J$=2.8$_{-0.3}^{+1.1}$ mag,  or A$_V$$\sim$9.9$_{-1.1}^{+3.9}$ mag, with a lower limit clipping of A$_J$=2.0 mag. \\
$\indent$Fig. \ref{fig:AGE_G48} (a) shows that the best fit was made using the 2.0 Myr cluster model, hence the derived age = 2.0$_{-0.8}^{+3.4}$ Myr. Offset fields have been used to estimate that there are 7.0$\pm$1.9 field stars contained within this final sample, representing a $\sim$21\% contamination; these have also been included in the simulations. Fig. \ref{fig:AGE_G48} (b) contains a dereddened CMD showing the positions of the 33 cluster candidates. Overlaid are the contours of the best-fit 2.2 Myr cluster model, and the majority of the stars are contained within the 100 sources per 0.025 mag x 0.5 mag contour. \\
$\indent$In comparison to G042.1268-0.6224, the cluster associated with G048.9897-0.2992 possesses more differential reddening, as the A$_J$ histogram of the former covers an extinction range of 2 mag, whereas the latter covers 3 mag, corresponding to a differential reddening of A$_V$$\sim$10.6 mag between cluster members. The similarities of the ages and the extinction distributions of both G048.9897-0.2992 and G042.1268-0.6224 are consistent with the idea that older clusters have had more time to disperse their natal clouds. The derived age for G048.9897-0.2992 is less certain than the age derived for G042.1268-0.6224.  Although the differential extinction is higher for G048.9897-0.2992, the main difference is likely to be that on average the photometry is slightly worse. \\
$\indent$As was done for G042.1268-0.6224, an apparent colour-magnitude has been produced for the 33 cluster candidates surrounding G048.9897-0.2992, presented in Fig. \ref{fig:G48_CMDnondered}. The 0.5, 2.0 and 5.0 Myr isochrones have been placed at the assumed cluster distance and reddened to the median extinction, A$_J$=3.1, using extinction values from an A0V reddening track. From inspection of this apparent CMD, it is clear that the age of the cluster is somewhere between 0.5 and 5.0 Myr. Using the traditional methods the derived age of the cluster has not been very well constrained, however the result is within agreement with the result derived in this paper. \\
$\indent$To determine the error in the age caused by the error in the assumed distance to the cluster, D = 5.6$\pm$0.5 kpc, the cluster models were reproduced assuming the full 1$\sigma$ error in the distance. Assuming a distance of 5.1 and 6.1 kpc, applying a 0.064 mag error cut and including seven (7.0$\pm$1.9) field stars in the simulations, the derived ages were 2.6$_{-1.1}^{+3.3}$ Myr and 1.6$_{-0.8}^{+3.0}$ Myr respectively, both are consistent with the previous result.

\subsubsection{G010.1615-0.3623}
\label{sec:age_det_W31}

\begin{figure}
\begin{center}
\begin{tabular}{c}
\resizebox{65mm}{!}{\includegraphics[angle=0]{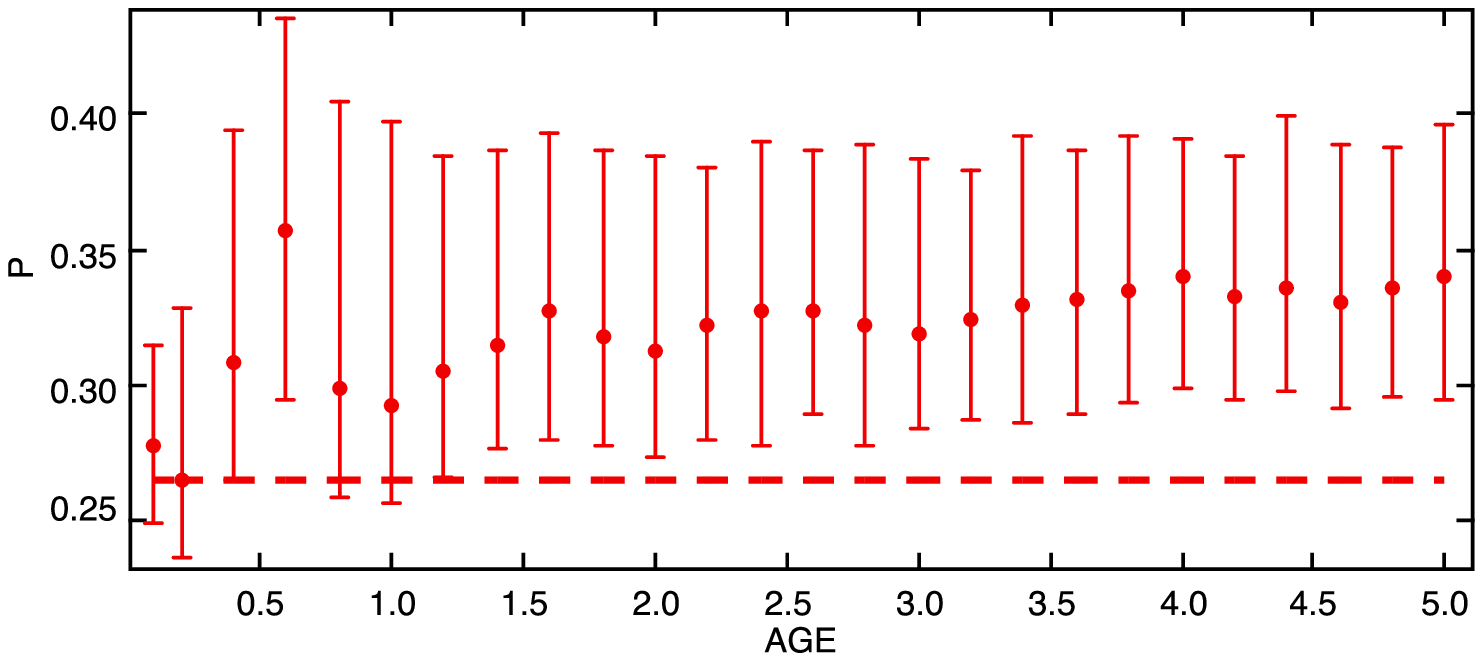}}\\
\resizebox{65mm}{!}{\includegraphics[angle=90]{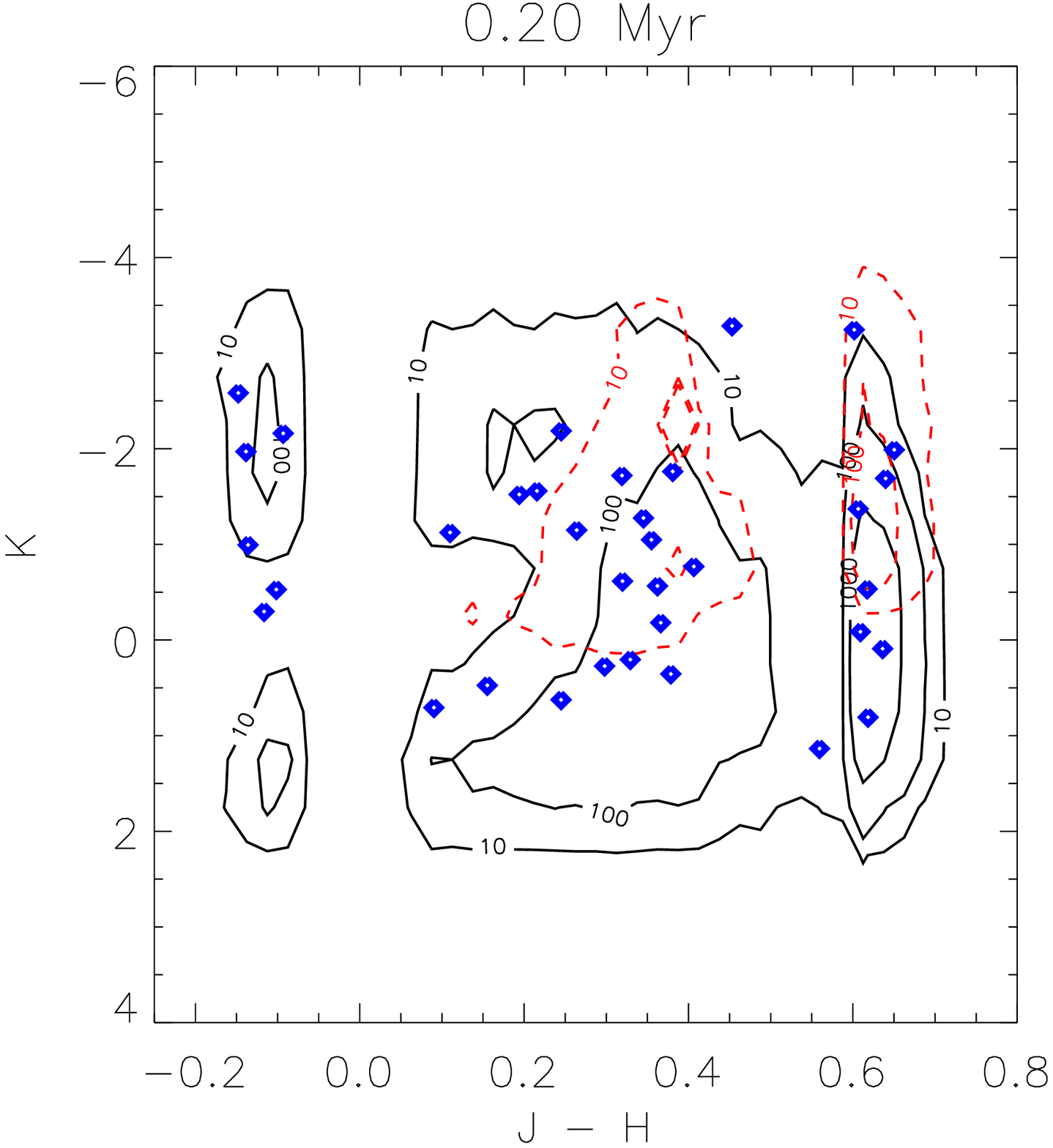}}
\end{tabular}
\caption[]{\small a (Top): Shows the minimisation of P (normalised to one) for the 35 cluster candidates extracted from the region containing the RMS source G010.1615-0.3623. P was mimimised at an age of 0.2 Myr. The presence of the 2 (2.0$\pm$0.8) field stars has also been modeled in the simulations. b (Bottom): A J-H vs K (intrinsic) dereddened CMD showing the position of the 35 cluster candidates, no binary stars have been included in the plot. Overlaid are both the solid contours of the 0.2 Myr cluster model and the dashed contours of the field population (provided by the $\Bes$ data), showing where the artificial data are above 10, 100, and 1000 per 0.025 mag x 0.5 mag regions, after the 0.076 mag photometric error cut has been applied. } 
\label{fig:AGE_G10} 
\end{center}
\end{figure}
\begin{figure}
\begin{center}
\begin{tabular}{c}
\resizebox{75mm}{!}{\includegraphics[angle=0]{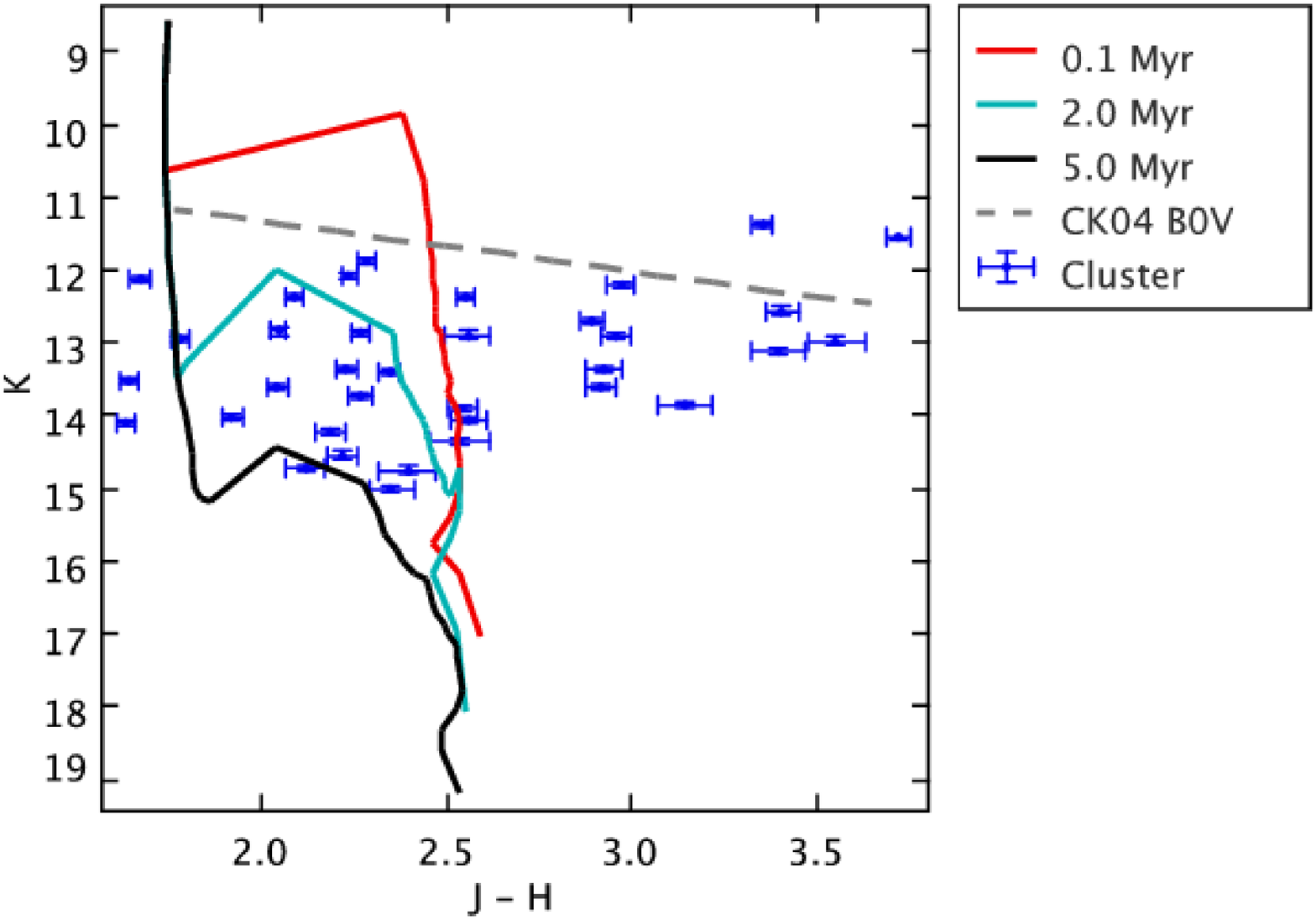}}
\end{tabular}
\caption[]{\small A J-H vs K (apparent colours and magnitude) CMD of the 35 cluster candidates (blue error bars) from W31. All foreground and IRXS stars have been removed from the sample, leaving only genuine cluster members and an estimated 2.0$\pm$0.8 field stars. The 0.1, 2.0 and 5.0 Myr isochrones are plotted as red, light blue and black lines respectively. Each has been placed at the assumed cluster distance and reddened to the median measured extinction (A$_J$=4.3 mag). Due to the highly differential reddening, the cluster candidates do not align well with the isochrones. For this reason it is only possible to say that the cluster is probably younger than 5 Myr. A CK04 B0V reddening track has also been plotted for reference to the direction of reddening.}
\label{fig:G10_CMDnondered}
\end{center}
\end{figure}

G010.1615-0.3623 is situated within a well known star-forming region, W31. As such it is well documented in the literature, and spectrophotometric distances have been derived for the region \citep[3.4$\pm$0.3 kpc,][]{blum01}. This distance shall be used to determine the age of the cluster. \citet{blum01} spectral type the six brightest stars and find four to be massive O stars. Such a high concentration of massive stars would suggest a large total number of stars. \citet{blum01} and \citet{furness10} also estimate the age of W31 using spectroscopic spectral typing.\\
$\indent$A stellar surface density map was used to select a $\sim$0.70 arcmin$^2$ region of the sky from which to extract a total of 86 stars, using a photometric error cut of 0.076 mag. After removal of 41 foreground stars and 5 IRXS candidates, 1 of which would have remained unidentified without GLIMPSE data, 40 cluster candidates remain. From these 40 sources there are a further 5 that have atypical stellar colours as they lie above the M0V reddening track even when their 1$\sigma$ errors are considered.
 It is highly probable that these sources are actually late-type bulge giant stars, as on an H-K vs J-H CCD (see Fig. \ref{fig:offsetVSclusterG48}) they are positioned around the M0III reddening track. As the cluster is located within close proximity to the Galactic centre, such stars may well be expected in the CCD and a similar star is also present in one of the offset fields. These 5 stars have therefore been removed from the final sample, leaving 35 cluster candidates. Offset fields have been used to estimate that there are 2.0$\pm$0.8 field stars, representing a $\sim$6\% contamination, contained within this final sample. These 35 cluster candidates have been dereddened to estimate the extinction distribution towards G010.1615-0.3623 to be A$_J$=4.3$_{-0.6}^{+1.4}$ mag, with a lower limit clipping of A$_J$=3.0 mag. \citet{blum01} make extinction measurements to 4 O-type stars and derive an average extinction of A$_J$=5.7$\pm$1.0 mag. This extinction distribution is consistent with the previously quoted result, however a true comparison cannot be made as the four O-type stars are saturated in UKIDSS. \\
$\indent$Fig. \ref{fig:AGE_G10} (a) shows the minimisation of P and the best fit was made using the 0.2 Myr cluster model, hence the derived age = 0.2$_{-0.2}^{+1.0}$ Myr. However, this minimisation did exclude the 0.8 Myr isochrone. Fig. \ref{fig:AGE_G10} (b) contains a dereddened CMD showing the positions of the 35 cluster candidates. Overlaid are the contours of the best-fit 0.2 Myr cluster model. The 0.20 Myr contours do not cover all of the data suggesting an age spread within W31. \citet{blum01} determine the age of four the O-type stars contained within the region and state that the stars are likely to be less than 1 Myr old. \cite{furness10} derive the age of the same four O-type stars, using new spectroscopy, to be $\sim$0.6 Myr. However they suggest that one star could be an unresolved binary, and if it were, the age of the cluster would be younger at $\sim$0.5 Myr. \\
$\indent$In comparison to G042.1268-0.6224 and G048.9897-0.2992, W31 is much more embedded, as the A$_J$ histogram of the former two have extinction ranges of 2 and 3 mag respectively, whereas the latter covers 5.5 mag, corresponding to a differential reddening of A$_V$$\sim$19 mag between cluster members. Due to the young age derived, the embedded nature of W31 is again consistent with the idea that the older clusters have had more time to disperse their natal clouds. An apparent colour-magnitude has been produced for the 35 cluster candidates extracted from W31, presented in Fig. \ref{fig:G10_CMDnondered}. The 0.1, 2.0 and 5.0 Myr isochrones have been placed at the assumed cluster distance and reddened to the median extinction, A$_J$=4.3, using extinction values from an A0V reddening track. Unlike for the two previous clusters, it is not possible to constrain an age for W31, using the traditional age determination methods with an apparent CMD, it is only possible to say that the cluster is probably younger than 5 Myr. This is due to the highly differential reddening between the cluster members. Finally, as the minimisation of P excluded the 0.8 Myr isochrone, this could be indicative of an age spread or even multi-epoch star formation. However more reliable photometry would be required to confirm this conclusion. \\
$\indent$To determine the error in the age caused by the error in the assumed distance to the cluster, D = 3.4$\pm$0.3 kpc, the cluster models were reproduced assuming the full 1$\sigma$ error in the distance. Assuming a distance of 3.1 and 3.7 kpc, applying a 0.076 mag error cut and including 2 field stars in the simulations, the derived ages were 0.2$_{-0.0}^{+0.0}$ Myr and 0.2$_{-0.0}^{+0.0}$ Myr respectively, both are consistent with the previous result. When considering the error in the distance, the results derived here are therefore consistent with both \citet{blum01} and \cite{furness10}. Furthermore, due to the young age derived, it is likely that star formation has yet to end.

\section{Conclusions}
\label{sec:conclusion}

Using a new Monte Carlo method ages have been derived for three different clusters by comparing real dereddened data to artificial data, that can be considered a good approximation to the real data. Stars have been dereddened to an intrinsic locus taking the inevitable problems for early and late type stars into account. Synthetic clusters, created using realistic photometry errors, selection effects and field star contamination, are treated in the same way and therefore account for the same dereddening issues.  \\
$\indent$It was not possible to analyse the most massive stars in the real clusters as they were saturated in UKIDSS. However, as a final test of the age determination method, it is possible to estimate the total luminosity of the real cluster and compare this value with the bolometric luminosities contained in the RMS survey from far-IR fluxes \citep{mottram11a}. Artificial clusters have been created that are one thousand times the size of each of the three real clusters. Each has the same age as those derived in this paper for the corresponding real clusters. The total luminosity of these artificial clusters has been determined, from the data available in each isochrone, and then scaled down by a factor of one thousand, thereby reducing the effect of any random variation. The calculated total luminosities are 10x10$^4$, 9x10$^4$ and 17x10$^4$ L$_{\odot}$, and correspond to 2.8, 2.4 and 1.5 times the measured RMS luminosities, for the clusters G042.1268-0.6224, G048.9897-0.2992 and G010.1615-0.3623  respectively. The RMS luminosities should represent a lower limit of the total cluster luminosity. This is because some of the optical/UV radiation will leak out and is not re-radiated in the far-IR. For this reason these values quoted here suggest that a reasonable approximation of the actual cluster luminosities has been performed. \\
$\indent$None of the three clusters show obvious signs of a large age spread within the resolution of the age determination process. However analysis the cluster containing G048.9897-0.2992 had a derived age of 2.0 Myr, this was despite the identification of both an MYSO and UCHII region in the RMS survey. These objects will have typical ages of 0.1 Myr \citep{mottram11b}, therefore this may suggest multi-epoch star formation rather than continual star formation over an extended period of time. To ascertain this, future work will model a star formation history as opposed to assuming a constant age.\\
$\indent$The age determination method presented in this paper provides a new way to determine the ages of embedded clusters when only photometric data are available. Results were found to be consistent with those determined using more traditional methods, however they have been more reliably determined. Where applicable, the results were also consistent with those derived via spectroscopy. Furthermore, unlike the more tradition photometric age determination methods, this new method can be applied to clusters where the extinction between members is highly differential. This method could be extended to clusters in the southern hemisphere where deeper VISTA data are becoming available \citep{minniti09}. The deeper data would also yield more reliable results. Finally, tests to this method could be performed using infrared multi-object spectroscopy. 

\appendix
\section{Figures for G048.9897-0.2992 and G010.1615-0.3623}
\label{sec:otherfigures}

\begin{figure*}
\begin{center}
\begin{tabular}{cc}
\resizebox{65mm}{!}{\includegraphics[angle=0]{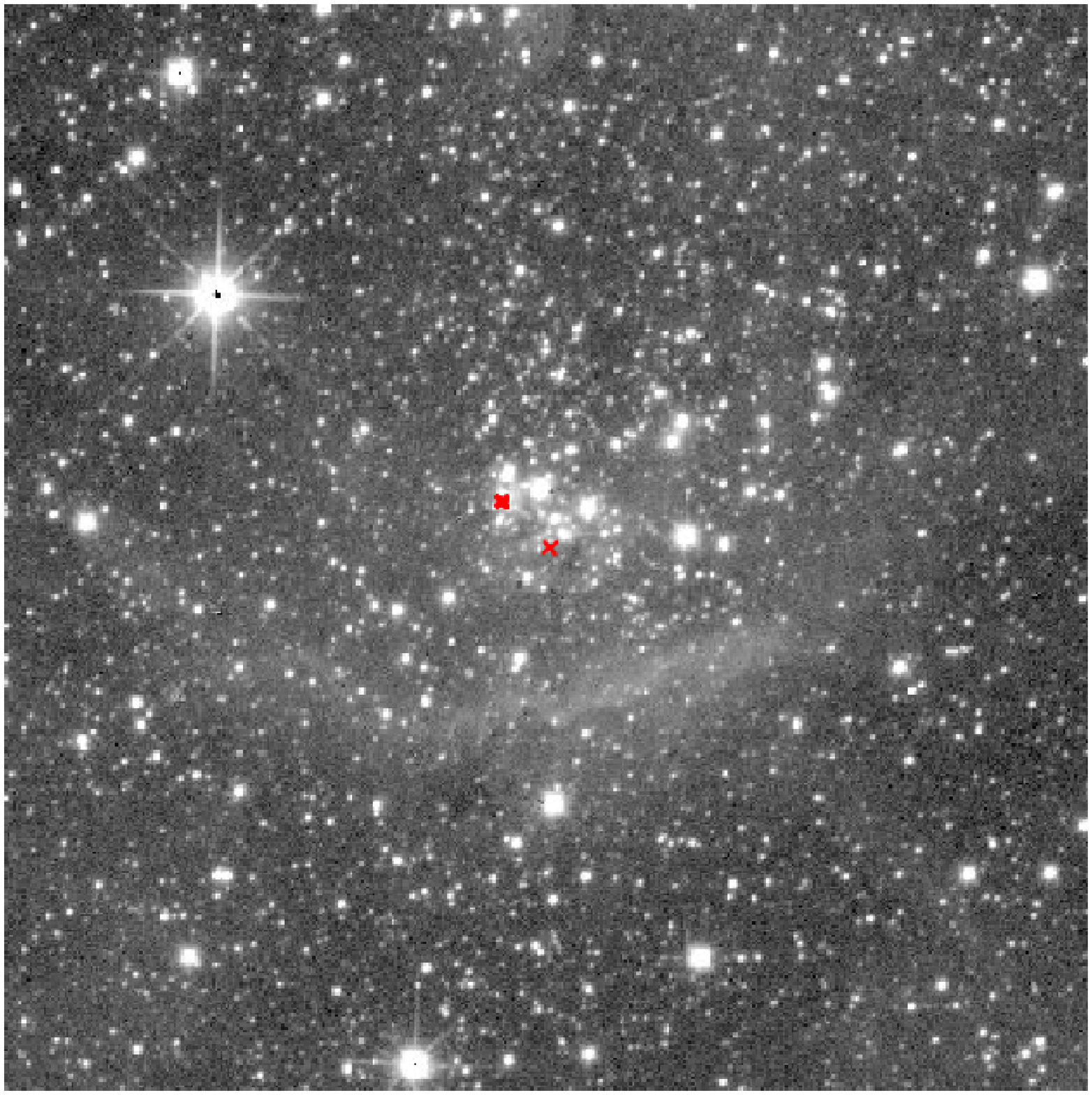}}&
\resizebox{65mm}{!}{\includegraphics[angle=0]{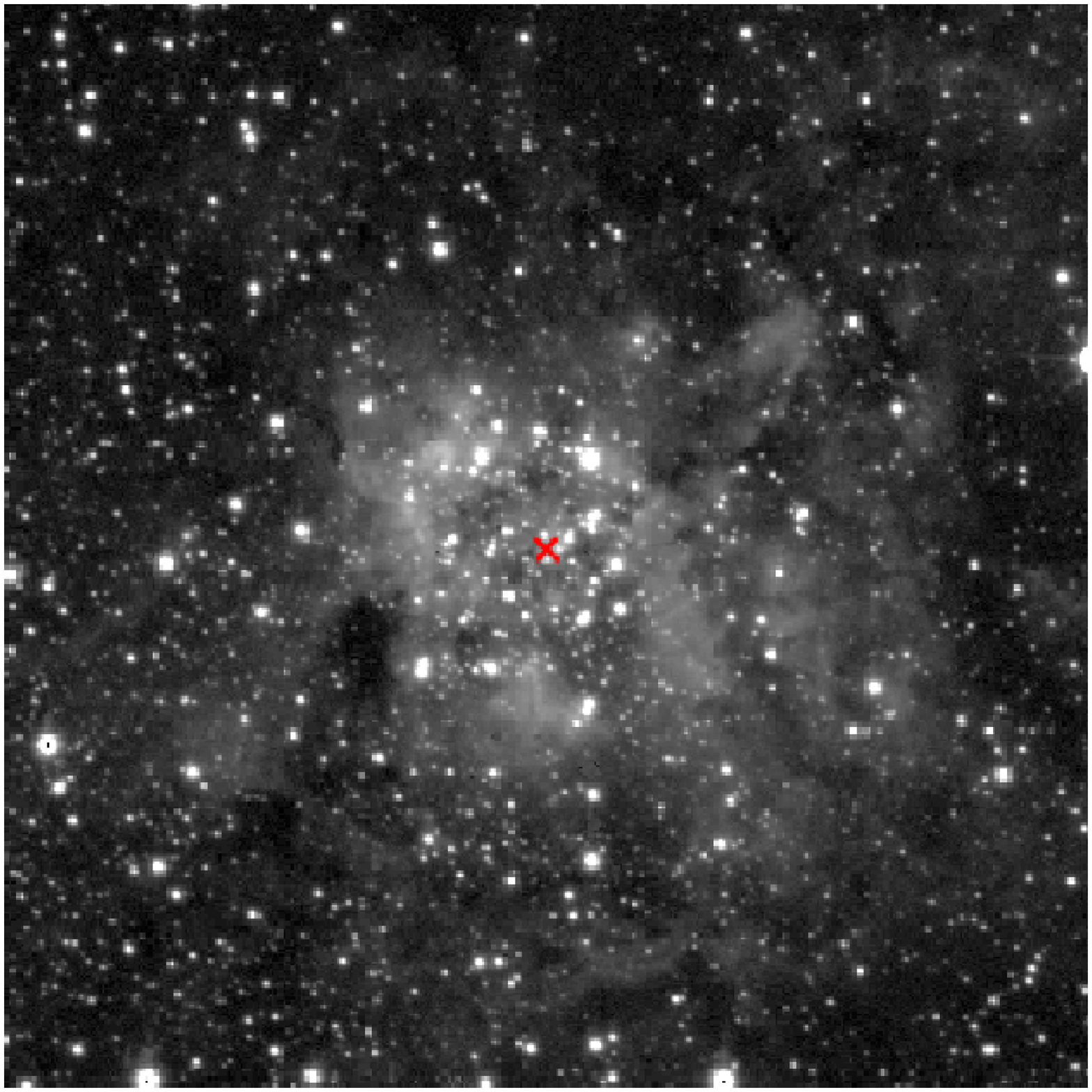}}
\end{tabular}
\caption[]{\small a (Left): A 3$\arcmin$x3$\arcmin$ UKIDSS K band image of the area of sky surrounding the RMS source G048.9897-0.2992 (indicated with crosses). North is up and east is left. As there is both an MYSO and UCHII region contained within this image, there are two crosses and the MYSO is the most northern cross. b (Right): Same as (a) except for the RMS source G010.1615-0.3623} 
\label{fig:G10.1615_SpatialMap} 
\end{center}
\end{figure*}

\begin{figure*}
\begin{center}
\begin{tabular}{cc}
\resizebox{75mm}{!}{\includegraphics[angle=90]{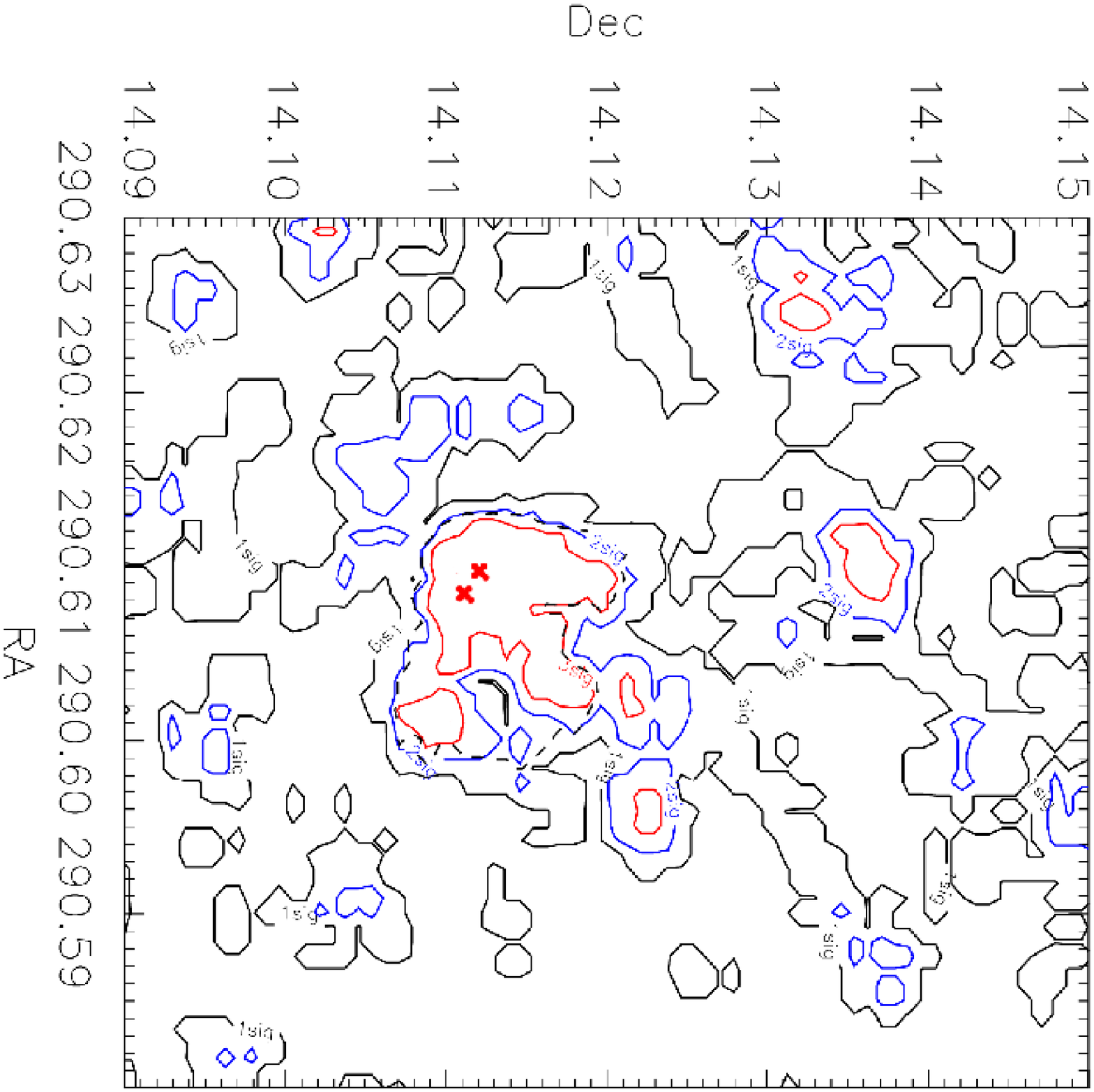}}&
\resizebox{75mm}{!}{\includegraphics[angle=90]{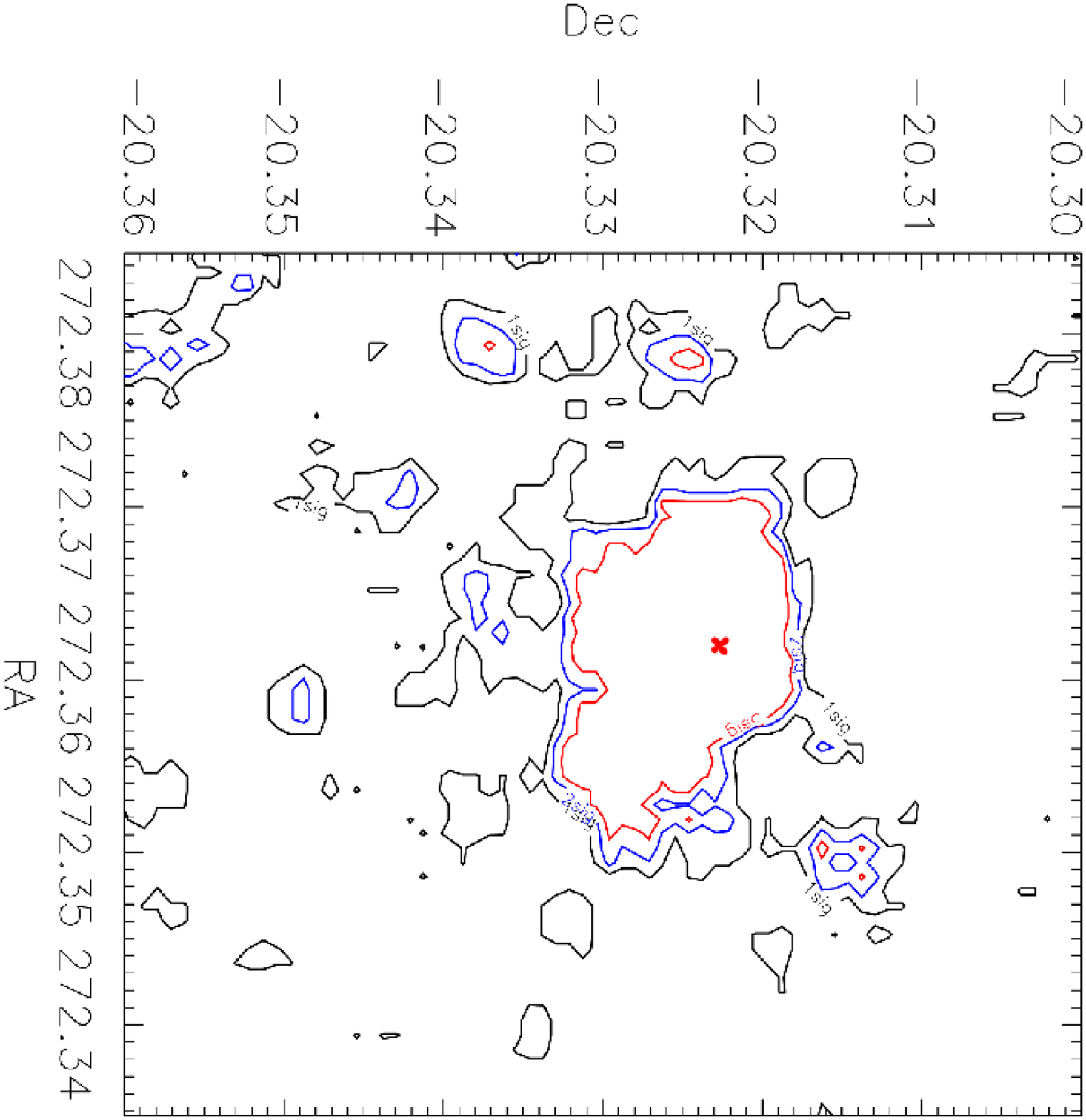}}
\end{tabular}
\caption[]{\small a (Left): A 3$\arcmin$x3$\arcmin$ stellar surface density map centred close to the RMS source G048.9897-0.2992 (indicated with crosses). Black, blue and red contour lines represent regions where star counts exceed the 1, 2 and 3$\sigma$ levels. The dashed black contour towards the centre of the map has been used to extract cluster candidates b (Right): Same as (a) except for the RMS source G010.1615-0.3623 and the closed 3$\sigma$ contour towards the centre of the map has been used to extract cluster
 candidates.} 
\label{fig:G048.9897_SpatialMap} 
\end{center}
\end{figure*}

\begin{figure*}
\begin{center}
\begin{tabular}{cc}
\resizebox{70mm}{!}{\includegraphics[angle=0]{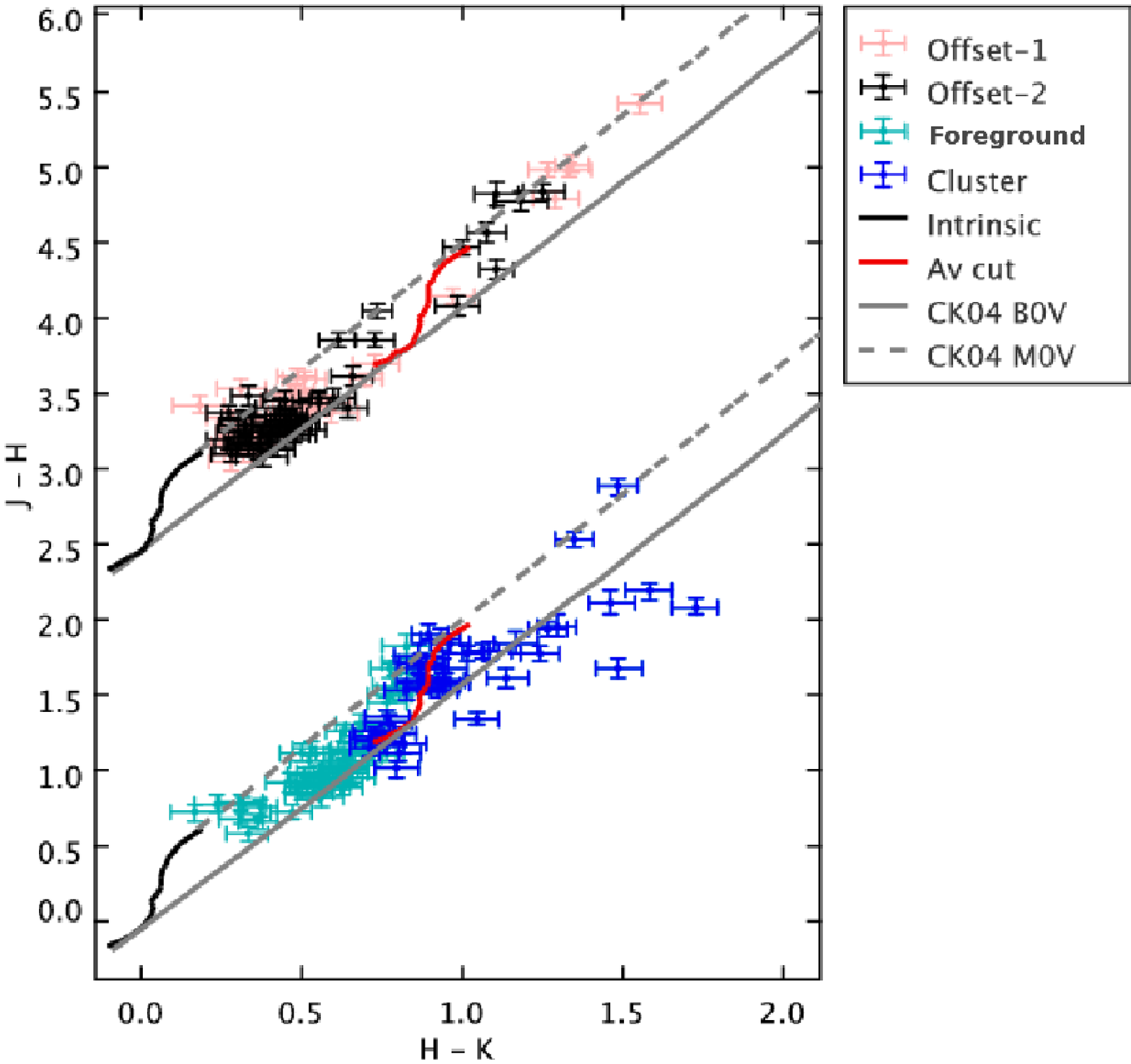}} &
\resizebox{70mm}{!}{\includegraphics[angle=0]{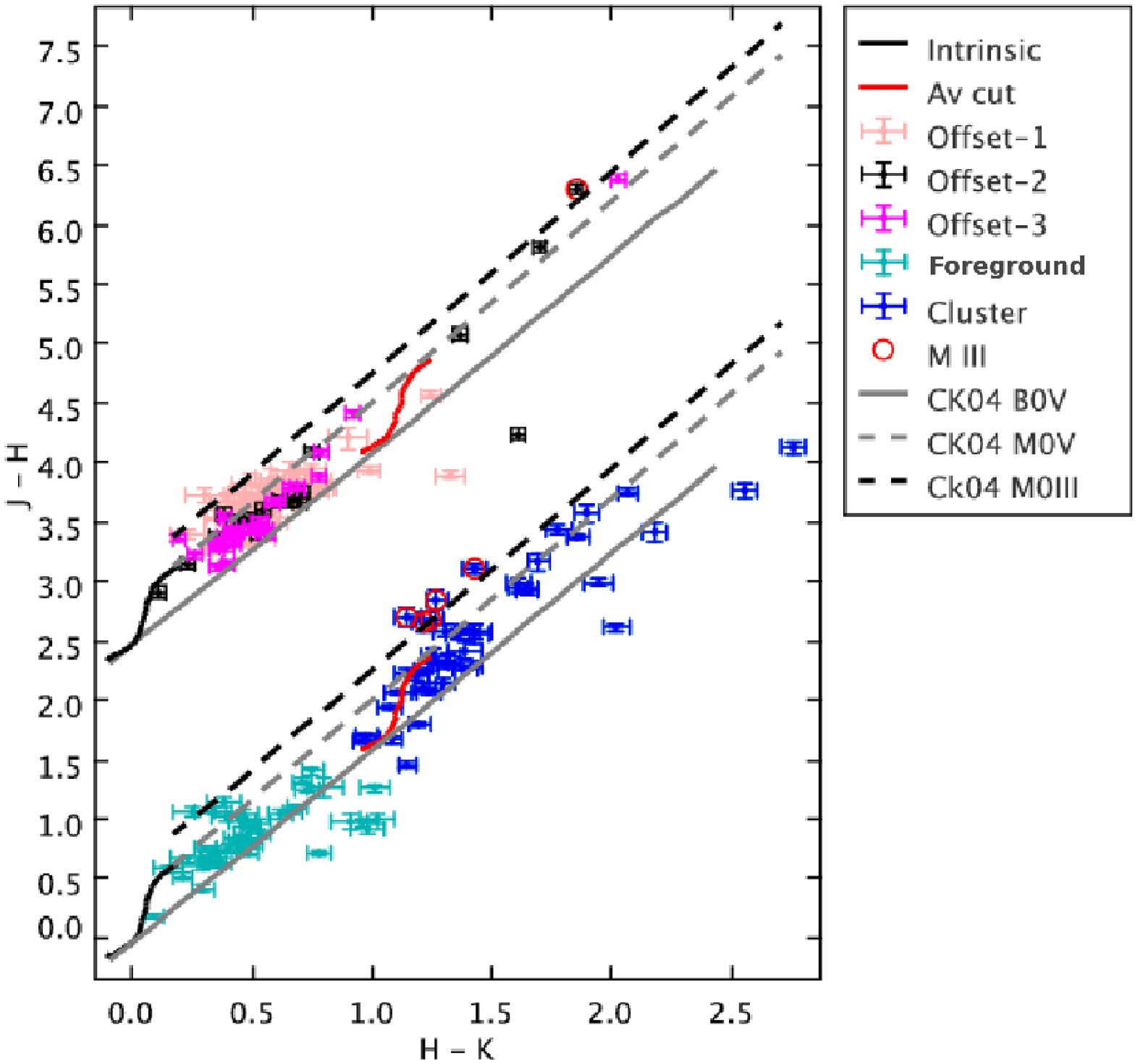}}
\end{tabular}
\caption[]{\small a (Left): An H-K vs J-H CCD of the 101 UKIDSS detections from within the 2$\sigma$ contour, surrounding the RMS source G048.9897-0.2992 (see Fig. \ref{fig:G048.9897_SpatialMap} (a)), have been plotted against stars from two offset fields that have been offset along the J-H axis by 2.5. An intrinsic colour curve has been reddened to A$_V$$\sim$11 \citep[this value has been obtained from the line of sight extinction measurements made in][]{stead10} and is used to split the cluster candidates, and offset fields, into two different populations. b (Right): Same as (a) except for the RMS source G010.1615-0.3623. The 86 cluster candidates, that have been extracted from within the 3$\sigma$ contour surrounding the RMS source G010.1615-0.3623 (see Fig. \ref{fig:G048.9897_SpatialMap} (b)), have been plotted alongside stars from three offset fields that are offset along the J-H axis by 2.5. An intrinsic colour curve has been placed at H-K=0.95 and used to split the cluster candidates, and offset fields, into two different populations. 5 cluster candidates and 1 offset field star possess atypical stellar colours, as they lie above the M0V reddening track. An M0III track has also been plotted suggesting that these stars are late-type giant stars (see text). Two of the offset field stars have colours consistent with IRXS sources and have therefore been removed from the sample.}
\label{fig:offsetVSclusterG48} 
\end{center}
\end{figure*}

\begin{figure*}
\begin{center}
\begin{tabular}{cc}
\resizebox{70mm}{!}{\includegraphics[angle=0]{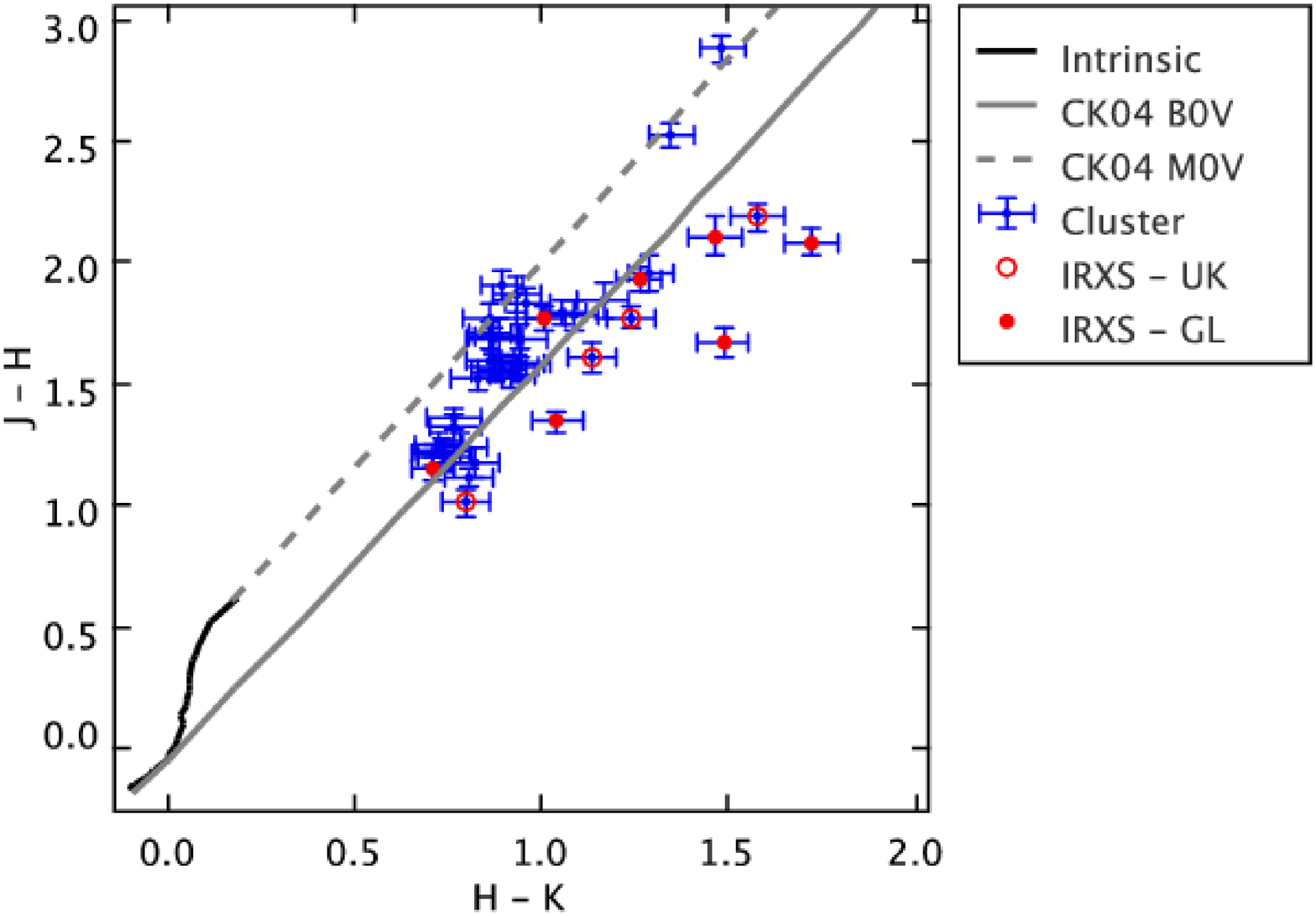}}&
\resizebox{70mm}{!}{\includegraphics[angle=0]{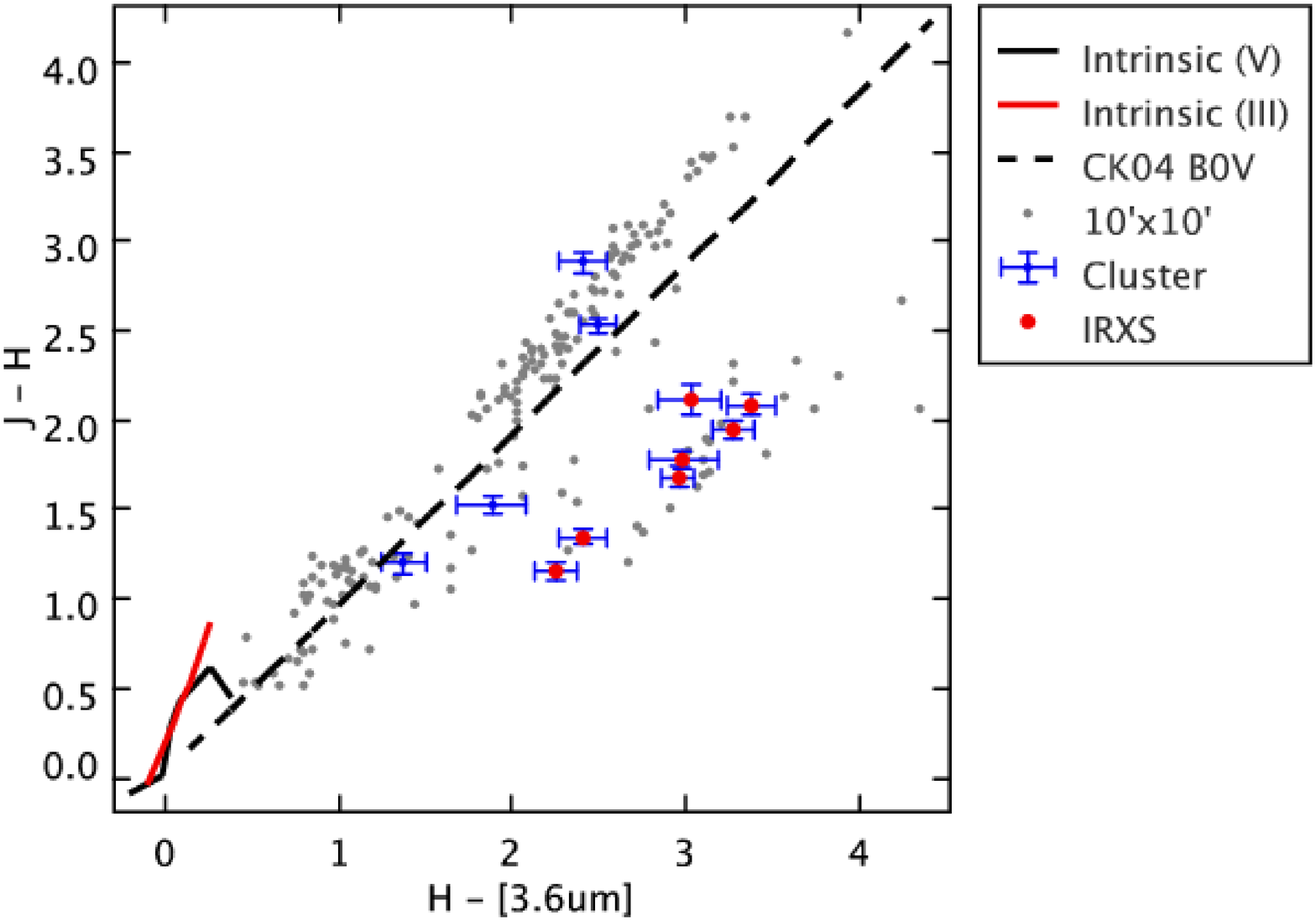}}
\end{tabular}
\caption[]{\small a (Left): An H-K vs J-H CCD of the remaining 44 cluster candidates surrounding G048.9897-0.2992 (blue error bars). Sources below the B0V reddening track are identified as IRXS candidates (open red circles). The 7 sources that were identified as IRXS candidates using GLIMPSE data in (b) are shown as closed red circles. b (Right): An H-[3.6$\mu$m] vs J-H CCD of the most reliable UKIDSS (err$\le$0.05 mag) and GLIMPSE (err$\le$0.08 mag) data (grey points), extracted from a 10$\arcmin$x10$\arcmin$ region centred on the RMS source, highlighting the general shape of the field population on a UKIDSS-GLIMPSE CCD. A total of 11 GLIMPSE sources were extracted from within the 2$\sigma$ contour (blue error bars), of these 11 sources, 7 stand below the B0V reddening track and have therefore been identified as IRXS candidates. Also plotted are the CK04 III and V intrinsic colours.}
\label{fig:IRXS_CCD_G48}
\end{center}
\end{figure*}

\begin{figure*}
\begin{center}
\begin{tabular}{cc}
\resizebox{70mm}{!}{\includegraphics[angle=0]{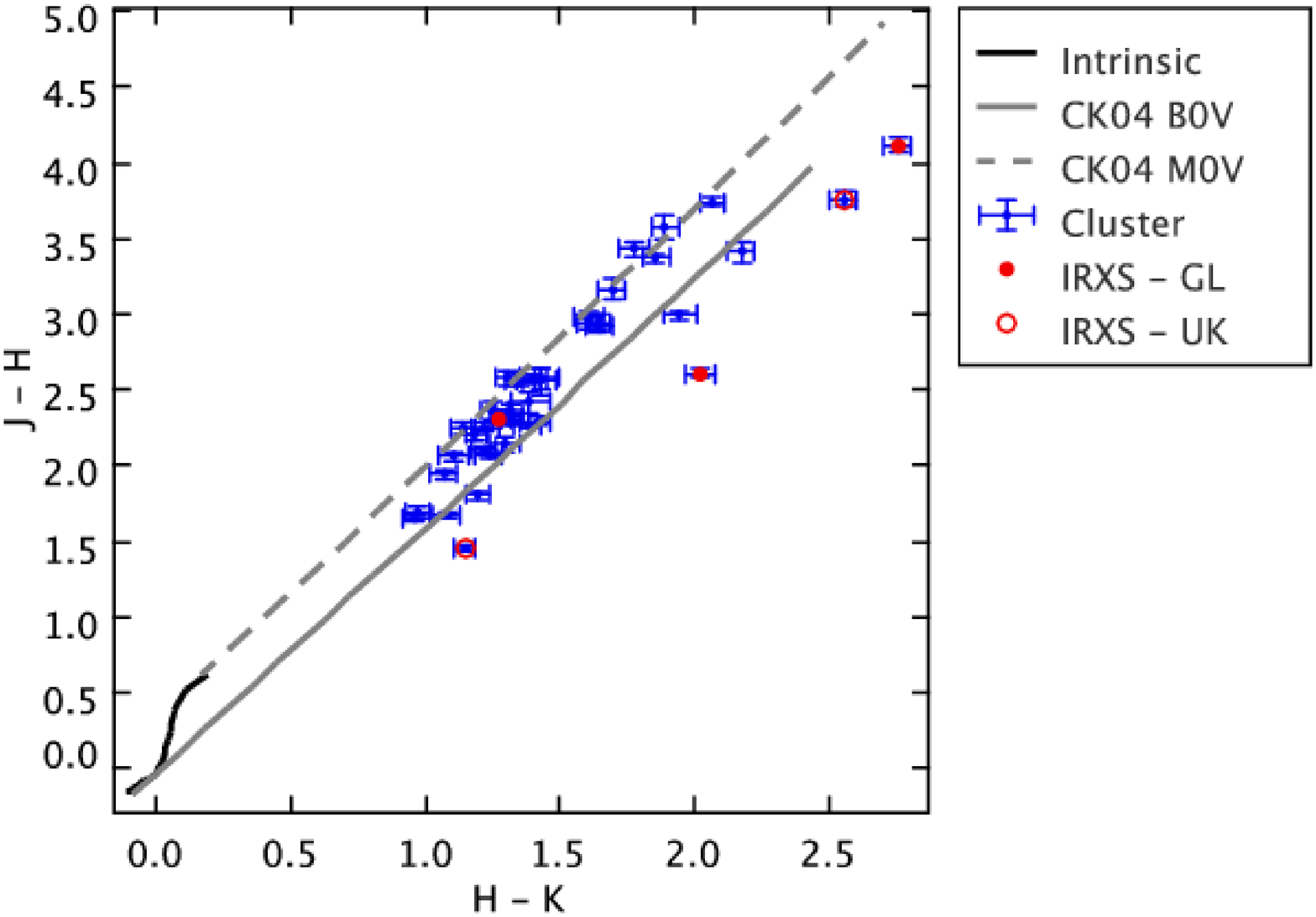}}&
\resizebox{70mm}{!}{\includegraphics[angle=0]{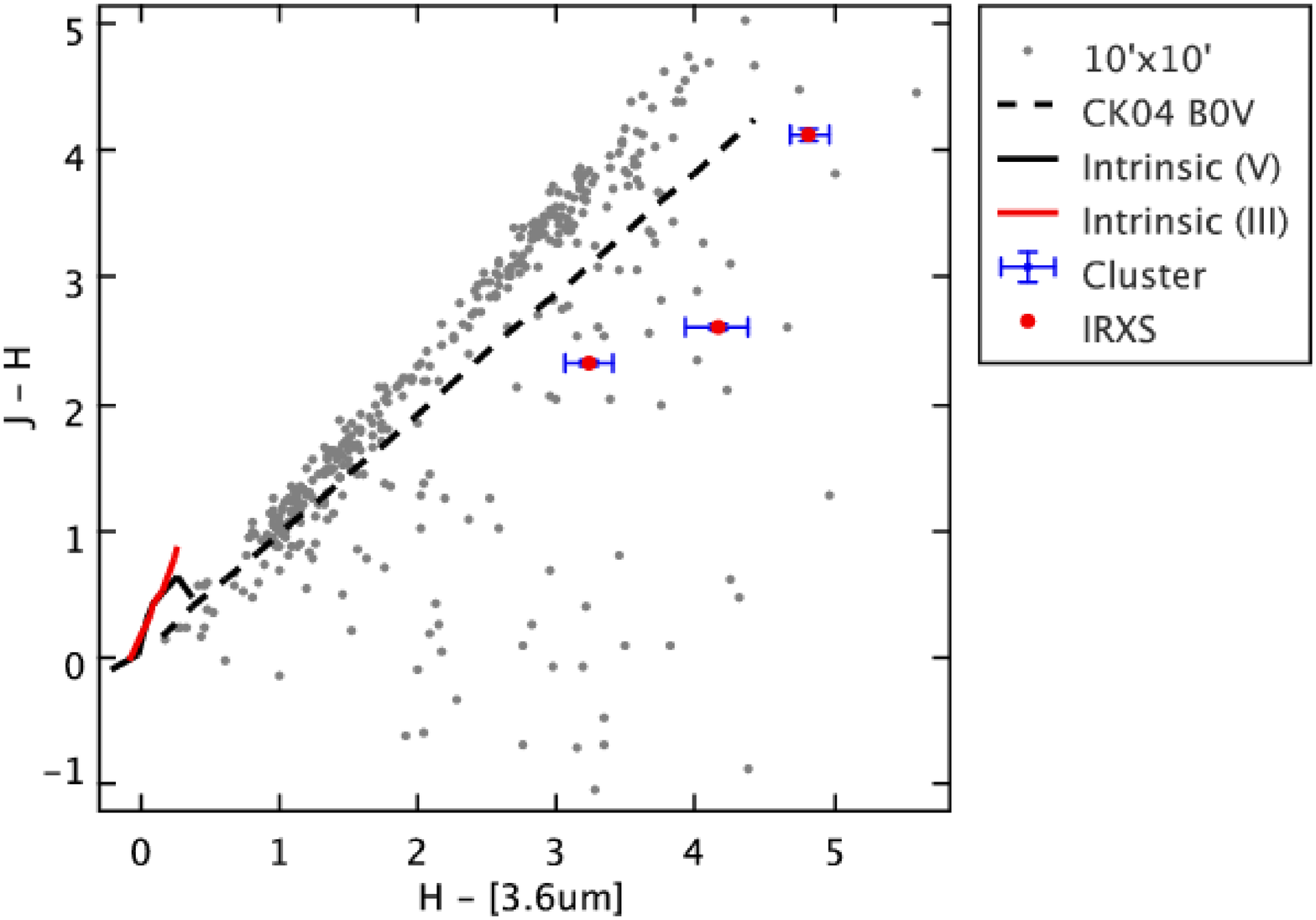}}
\end{tabular}
\caption[]{\small a (Left): An H-K vs J-H CCD of the remaining 40 cluster candidates surrounding G010.1615-0.3623 (blue error bars). Sources below the B0V reddening track are identified as IRXS candidates (open red circles). The 3 sources that were identified as IRXS candidates using GLIMPSE data in (b) are shown as closed red circles. b (Right): An H-[3.6$\mu$m] vs J-H CCD of the most reliable UKIDSS (err$\le$0.05 mag) and GLIMPSE (err$\le$0.08 mag) data (grey points), extracted from a 10$\arcmin$x10$\arcmin$ region centred on the RMS source, highlighting the general shape of the field population on a UKIDSS-GLIMPSE CCD. A total of 3 GLIMPSE sources were extracted from within the 3$\sigma$ contour (blue error bars). All of these sources stand below the B0V reddening track and have therefore been identified as IRXS candidates. Also plotted are the CK04 III and V intrinsic colours.}
\label{fig:IRXS_CCD_G10}
\end{center}
\end{figure*}

As mentioned in the figure captions.

\appendix
\section{The New Main Sequence Intrinsic Colours}
\label{sec:Appendix}

Tables \ref{tabz:mainsequence} contain the new main sequence intrinsic colours that have been derived in this paper. 

\begin{table*}
\centering
\caption[]{Main Sequence Intrinsic Colours: O8V to M0V}
\begin{tabular}{ c | c c c c c c c}
\hline
Spectral Type&\# of stars&H-K&J-H&H-K&J-H&e(H-K)&e(J-H)   \\
&per subtype&(UKIDSS)&(UKIDSS)&(2MASS)&(2MASS)&& \\
\hline
O8V	&6	&	--0.10	&	--0.19	&	--0.10	&	--0.21	&	0.017	&	0.019	\\
O9V	&23	&	--0.10	&	--0.15	&	--0.09	&	--0.17	&	0.009	&	0.011	\\
B0V	&24	&	--0.10	&	--0.17	&	--0.09	&	--0.19	&	0.008	&	0.010	\\
B1V	&54	&	--0.09	&	--0.16	&	--0.08	&	--0.17	&	0.006	&	0.008	\\
B2V	&105	&	--0.06	&	--0.14	&	--0.06	&	--0.15	&	0.004	&	0.005	\\
B3V	&101	&	--0.05	&	--0.12	&	--0.05	&	--0.13	&	0.004	&	0.005	\\
B4V	&35	&	--0.04	&	--0.10	&	--0.04	&	--0.11	&	0.007	&	0.008	\\
B5V	&125	&	--0.03	&	--0.10	&	--0.03	&	--0.11	&	0.004	&	0.004	\\
B6V	&70	&	--0.02	&	--0.09	&	--0.02	&	--0.10	&	0.006	&	0.006	\\
B7V	&85	&	--0.02	&	--0.08	&	--0.02	&	--0.09	&	0.005	&	0.006	\\
B8V	&240	&	--0.02	&	--0.08	&	--0.01	&	--0.09	&	0.003	&	0.003	\\
B9V	&351	&	0.00	&	--0.06	&	0.00	&	--0.06	&	0.002	&	0.003	\\
A0V	&195	&	0.00	&	--0.04	&	0.00	&	--0.04	&	0.003	&	0.004	\\
A2V	&353	&	0.02	&	--0.02	&	0.02	&	--0.02	&	0.002	&	0.003	\\
A5V	&215	&	0.03	&	0.02	&	0.03	&	0.03	&	0.003	&	0.004	\\
F0V	&232	&	0.04	&	0.09	&	0.04	&	0.10	&	0.003	&	0.004	\\
F2V	&245	&	0.03	&	0.13	&	0.04	&	0.14	&	0.003	&	0.004	\\
F5V	&231	&	0.05	&	0.16	&	0.05	&	0.18	&	0.003	&	0.004	\\
F8V	&146	&	0.05	&	0.19	&	0.06	&	0.21	&	0.004	&	0.005	\\
G0V	&115	&	0.06	&	0.22	&	0.07	&	0.25	&	0.005	&	0.006	\\
G2V	&95	&	0.06	&	0.24	&	0.07	&	0.27	&	0.005	&	0.006	\\
G5V	&141	&	0.06	&	0.28	&	0.07	&	0.31	&	0.005	&	0.006	\\
G8V	&171	&	0.06	&	0.35	&	0.07	&	0.38	&	0.004	&	0.005	\\
K0V	&136	&	0.07	&	0.37	&	0.08	&	0.40	&	0.005	&	0.006	\\
K2V	&47	&	0.08	&	0.40	&	0.09	&	0.44	&	0.008	&	0.010	\\
K5V	&20	&	0.12	&	0.50	&	0.13	&	0.55	&	0.014	&	0.019	\\
M0V	&12	&	0.19	&	0.60	&	0.21	&	0.67	&	0.015	&	0.019	\\
\hline
\label{tabz:mainsequence}
\end{tabular}
\end{table*}

\section*{Acknowledgements}

We would like to thank our referee M.S. Bessel who greatly contributed to this paper through his expertise with infrared intrinsic colours. Bessel investigated the intrinsic colours of different isochrones using correct log g to show that it only affects M stars. This work is based in part on
data obtained as part of the UKIRT Infrared Deep Sky Survey. We made use of the VizieR service (http://vizier.u-strasbg.fr/viz-bin/VizieR) to obtain 2MASS data and the M06 extinction distributions. (The reddening tracks used in this paper can be downloaded at www.ast.leeds.ac.uk/RMS/ReddeningTracks/)

\bibliographystyle{mn2e}
\bibliography{Joey}{}

\bsp

\label{lastpage}

\end{document}